%% file: main.tex
\pdfoutput=1
\documentclass[10pt]{article}

\usepackage[T1]{fontenc}
\usepackage[a4paper,margin=25mm]{geometry}
\usepackage[numbers,sort&compress]{natbib}
\usepackage{authblk}

\usepackage{graphicx}%
\usepackage{multirow}%
\usepackage{amsmath,amssymb,amsfonts}%
\usepackage{amsthm}%
\usepackage{mathrsfs}%
\usepackage[title]{appendix}%
\usepackage{xcolor}%
\usepackage{textcomp}%
\usepackage{manyfoot}%
\usepackage{booktabs}%
\usepackage{algorithm}%
\usepackage{algorithmicx}%
\usepackage{algpseudocode}%
\usepackage{listings}%
\usepackage{float}
\usepackage{hyperref}
\usepackage{cleveref}
\usepackage{array}
\usepackage{longtable}
\usepackage{ragged2e}
\usepackage{makecell}
\usepackage{tcolorbox}
\usepackage{mdframed}
\usepackage[table]{xcolor}
\usepackage{subcaption}

\hypersetup{colorlinks=true,linkcolor=blue,citecolor=blue,urlcolor=blue}
\makeatletter

\renewcommand{\abstract}[1]{\gdef\dts@abstract{#1}}
\newcommand{\keywords}[1]{\gdef\dts@keywords{#1}}
\newcommand{\printabstract}{%
  \begin{dtsabstractenv}\dts@abstract\par\medskip
    \noindent\textbf{Keywords:} \dts@keywords
  \end{dtsabstractenv}}
\makeatother
\renewcommand{\onecolumn}{}
\renewcommand{\twocolumn}[1][]{#1}

\lstdefinelanguage{oml}{
  morekeywords={instance, concept, relation, entity, aspect, scalar, property, from, to, forward, reverse, domain, range, functional},
  sensitive=true,
  alsoletter={:,_,-},
  morecomment=[l]{//},
  morestring=[b]",
  morekeywords=[2]{DTDFVocab:Service, DTDFVocab:Enabler, DTDFVocab:Model, DTDFVocab:Data, DTDFVocab:Input, DTDFVocab:Standardization, DTDFVocab:Insight, DTDFVocab:Action, DTDFVocab:DataTransmitted, DTDFVocab:SensingComponent, DTDFVocab:ActingComponent, Service, ProvidedThing,  Enabler, Model, Data, Input, Standardization, DataTransmitted, SensingComponent, ActingComponent, VirtualToPhysical, ConnectionComponent, DTComponent, TimeScaleThing, AutomaticEnabledThing, base:DescribedThing},
  morekeywords=[3]{DescribedThing, IsCommEnabler, base:desc, xsd:boolean, desc, enables, enabledBy, producedData, producedFrom, inputTo, hasInput, DTDFVocab:enables, DTDFVocab:enabledBy, DTDFVocab:inputTo, DTDFVocab:fromData, DTDFVocab:IsAutomatic, DTDFVocab:hasInput, DTDFVocab:producedFrom, DTDFVocab:producedData, provides, providedBy, DTDFVocab:provides, DTDFVocab:atStage, asData, fromData},
  morekeywords=[4]{<, [, ], :},
  literate={,}{{,}}{1}
}

\newtcolorbox{resultbox}[1][]{
  colback=gray!10,
  colframe=black,
  fonttitle=\bfseries,
  title=#1,
  sharp corners,
  boxrule=0.5pt
}

\newmdenv[
  topline=false,
  bottomline=false,
  rightline=false,
  linewidth=3pt,
  linecolor=gray,
  innerleftmargin=10pt,
  innertopmargin=0pt,
  innerbottommargin=0pt
]{leftbarbox}

\definecolor{highgreen}{RGB}{210, 245, 210}
\definecolor{lowred}{RGB}{255, 215, 215}
\definecolor{goodblue}{RGB}{220, 230, 255}
\definecolor{badorange}{RGB}{255, 240, 220}

\newcommand{\best}[1]{\cellcolor{goodblue}\textbf{#1}}

\title{DTScribe: Generating Models for Digital Twin Reporting using Local Large Language Models}

\author[1,*]{K\'{e}rian Fiter}
\author[2]{Angelica Portocarrero}
\author[1]{Bentley Oakes}

\affil[1]{Département de Génie Informatique et Génie Logiciel, Polytechnique Montr\'{e}al, 2500, chemin de Polytechnique, Montréal, H3T 1J4, Québec, Canada}
\affil[2]{Facultad de Ingeniería y Ciencias Básicas, Universidad Autónoma de Occidente, Carrera 25 \#115-85, Santiago de Cali, 760030, Colombia}
\affil[*]{Corresponding author: \href{mailto:kerian.fiter@polymtl.ca}{kerian.fiter@polymtl.ca}}
\affil[ ]{Contributing authors: \href{mailto:maria_a.portocarrero@uao.edu.co}{maria\_a.portocarrero@uao.edu.co}; \href{mailto:bentley.oakes@polymtl.ca}{bentley.oakes@polymtl.ca}}

\date{}

\begin{document}

\input{text/abstract}
\maketitle
\printabstract

\input{text/intro}

\input{text/background}

\input{text/approach}

\input{text/metrics}

\input{text/evaluation}

\input{text/results}

\input{text/discussion}

\input{text/related}

\input{text/conclusion}

\input{text/acknowledgment}

\bibliography{main}

\input{text/appendix}

\end{document}

%% file: text/abstract.tex
\abstract{
    The development of a digital twin brings together many stakeholders and technical components. We have previously investigated the reporting of the conceptual architecture for these digital twins as an artifact for discussion. However, we find that the activity of modeling the digital twin for visualization and consensus-building can be quite tedious and time-consuming.
    
    In this article, we explore the automation of model generation using large language models, targeting the model required for digital twin reporting. Our approach is designed to ingest digital twin documents and output both the conceptual architecture visualization and a reporting page for that digital twin. We first extract from the input the key details aligning with our twenty-one-characteristic Digital Twin Description Framework described in previous work. Then, we generate a description model in the Ontological Modeling Language, representing the digital twin as a constellation of models, data, enablers, and services. We utilize refinement loops with large language models acting as judges and ontological reasoning to increase artifact quality as measured by our metrics.
    
    We investigate how architectural and parameter choices affect artifact quality, such as how refinement loops enable local large language models to produce sufficiently high-quality results comparable to larger models. We also detail insights and lessons learned useful for practitioners, such that they can integrate our techniques to accelerate their digital twin reporting or for other model generation tasks. These insights include the use of medium-sized local models (around 20 billion parameters), the avoidance of large language models as judges and clustering for characteristic extraction, and the recommendation of at least two reasoner loops for Ontological Modeling Language refinement.
}

\keywords{Digital Twins, ontologies, ontology engineering, Large Language Models, reporting, Ontological Modeling Language, model generation, ontology refinement loop}

%% file: text/intro.tex
\section{Introduction}
\label{sec:intro}

Digital Twins (DTs) are digital representations of physical systems, enabling monitoring, insight generation, and control of their physical counterparts through real-time, bidirectional data exchange~\cite{madni2019leveraging}. Their use spans multiple industries, from manufacturing to construction, healthcare, city planning, and robotics. Academic research is also increasingly engaging with DTs, recognizing them as a rapidly expanding area of study, although many challenges remain~\cite{combemale2025engineering}. 
In particular, DTs are complex software systems often integrated across entire life-cycles, from design and manufacturing to operation and decommissioning. DTs and their associated physical systems---which we will call the Physical Twin (PT)---also tend to continually evolve~\cite{alskaif2025evolution}, adding to the complexity of those systems.

To manage this complexity, stakeholders require continuous, detailed reporting of their characteristics and capabilities to understand the structure and behavior of DTs at any time~\cite{gilSystematicReportingFramework2024}. DT reporting remains a time-consuming, unstructured process, hindering experts from reaching consensus on definitions and non-experts from easily understanding the system~\cite{Oakes2021improving,goffi2025beerdt}.
A common system view facilitates interdisciplinary collaboration, improves cross-team communication, and reduces technical ambiguity.

\subsection{Challenges}

In our previous work, we introduced DTInsight~\cite{fiter2025dtinsight}, a tool to assist DT reporting. DTInsight enables explicit reporting through ontology-based representations, interactive reporting via conceptual architecture generation combined with data monitoring and 3D visualization, and continuous reporting by automatically generating reporting pages within a CI/CD pipeline. We proposed modeling DTs using our twenty-one characteristic DT Description Framework (DTDF)~\cite{gilSystematicReportingFramework2024}, formalized as an ontology. A Godot engine–based visualizer then queries this model to produce a conceptual visualization of the DT’s structure, of its data flows, and of selected aspects of its behavior.

However, through our discussions with DT practitioners and researchers, there are still numerous barriers to widespread use of DTInsight and DT reporting in general. More precisely, we aim to address the following challenges:

\paragraph{Challenge 1: Slowness and difficulty of modeling domain knowledge for reporting}

Small and Midsize Enterprises (SMEs) and researchers have limited experience and time to devote to domain knowledge modeling---and its formal, logic-based subset called Ontology Engineering (OE)---as it is slow, labor-intensive, and reliant on expert modelers~\cite{jin2024evaluation,burgueno2025automation,chenAutomatedDomainModeling2023a,noy1997state}.

Furthermore, while recent research has focused on Ontology Learning (OL)~\cite{du2024short,saeedizade2024navigating,fathallah2024neon,lippolis2025ontology,zhang2024ontochat,babaei2023llms4ol,shimizu2025accelerating} (the automatic discovery of a domain's \textit{ontological structure}), little work has been done on \textit{instance model generation}~\cite{pan2025llm} (the automatic generation of \textit{ontology instances} conforming to a certain metamodel).

\paragraph{Challenge 2: Privacy and sovereignty concerns}

DT reporting capabilities are often limited by data privacy and sovereignty requirements in industrial sectors~\cite{gil2025fmi,jacoby2021approach}. DT documentation often contains sensitive or proprietary data that cannot be processed through cloud-based infrastructures. Consequently, there is a need for solutions that operate \textit{locally} under moderate hardware constraints, to align with the typical resources and technical expertise available in SMEs~\cite{imanifard2025smartgridagent}.

Here, ``\textit{locally}'' refers to processing documentation on organizational premises, including desktops, laptops, and on-premise servers, favoring a `fog-based' rather than `cloud-based' approach.

While our intended deployment relies on local, privacy-preserving infrastructure, this paper's empirical evaluation used cloud resources to achieve statistical significance across numerous experiments. However, deploying open-weight models allowed us to scale the evaluation while mirroring the capabilities of models executing on-premise in real-world deployments.

\paragraph{Challenge 3: Closing the gap between static documents and continuous reporting for non-technical personnel}

To deliver insights to diverse stakeholders throughout the DT life-cycle, reporting must be accessible to non-technical personnel. This necessitates an architecture capable of generating multi-view representations~\cite{sommaTwinArchDigitalTwin2026}, allowing users with varying expertise to interpret results within agile processes in a TwinOps style~\cite{huguesTwinOpsDevOpsMeets2020}.

While we partially addressed these requirements through DTInsight and the DTDF, a significant gap remains between static project documentation and the dynamic generation of reports. Specifically, current methods struggle to align reports with evolving DT data and processes~\cite{alskaif2025evolution}, particularly when reports are not treated as first-class artifacts within CI/CD pipelines.

\subsection{Proposed Approach}

We present the \textit{DTScribe} approach to these challenges:

\begin{description}
    \item [Addressing Challenge 1]: We tackle the slowness and difficulty of modeling by \textit{leveraging Large Language Models (LLMs) to extract as much of the twenty-one DTDF characteristics as possible from unstructured user-provided documents}. We use those characteristics to then generate an OML description model within an LLM-based ontology repair loop.\\
    \item [Addressing Challenge 2]: We tackle the privacy and sovereignty concerns by targeting \textit{open-source LLMs running on constrained, local infrastructure}. Unlike larger proprietary, cloud-hosted LLM services that rely on extensive context windows to process documents, our approach optimizes performance for smaller, privacy-preserving models. We achieve this through LLM-based evaluation for characteristic extraction and LLM-based ontology repair for OML description quality.\\
    \item [Addressing Challenge 3]: We target non-domain experts in DT reporting by \textit{generating an ontology} compatible with the DTInsight tool~\cite{fiter2025dtinsight}. Finally, we tackle CI/CD by \textit{making it easy to integrate into the continuous pipeline} of DTInsight.
\end{description}

While academic papers contain human-readable structures, they lack the formal semantics needed to automate DT reporting. Given that industrial DT data is often fragmented, using academic papers establishes an initial extraction and formalization baseline before tackling unstructured industrial datasets.

\paragraph{Novelty and Impact}

Our first novelty is the automatic translation of unstructured documents into formal models for DT reporting with privacy-respecting and open-source LLMs to facilitate domain knowledge modeling. Specifically, we propose a novel technique for generating OML description instance models that utilizes an OML refinement and repair mechanism. This allows our approach to match the output quality of larger LLMs.
The generated ontology integrates with DTInsight to provide continuous DT reporting to non-technical personnel. We also define quality metrics for this process, enabling the evaluation of our model generation approach.

The primary impact of this research is to provide a robust, systematic methodology and tooling that enable users to derive an initial DT conceptual model from artifacts such as documentation. This model then serves as the foundation for reporting throughout the DT life-cycle. That is, the user (or other LLM agents) could make smaller edits to this model to keep it aligned with the DT as it evolves. 

\subsection{Research Questions}

This papers answers the following Research Questions (RQs):

\begin{description}
    \item[RQ1:] What is a \textbf{technical architecture with local LLMs} to enable the extraction of DT characteristics and the generation of valid OML description models?
    \item[RQ2:] What \textbf{metrics and methodologies} are required to effectively evaluate completeness and accuracy for LLM-extracted DT characteristics?
    \item[RQ3:] To what extent do iterative reasoner-guided repair and LLM-as-a-judge \textbf{improve the structural validity of the generated OML models} compared to zero-shot extraction?
\end{description}

\subsection{Contributions and Structure}

This paper makes the following contributions:

\begin{enumerate}
    \item \textbf{Automated Reporting Framework:} A novel approach (available on GitHub\footnote{\url{https://github.com/oakeslabmtl/DTScribe}}) for translating narrative-driven unstructured documents containing architectural descriptions into formal and queryable OML instance models using local and privacy-respecting LLMs.
    \item \textbf{Feedback Architecture:} A technical architecture that employs iterative refinement loops to overcome the limitations of smaller, local models. These loops are 1) LLM-as-a-judge for characteristic refinement and, 2) reasoner-guided repair for ontology generation.
    \item \textbf{Empirical Evaluation:} An evaluation of the approach across multiple LLMs, analyzing the trade-offs between model size, deployment modality, and the quality of the generated reports and models.
    \item \textbf{Tool Integration:} The integration of the generated artifacts into the DTInsight tool to bridge the gap between static documentation and continuous DT reporting.
\end{enumerate}

An example of a reporting page generated by our approach for the exemplar paper~\cite{Ramdhan2025EngineeringAD} can be found here\footnote{\url{https://oakeslabmtl.github.io/dtdf-dtscribe/}}.

This article provides actionable insights for researchers and practitioners making architectural decisions regarding local LLMs for DT reporting. Specifically, we evaluate the critical trade-offs between computational cost and extraction accuracy, highlighting the surprising pitfalls of document clustering and the efficacy of iterative, reasoner-guided repair. Consequently, we provide concrete recommendations on optimal model size, the use of LLM-as-a-judge mechanisms, and strategies for reliable OML generation.

The remainder of this paper is organized as follows. \Cref{sec:background} reviews the necessary background. \Cref{sec:approach} details our proposed approach. \Cref{sec:metrics} defines the evaluation metrics, while \Cref{sec:evaluation} describes the experimental setup. \Cref{sec:results} presents and analyzes the empirical results. \Cref{sec:discussion} discusses implications, limitations, and threats to validity. \Cref{sec:related_work} situates our contributions within the broader literature. Finally, \Cref{sec:conclusion} concludes the paper.

%% file: text/background.tex
\section{Background}
\label{sec:background}

This section discusses the key concepts and technological foundations underlying DT construction, reporting, and ontology-based modeling. We focus on the emerging role of Large Language Models (LLMs) and Retrieval-Augmented Generation (RAG) in supporting these processes.

\subsection{DT Reporting with DTInsight}

DT Engineering (DTE) increasingly relies on Model-Driven Engineering (MDE)~\cite{michaelModelDrivenEngineeringDigital2025}, defined as ``the use of software models to design, analyze, generate, and document software''~\cite{liebel2024human}. Inspired by the use of ontologies for DT construction and architectural definition~\cite{karabulut2024ontologies,ZhengEmergence2022,oakes2024towards}, we propose leveraging ontologies for DT reporting using LLMs. This approach aligns DTE with formal ontological representations and the emerging automation capabilities enabled by Foundation Models (FM) and Generative AI~\cite{luFoundationModelsSoftware2025,liChatTwinAutomatedDigital2023,Gebreab2024DTGenAI,giabbanelli2025over,giabbanelli2026guidelargelanguagemodels,cheng2025generative}.

\paragraph{DT Reporting}

This convergence of models, ontologies, and AI~\cite{marcus2020next} contrasts with the current state of \textit{DT reporting}, that we define as the systematic documentation and communication of a DT's architecture, capabilities, operational state, and life-cycle to stakeholders. DTs remains under-specified~\cite{gilSystematicReportingFramework2024}, and especially lack continuous reporting of the full DT life-cycle (TwinOps~\cite{huguesTwinOpsDevOpsMeets2020}).
Existing standards and frameworks have attempted to structure DT descriptions, but often with differing concerns. For instance, ISO 23247~\cite{iso2021iso,heithoff2025method} provides a reference architecture for manufacturing DTs, recognizing reporting as a functional entity, but lacking details. Similarly, a DT description mapping to the Asset Administration Shell (AAS) offers a standardized digital representation for assets~\cite{Oakes2021improving,oakes2021digital}, but is biased toward operational descriptions. In contrast, our approach relies on the DT Description Framework (DTDF)~\cite{gilSystematicReportingFramework2024}, which provides a domain-agnostic foundation to bridge the communication gap between diverse stakeholders.\\

\paragraph{DT Description Framework (DTDF)}

To structure this domain, the DT Description Framework~\cite{gilSystematicReportingFramework2024} characterizes DTs across twenty-one dimensions. These range from functional aspects like \textit{DT services (C6)}, which defines capabilities such as simulation or monitoring, to lifecycle management via the \textit{Twinning process and DT evolution (C13)}, and governance concerns regarding \textit{Data ownership and privacy (C19)}. The framework facilitates consensus among diverse stakeholders enabling them to define each system aspect in a few targeted sentences and \textit{constellation diagram}~\cite{goffi2025beerdt}.

First introduced by Oakes \textit{et al.}~\cite{Oakes2021improving}, the \textit{DT constellation} is conceptually similar to a traditional software component-and-connector diagram. It serves as a specialized architectural view that maps the DT, illustrating the digital and physical entities (components) and their data exchanges (connectors). This conceptual structure was subsequently formalized into an ontology, as detailed in the following sections.

\paragraph{DTInsight}

\begin{figure*}[htbp]
    \centering
    \includegraphics[width=1\linewidth]{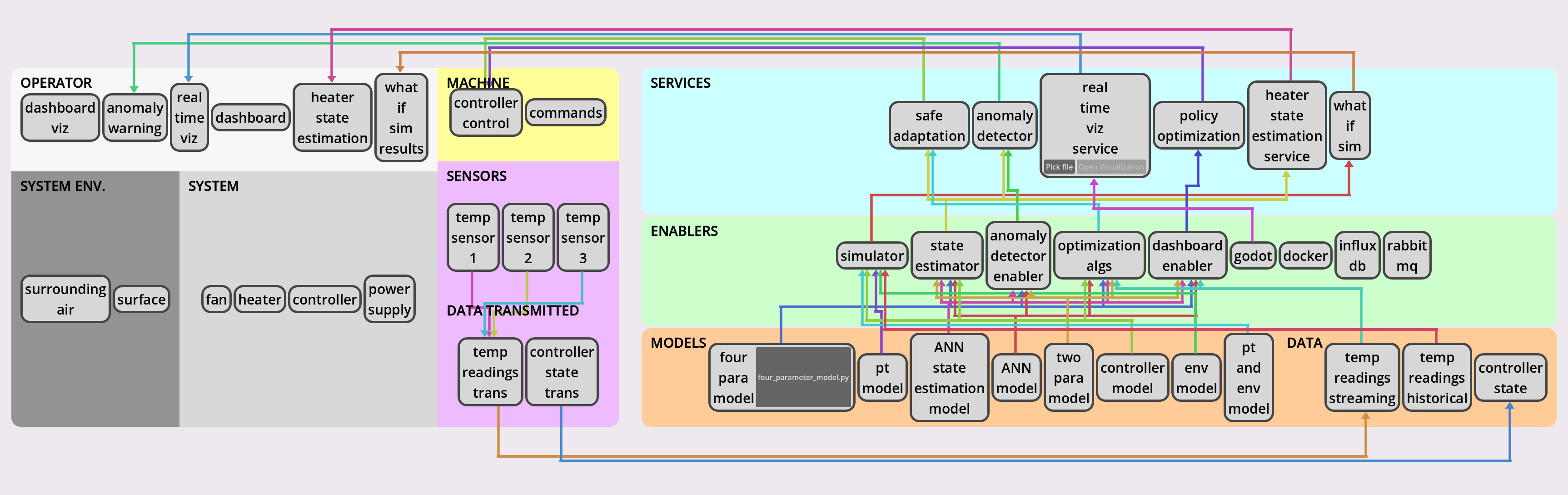}
    \caption{The incubator~\cite{feng2021incubator} DT \textit{constellation}~\cite{gilSystematicReportingFramework2024} automatically generated by DTInsight from the (hand-built) OML description. Interactive reporting page available at \url{https://oakeslabmtl.github.io/DTDF/}.}
    \label{fig:dt_constellation}
\end{figure*}

Our tool DTInsight~\cite{fiter2025dtinsight} addresses the lack of structured, automated reporting by leveraging ``the symbiotic relationship between MDE and DevOps''~\cite{escottJidokaAutomationHuman2025}.
It integrates ontology-based DT descriptions into a continuous reporting pipeline to generate two primary artifacts: a web-based reporting page\footnote{\url{https://oakeslabmtl.github.io/DTDF}} and a visual \textit{DT constellation} (\Cref{fig:dt_constellation}). These artifacts are designed to serve as an initial draft for discussion and iterative modification, enabling stakeholders to grasp the current characteristics and conceptual architecture of the DT. The interactive constellation allows users to explore data flows between components, view underlying scripts, and visualize real-time sensor data through graphs or 3D models.

By integrating DTInsight into a continuous TwinOps pipeline, the generated reporting artifacts are not treated as static but can be automatically regenerated as the system evolves over time. This ensures that dynamic updates to the DT’s underlying architecture, newly added components, or modified data flows are continuously reflected in the reporting model, keeping the documentation synchronized with the operational twin.

DTInsight models the DTDF within an ontology (publicly available here\footnote{\url{https://github.com/oakeslabmtl/DTDF}}) written in the Ontological Modeling Language (OML)~\cite{elaasarOpenCAESARBalancingAgility2023}, which is described later. If the generated OML description conforms to the DTDF vocabulary, the tool automatically generates the \textit{DT constellation} and the reporting page.
However, the creation and maintenance of the underlying DTDF ontology currently remain largely manual and effort-intensive, limiting the scalability and broader adoption of ontology-driven DT reporting.

\paragraph{Running Example: Incubator Case Study}

Following the DTInsight article~\cite{fiter2025dtinsight}, we adopt the incubator DT case study~\cite{feng2021incubator} as a running example throughout this work. The objective of this system is to regulate the temperature within an enclosed box by controlling a heater, based on measurements collected from multiple temperature sensors and a corresponding thermal model. 

In DTInsight, this case study was selected due to the simplicity of its control objective combined with the complexity of its behavior, which covers multiple domains including electrical, thermal, and mechanical. \Cref{fig:dt_constellation} illustrates the DTInsight constellation representing the incubator DT.

The original publication of the case study spans eighteen pages and covers multiple aspects of the DT's realization such as hardware components, software architecture, model definition, calibration procedures, and operational goals. Notably, this description is provided without relying on any standardized reporting framework. This lack of structure makes the incubator DT an especially suitable candidate for our study, as it allows us to evaluate the feasibility of automatically deriving a DT reporting model in accordance with the DTDF.

\subsection{Ontologies and OML}

An ontology is ``an explicit specification of a conceptualization''~\cite{gruber1993translation}.
Here, we write our ontology in the Ontological Modeling Language (OML) using the openCAESAR platform~\cite{elaasarOpenCAESARBalancingAgility2023} from NASA’s Jet Propulsion Laboratory (JPL).

OML is increasingly being considered for Model-Based System Engineering (MBSE)~\cite{humphries2025integrated}, especially in the case of DTE~\cite{oakes2024towards}. This is because OML can combine the expressiveness of formal ontologies with practical MBSE workflow integration, providing precise semantic definitions and automated consistency checking while enabling both human-readable specifications and machine-processable reasoning. 

OML provides a Domain-Specific Language (DSL) for ontology modeling, distinguishing between two complementary layers: an OML \textit{vocabulary}, which defines domain concepts, constraints, and their relations (similar to an M1 metamodel), and an OML \textit{description}, which instantiates these concepts using the vocabulary (similar to a M0 model). A slice of the DTDF OML vocabulary is presented in \Cref{lst:dtdf_vocab_oml}, and its instantiation as an OML description on \Cref{lst:dtdf_desc_oml_error}.

\begin{figure}[tbh]
    \begin{lstlisting}[language=oml, caption=DTDF OML vocabulary example., label=lst:dtdf_vocab_oml]
    // C2: Acting Components
	concept ActingComponent < ConnectionComponent, TimeScaleThing
    // C3: Sensing Components
	concept SensingComponent < ConnectionComponent, TimeScaleThing
    // C4: Physical to Virtual Interactions
	concept DataTransmitted < ConnectionComponent, TimeScaleThing, AutomaticEnabledThing
	relation entity ProducedData [
		from SensingComponent
		to DataTransmitted
		forward producedData
		reverse producedFrom
	]
\end{lstlisting}
\end{figure}

Our model generation approach thus touches on \textit{Ontology Learning} (OL), which is concerned with the (semi-)automatic discovery and extraction of ontological structures from raw data. The advent of LLMs has prompted substantial research interest in leveraging them to advance OL~\cite{du2024short,saeedizade2024navigating,fathallah2024neon,lippolis2025ontology,zhang2024ontochat,babaei2023llms4ol,shimizu2025accelerating}.

However, our research focuses on a similar yet distinct area: \textit{instance model generation}~\cite{wu2012metamodel}. While OL typically revolves around the discovery of conceptual structures from data, instance model generation assumes the existence of a metamodel that formally represents the abstract system structure. The goal is then to generate instances that match the structure and constraints of the metamodel. As with OL, there has been interest in using LLMs for instance model generation~\cite{pan2025llm}, but to a lesser degree.

In our work, the pre-existing metamodel is the DTDF~\cite{gilSystematicReportingFramework2024} OML vocabulary, which has been formally represented in DTInsight~\cite{fiter2025dtinsight}---with the full vocabulary publicly available online\footnote{\url{https://github.com/oakeslabmtl/DTDF}}. Here, the \textit{instance model} is the DTDF OML description.

\subsection{Large Language Models}

Large Language Models (LLMs) are transformer-based neural networks trained on vast text corpora, enabling them to perform a wide range of natural language tasks such as question answering, code generation, translation, and text summarization~\cite{patil_2024_llm_review}. However, their knowledge remains limited to the scope of the data available during training, inheriting not only factual information but also the potential biases and misconceptions present in those sources. Moreover, when processing long text sequences, LLMs may omit important contextual details, leading to incomplete or incorrect responses~\cite{kumar2024llmsurvey}. 

To address these limitations, several methods have been adopted. \textit{Retrieval Augmented Generation} (RAG)~\cite{gao2023retrieval} supplements LLMs with external domain knowledge by retrieving relevant information from a vector database at inference time. In RAG, documents are split into smaller passages (chunks), embedded into vector representations, and stored in a vector database. During inference, a user query is embedded and the $k$ most semantically similar chunks are retrieved and appended to the model’s context. 

RAG thus allows the LLM to ground its response on up-to-date, domain-specific information. This method is effective for Information Extraction (IE) tasks~\cite{zheng2023survey}, where relevant data may be spread across multiple sections of text, and for Aspect-Based Summarization (ABS), which condenses unstructured text into focused summaries on specific topics~\cite{korkankar-etal-2024-aspect}. Combining LLMs with embedding-based retrieval can reduce hallucinations by grounding the model’s responses in actual, relevant content, improving both extraction accuracy and the quality of summaries~\cite{feng2025aspect}.

LLMs often face difficulties when a prompt contains multiple rules, multi-step reasoning, and strict structured output requirements. To guide LLMs, the \textit{prompt chaining} technique is commonly employed to decompose a workflow into sequential, narrower prompts. This especially improves output quality for smaller models with a small context window. Similarly, the \textit{Chain-of-Thought} (CoT) prompting technique has been shown to improve output quality~\cite{wei2022chain} by explicitly structuring complex prompts as a step-by-step reasoning process.
However, recent research into LLMs has shown that both techniques are becoming outdated as even small LLMs now support large context windows and directly implement reasoning in their architecture~\cite{zelikman2024quiet}.

%% file: text/approach.tex
\section{Approach}\label{sec:approach}

This section details our approach, the objective of which is to ingest user documents and produce a DTDF description model for automatically generating the \textit{DT constellation} figure in DTInsight.

\begin{figure*}[htbp]
    \centering
    \includegraphics[width=1\linewidth]{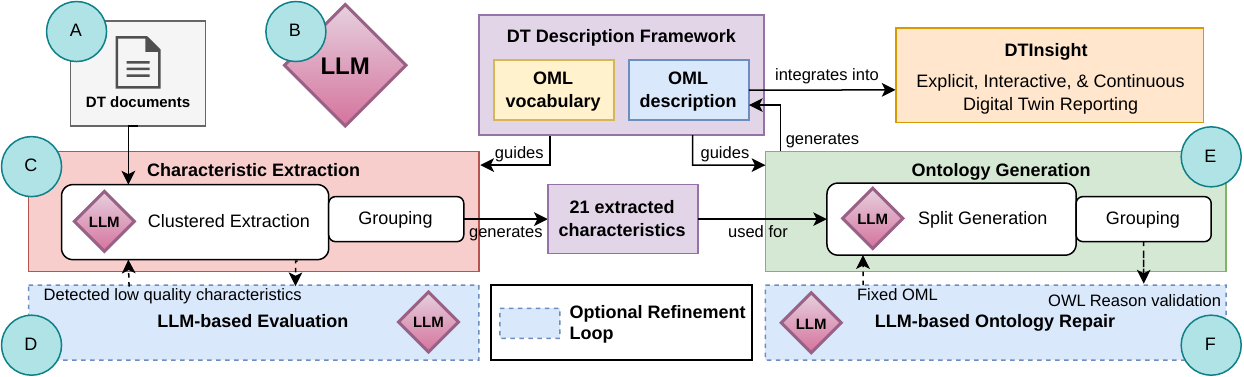}
    \caption{Technical architecture of our DTScribe approach.}
    \label{fig:architecture}
\end{figure*}

As illustrated in \Cref{fig:architecture}, the technical process begins with user-provided DT documents (A), from which an LLM (B) extracts characteristics (C) guided by the DTDF. These characteristics are fed to another LLM for ontology generation (E). To enhance output quality, we integrate two optional refinement loops: an \textit{LLM-as-a-judge} evaluation loop (D) to iteratively improve extracted characteristics, and an ontological reasoning loop (F) to repair and validate the generated OML.

\textbf{Proposed Workflow.} From the user's perspective, the operation is a three-step workflow: \begin{enumerate} \item \textbf{Setup:} The user deploys the local LLM(s), the DTInsight tool, and our reporting component. \item \textbf{Execution:} The user selects the relevant source artifacts (text files) describing the DT and initiates the reporting process. \item \textbf{Output:} The system automatically generates a DT report aligned with the reporting framework, which is then visualized in DTInsight. \end{enumerate}

Our work is grounded in the incubator DT running example described in \Cref{sec:background}, providing sample input and output throughout this section.

\subsection{Component A: DT Documents}

Our pipeline supports the ingestion of standard text files, including \verb|.docx| and \verb|.txt|. However, for this article, we limit our processing to PDFs of academic papers that describe a DT exemplar, as listed in \Cref{sub:exemplars}.

We have three assumptions about these documents: 1) they contain a description of a DT, 2) they do not explicitly use DTDF terminology so that our characteristic extraction is unbiased, and 3) they are devoid of malicious prompt injections that could tamper with our extraction process. 

\paragraph{Document pre-processing}

Raw user-provided documents are rarely suitable for direct ingestion by LLMs. This is particularly true for smaller models, which can easily be sidetracked by formatting errors and superfluous information. As such, a pre-processing step is crucial to ensure the model receives a clean representation of the data.

To achieve this, we ingest each document and convert their text content into Markdown, preserving the original structural hierarchy of sections, headings, and lists. We automatically remove the ``References'' section from academic papers, as initial experiments indicated it degraded retrieval process quality. We preserve sentence and paragraph boundaries during chunking and embed the segments into a vector database for use in clustered characteristic extraction.

As detailed in \Cref{subsec:lessons_learned}, in future work we plan to expand this workflow to include multimodal processing, such as converting figures to text to enhance retrieval context.

\subsection{Component B: LLM}

The LLM is the core component for two downstream tasks: (i) extracting a structured summary of DTDF characteristics and (ii) generating an OML description from this summary. The LLM component is treated as a configurable module, allowing us to experiment with different models, enabling systematic comparison without altering the pipeline logic.

Because no single model is optimal across both tasks required, we leverage six LLMs that vary in architecture, scale and deployment modality. These differences carry trade-offs in latency, cost and reliability; for instance, cloud hosted models can offer lower latency and access to higher parameter sizes, however their reliability depends on network stability and external service constraints. On the other hand, locally available models allow for higher configuration and privacy controls, but remain limited by available hardware.

Since each task has its specific requirements (evidence-based and specific summarization for extraction, and strict syntactic/semantic validity for OML generation), these task-related constraints are considered alongside the practical trade-offs while reporting results for six model alternatives under a common experimental protocol. The final model choices are determined by the performance across both dimensions and the quality of outputs for both tasks.

\subsection{Component C: Characteristic Extraction}

This component extracts DTDF characteristics from user-provided documents with LLMs, which are then used for the subsequent OML generation.

\subsubsection{Clustered Extraction with RAG}

We investigated a clustered extraction strategy to test its impact on the semantic alignment between the retrieved documents and the target characteristics. While extraction could technically be performed individually for each characteristic or monolithically for all twenty-one, our initial hypothesis was that clustering would offer an advantage in \textit{retrieval precision} and \textit{context management}.
By grouping semantically related characteristics, we theorized that we could formulate more targeted retrieval queries (e.g., focusing solely on ``Orchestration'' terms rather than broad system descriptions). This aimed to provide the LLM with context dense with relevant evidence while minimizing the noise of unrelated text chunks. Furthermore, by extracting a small cluster of items, we expected the model to maintain focus, reducing the hallucinations or omissions often observed when attempting to extract a large number of complex variables in a single pass.
However, as detailed in our evaluation (\Cref{sec:evaluation}), empirical results demonstrated that this clustered approach degraded extraction accuracy. We present the methodology here to showcase the full pipeline architecture and to document the testing and rejection of this hypothesis.

We clustered the twenty-one characteristics based on semantic similarity by applying the k-means algorithm to the embeddings of their definitions (found in ~\cite{gilSystematicReportingFramework2024}), resulting in six groups: \textit{Purpose~($\mathcal{CC}_1$), Orchestration~($\mathcal{CC}_2$), Components~($\mathcal{CC}_3$), Connectivity~($\mathcal{CC}_4$), Life-cycle~($\mathcal{CC}_5$),} and \textit{Governance~($\mathcal{CC}_6$)}.
\Cref{tab:characteristics_combined} (in the appendix) details the distribution of characteristics within all clusters, along with their respective extraction prompts. For example, the \textit{Orchestration} Cluster ($\mathcal{CC}_2$) contains four characteristics: Twinning time-scale ($C_7$), Multiplicities ($C_8$), DT constellation ($C_{12}$) and Horizontal integration ($C_{18}$). An extracted description example for this cluster is showcased on~\Cref{tab:cluster2_descriptions}.

Each cluster reuses a CoT prompt template (detailed in \Cref{lst:extraction_guide}), supplemented with the corresponding descriptions of characteristics to retrieve (detailed in \Cref{tab:characteristics_combined}) and semantically relevant document chunks retrieved using embeddings of tailored keywords (detailed in \Cref{tab:cluster_queries}). The output is a structured JSON object mapping characteristic names to their extracted descriptions, which are then aggregated to serve as the input for the subsequent ontology generation component.

\begin{table*}[htbp]
    \caption{Example Characteristic Extraction of the Orchestration Cluster ($\mathcal{CC}_2$) by gpt-oss-120b, describing Twinning time-scale ($C_7$), Multiplicities ($C_8$), DT constellation ($C_{12}$) and Horizontal integration ($C_{18}$) for the incubator case study.}
    \label{tab:cluster2_descriptions}
    \centering
    \small
    \renewcommand{\arraystretch}{1.25}
    \begin{tabular}{@{}p{0.02\linewidth}p{0.94\linewidth}@{}}
        \toprule
         & \textbf{Extracted Description} \\
        \midrule
        $C_{7}$ & 
        The DT synchronises with the Physical Twin through an event‑driven RabbitMQ message bus. Sensor readings are streamed from the low‑level driver to the broker and consumed by DT components (visualisation dashboard, Kalman filter, calibration module) in near‑real‑time, enabling sub‑second latency for monitoring and control. Calibration can be triggered on‑demand, and co‑simulation can run on historical or live data streams.
        \\

        $C_{8}$ & 
        The architecture defines a single Digital Twin instance associated with one Physical Twin (the incubator). The hierarchy consists of the Physical Twin, a low‑level driver, a controller, and the Digital Twin which contains the simulation models, calibration and monitoring services. No decentralized or multiple‑twin coordination is described.
        \\

        $C_{12}$ & 
        The system follows a message‑driven component architecture. Core components are: Low‑level driver (sensor/actuator abstraction), Controller (bang‑bang logic), Communication Server (RabbitMQ), Time‑Series Database, Dashboard, Calibration service, Kalman‑filter based anomaly detector, and co‑simulation engine. RabbitMQ provides synchronization and decoupled communication; services subscribe to relevant topics, process data, and publish results, forming an orchestrated DT constellation.
        \\

        $C_{18}$ & 
        Integration with external systems is achieved via the RabbitMQ message broker, which exposes standard AMQP interfaces for data exchange. Sensor data and control commands are published/subscribed over this broker, allowing other enterprise applications or additional Digital Twins to connect. The time‑series database and dashboard consume the same streams, providing interoperability through common messaging protocols. No specific enterprise standards (e.g., OPC UA) are mentioned.
        \\
        \bottomrule
    \end{tabular}
\end{table*}

\paragraph{Prompting details}

Our extraction prompt follows an instruction-driven structure inspired by CoT prompting to decompose the task~\cite{patil_2024_llm_review}. As mentioned before, LLMs often struggle to reliably execute long, multi-constraint tasks in a single step. This component follows the same design principle of task decomposition adopted throughout the pipeline.

\input{listings/cluster_extraction}

The prompt in Listing~\ref{lst:extraction_guide} has four main sections. First, it defines the role of the model as an expert in Digital Twin systems and ontology modeling, which encourages technical specificity and will guide the LLM towards domain-appropriate terminology acquired during its pretraining. Second, it provides the task inputs: (i) the retrieved context documents for the current cluster, (ii) the list of target characteristics along with their definitions, and (iii) optional judge feedback. Judge feedback is only included during refinement iterations (see \Cref{sec:llm-based-eval}); otherwise, this field is left empty.

Third, the prompt specifies a three-phase procedure. In the \emph{reasoning phase}, the model is instructed to inspect the retrieved documents, recognize relevant evidence, and infer the use-case context, allowing for generalization across different domains. In the \emph{extraction phase}, it must produce characteristic descriptions grounded solely in the provided documents, explicitly mark missing evidence with \textit{Not in Document}, and incorporate judge feedback to refine low quality output. In the \emph{validation phase}, the model is instructed to ensure that all statements are evidence-based, technically correct, complete according to the requested characteristics, and consistent with the judge feedback when provided.

Finally, strict formatting constraints are described, instructing the model must output only valid JSON, with no additional text. We enforce this constraint using LangChain’s structured output parsing, ensuring schema compliance and preventing additional content from appearing in the final extraction output.

\subsubsection{Component D: LLM-based Evaluation}
\label{sec:llm-based-eval}

LLM generated extractions exhibit inherent variability due to sampling randomness and sensitivity to prompts. Consequently, the same characteristic might be correctly extracted in one run and fail in another. Therefore, an automated mechanism for assessing extraction quality is required to ensure that downstream OML generation operates on accurate and evidence-based inputs, increasing its success rate and accuracy.

To achieve this, we adopt an LLM-as-a-judge framework employing a secondary LLM. This evaluator examines each of the provided characteristics and provides feedback, approximating human judgment to guide refinement. To mitigate output randomness, we implement an iterative loop of \textit{extraction}, \textit{evaluation}, and \textit{revision}. This process terminates when the characteristics achieve a sufficient quality score or when a maximum retry limit is reached. This type of refinement strategy pipeline employing repeated judgment–revision cycles has been shown to improve the reliability of LLM evaluations~\cite{gu2025surveyllmasajudge}.

For the evaluation role, we employ GLM-4.7 as 1) large parameter size LLMs have been shown to provide higher-quality judgments~\cite{crupi2025effectiveness}; and 2) utilizing a distinct model mitigates self-enhancement bias, a known risk when the same LLM performs both generation and evaluation~\cite{gu2025surveyllmasajudge}. While our current implementation utilizes a cloud-hosted evaluator to maximize judgment quality and improve reproducibility, the architecture is modular. Users prioritizing data sovereignty can substitute a local model or bypass the step entirely without disrupting the core infrastructure.

\Cref{alg:char_eval} outlines the process for sequentially evaluating and refining characteristics after each cluster extraction. It proceeds in four steps:
\begin{enumerate}
    \item \textbf{Extract characteristics:} The LLM generates the initial set of characteristics extraction ($\mathcal{CC}_i'$) from the source documents ($D$) and the cluster of characteristics with their definitions ($\mathcal{CC}_i$).
    \item \textbf{Evaluate quality and filter:} The judge assesses each characteristic on a 1--5 Likert scale (detailed in \Cref{lst:scoring_scale}). However, if the current iteration is a retry, the corresponding characteristics in $P$ will be overwritten over the low-quality ones before judging the extraction ($\mathcal{CC}_i'$). Characteristics scoring below a threshold of 4 are isolated into a low-quality set ($LQ$), while high-quality items are preserved, added to the ${P}$ set. 
    \item \textbf{Update sets:} If the current iteration is a retry, $P$ and $LQ$ sets are updated based on new $E_{q}$ values, preserving the characteristics that reached an acceptable quality ($E_{q}(C) \geq 4$).
    \item \textbf{Stop or refine:} If there are remaining low-quality characteristics ($LQ \neq \emptyset$) and there are still retries available, the loop will continue and the system attempts to refine the cluster by re-processing the extraction and preserving high quality items in the next iteration until the retry limit or acceptable quality is reached.
\end{enumerate}

The evaluation score is guided by examples of Very Poor (1), Fair (3), and Very Good (5) characteristics, provided alongside the full source document to assess factual consistency~\cite{huyen_ai_engineering_2025}, specificity and relevance. We enforce a strict JSON output format containing the characteristic name, reasoning, and score. This design ensures that judgments remain grounded solely in the provided evidence, reducing the risk of hallucinations.

\begin{algorithm}[tbh]
    \caption{Iterative LLM-Based Refinement for One Cluster $\mathcal{CC}_i$}
    \label{alg:char_eval}
    \begin{algorithmic}[1]
        \Require Characteristics Cluster $\mathcal{CC}_i$, documents $D$, max retries $r_{\max}$, LLM
        \State $r \gets 0$  \Comment{retry count}
        \State $P \gets \emptyset$ \Comment{preserved set}
        \State $LQ \gets \emptyset$ \Comment{low quality set}
        
        \While{true}
            \Statex 1. \textbf{Extract characteristics}
            \State $\mathcal{CC}_i' \gets \textsc{Extract}(\mathcal{CC}_i, D)$
            
            \Statex 2. \textbf{Evaluate quality and filter}
            \If{$r > 0$} %
                \State $\mathcal{CC}_i' \gets \left(\mathcal{CC}_i' \setminus LQ\right) \cup P$
            \EndIf
            \State $\{E_{q}(C) \mid C \in \mathcal{CC}_i'\} \gets \textsc{Judge}(\mathcal{CC}_i', D)$
        
            \If{$r = 0$} %
                \State $LQ \gets \{ C \in \mathcal{CC}_i' \mid E_{q}(C) < 4 \}$
                \State $P \gets \{ C \in \mathcal{CC}_i' \mid E_{q}(C) \geq 4 \}$
                
            \Statex 3. \textbf{Update sets}
             \Else{} %
                \For{each $C \in LQ$}
                    \If{$E_{q}(C) \geq 4$}
                        \State $P \gets P \cup \{C\}$
                        \State $LQ \gets LQ \setminus \{C\}$
                    \EndIf
                \EndFor
            \EndIf
            
            \Statex 4. \textbf{Stop or refine}
            \If{$LQ = \emptyset$}
                \State \Return $\{(C,E_{q}(C)) \mid C \in P\}$
            \ElsIf{$r = r_{\max}$}
                \State \Return $\{(C,E_{q}(C)) \mid C \in P \cup LQ\}$
            \Else
                \State $r \gets r + 1$
            \EndIf
            \color{black}
        \EndWhile
    \end{algorithmic}
\end{algorithm}

\begin{figure}[tbh]
    \begin{lstlisting}[language=, caption=LLM characteristic scoring scale guide., label=lst:scoring_scale]
1 - Very poor
- "Not in Document" or states the characteristic is missing when the source documents DO mention relevant information.
- Vague or generic information.
- Uses speculative language such as "likely", "possibly" instead of directly using the text.
- No relevance to the specific system mentioned in the Source Documents.

2 - Poor
- Very general definition.
- Not specific to the system mentioned in the Source Documents.
- Lacks technical content.

3 - Fair
- Relevant to the specific system.
- Some correct context, but missing key technical details that are present in the Source Documents.

4 - Good
- Correct and specific.
- Contains relevant technical details.
- Could still be expanded with more depth.

5 - Very good
- "Not in Document" or states the characteristic is missing and the Source Documents DO NOT contain relevant information.
- Highly specific and detailed.
- Provides in-depth technical explanation of the characteristic's behavior within the system.
- Strongly grounded in the provided sources.
\end{lstlisting}
\end{figure}

\subsection{Component E: Ontology Generation}
\label{sec:ontology_gen}

The goal of this framework component is to generate a valid OML description from the set of natural language DTDF characteristics using an LLM. To do so, it a) splits characteristics for OML instance generation, and b) runs an iterative LLM-based ontology repair loop.

This pipeline uses an ontology to enable automated consistency checks. By formalizing the extracted natural language characteristics into a structured OML description, the framework can verify the logical and semantic integrity of the generated models. This ensures that DT characteristics are free from structural contradictions and adhere to defined domain rules.

\subsubsection{Split Generation}

Smaller LLMs struggle to maintain structural coherence. As such, their context window must be optimized to minimize context noise that could disrupt the current LLM objective. We thus propose \textit{split generation} of the characteristics by only tasking the LLM to generate OML instances that cannot just be constructed programmatically. This aims to reduce context length, lower inference cost, and simplify the generation task.

As such, we categorized the DTDF characteristics into a) \textit{component characteristics} $C_{\text{component}}$, which are multi-attribute, interdependent, and composed of multiple instances, and b) \textit{description characteristics} $C_{\text{description}}$, which are single-attribute and independent. Component-based OML instances are LLM-generated, guided by characteristics, while simple description-based ones are programmatically constructed.

For example, $C_{10}$ and $C_{11}$ are component characteristics that respectively describe the models/data and tooling/enablers in the system. These are represented by the boxes connected by arrows in \Cref{fig:dt_constellation}. Oftentimes, data or models are used by those tools or enablers, which make them adhere to the definition of a component characteristic. 

Conversely, description characteristics like $C_9$ describe the life-cycle stages of the DT and system, which have no representation in our DT constellation. Thus, $C_9$ is best kept as a textual-only description in the characteristic table report. Other examples of description characteristics are shown in Table~\ref{tab:cluster2_descriptions}.

\Cref{eq:component_sets} showcases the two characteristic sets and their assigned characteristics.

\begin{equation}
    \label{eq:component_sets}
    \begin{aligned}
        C_{\text{component}}   & = \{C_2, C_3, C_4, C_6, C_{10}, C_{11}, C_{17}\},       \\
        C_{\text{description}} & = \{C_1,\dots,C_{21}\} \setminus C_{\text{component}}.
    \end{aligned}
\end{equation}

Simply providing the LLM with the DTDF OML vocabulary resulted in too many syntactic and semantic mistakes, especially at lower parameter size. As such, we employed CoT prompting to provide the LLM with curated rules and guides. We modified the prompt until our preliminary experiments gave results of sufficiently-high ontology generation quality.
The prompt provides a list of characteristics to extract or not, naming conventions, a syntax guide, rules for generating the OML instances, and a characteristic validation checklist. \Cref{lst:dtdf_guide_oml} showcases the LLM syntax guide provided to the LLM to generate the component characteristics.

\begin{figure}[tbh]
    \begin{lstlisting}[language=oml, caption=LLM syntax guide for OML generation of the DTDF description model using component characteristics., label=lst:dtdf_guide_oml]
// C2: Acting Components
instance <ActuatorName> : DTDFVocab:ActingComponent [
    base:desc "<specific description from characteristics>"]
// C3: Physical sensing components
instance <SensorName> : DTDFVocab:SensingComponent [
    base:desc "<specific description from characteristics>"
    DTDFVocab:producedData <DataTransmittedName>]
// C4: Physical-to-virtual interaction
instance <DataTransmittedName> : DTDFVocab:DataTransmitted [
    DTDFVocab:producedFrom <SensorName1>, <SensorName2>]
// C10: Models/Data
instance <ModelName> : DTDFVocab:Model [
    base:desc "<specific model description from characteristics>"
    DTDFVocab:inputTo <EnablerName1>, <EnablerName2>]
instance <DataName> : DTDFVocab:Data [
    base:desc "<specific data description from characteristics>"
    DTDFVocab:inputTo <EnablerName1>, <EnablerName2>
    DTDFVocab:fromData <DataTransmittedName>]
// C11: Enablers
instance <EnablerName> : DTDFVocab:Enabler [
    base:desc "<specific enabler description from characteristics>"
    DTDFVocab:enables <ServiceName1>, <ServiceName2>]
// C6: Services
instance <ServiceName> : DTDFVocab:Service [
    base:desc "<specific service description from characteristics>"
    DTDFVocab:provides <InsightName1>, <ActionName2>]
// C17: Insights/Actions
instance <InsightName> : DTDFVocab:Insight [
    base:desc "<specific insight description from characteristics>"]
instance <ActionName> : DTDFVocab:Action [
    base:desc "<specific action description from characteristics>"
    DTDFVocab:IsAutomatic <true|false>]
\end{lstlisting}
\end{figure}

\subsubsection{Component F: LLM-based Ontology Repair}

Despite the constrained generation prompts, the initial OML output rarely satisfies all syntactic and semantic constraints. Common errors include \textit{hallucinated references} (linking to instances that were never defined), \textit{type mismatches} (assigning a property to the wrong entity type), and \textit{syntax violations} (malformed OML structures). Rather than discarding these outputs or requiring manual user intervention, we leverage the deterministic error reporting of an ontology reasoner to guide the LLM in a self-correction loop~\cite{giabbanelli2026guidelargelanguagemodels}.

\begin{algorithm}[htbp]
    \caption{Iterative Refinement of LLM-Generated OML Descriptions}
    \label{alg:oml_valid}
    \begin{algorithmic}[1]
        \Require Characteristics $C_{\text{component}}$, $C_{\text{description}}$, Max retries $R_{\max}$
        \Ensure Validated Ontology $\mathcal{O}$ or Failure
        
        \Statex 1. \textbf{Split Generation}
        \State $\mathcal{O}_{\text{desc}} \gets \textsc{StaticGen}(C_{\text{description}})$
        \State $\mathcal{O}_{\text{comp}} \gets \textsc{LLMGen}(C_{\text{component}})$
        
        \Statex 2. \textbf{Integration and Validation}
        \State $r \gets 0$
        \While{$r \le R_{\max}$}
        \color{black}
            \State $\mathcal{O} \gets \textsc{Integrate}(\mathcal{O}_{\text{comp}}, \mathcal{O}_{\text{desc}})$
            \State $(\textit{valid}, \textit{report}) \gets \textsc{ReasonerCheck}(\mathcal{O})$
            
            \If{$valid$}
                \State \textsc{DeployToFuseki}($\mathcal{O}$)
                \State \Return $\mathcal{O}$
            \EndIf
            
            \Statex 3. \textbf{Error Injection and Repair}
            \State $\mathcal{O}_{\text{comp}} \gets \textsc{InjectErrors}(\mathcal{O}_{\text{comp}}, \textit{report})$
            \State $\mathcal{O}_{\text{comp}} \gets \textsc{LLMRepair}(\mathcal{O}_{\text{comp}}, C_{\text{component}})$
            \State $r \gets r + 1$\color{black}
        \EndWhile
        \State \Return \textbf{Failure}
    \end{algorithmic}
\end{algorithm}

\Cref{alg:oml_valid} details this iterative generation and repair process, that :

\begin{enumerate}
    \item \textbf{Split Generation:} The system programmatically constructs the description ontology ($\mathcal{O}_{\text{desc}}$) from $C_{\text{description}}$ and prompts the LLM to generate the component ontology ($\mathcal{O}_{\text{comp}}$) from $C_{\text{component}}$.
    \item \textbf{Integration and Validation:} The system merges these fragments and executes the OWL Reason\footnote{\url{https://github.com/opencaesar/owl-tools/tree/master/owl-reason}} ontology reasoner. This tool validates type satisfiability, logical consistency, and reference integrity~\cite{elaasarOpenCAESARBalancingAgility2023}. If validation succeeds, the ontology is deployed to the Fuseki server. If validation fails, the system triggers the error injection phase.
    \item \textbf{Error Injection and Repair:} To facilitate repair, we capture the reasoner's output and inject the specific error messages directly into the OML description model as context-aware comments immediately following the problematic lines. For example, if the model references an undefined entity \texttt{SensorX}, the line is annotated with \texttt{// TODO: Fix the error on this line using the following error message: `Couldn\textquotesingle t resolve reference to NamedInstance \textquotesingle SensorX\textquotesingle'}. Other reasoner outputs (such as the \verb|reasoning.xml| file), are directly ingested into the prompt. The LLM is then prompted to repair the code using these annotations as a guide. This cycle repeats until the ontology is valid or the maximum retry limit ($R_{\max}$) is exhausted. 
    
    \Cref{lst:dtdf_desc_oml_error} shows an error-injected OML description. The first error misspells \textit{SensingComponent} as \textit{SensorComponent}, the second couldn't link to \textit{ThermalSensorDataTransmitted} because its current instance is not related to a vocabulary concept, and the third error misspells the instance name (\textit{ThermalSensor}).
\end{enumerate}

Merging $\mathcal{O}_{\text{comp}}$ and $\mathcal{O}_{\text{desc}}$ creates a potential blind spot during the repair step: because reasoner errors are only injected into $\mathcal{O}_{\text{comp}}$, the LLM could generate components incompatible with the descriptions. However, since their respective ontology concepts are independent, we assume no interaction errors arise between programmatically and LLM-generated instances.

\begin{figure}[htbp]
    \begin{lstlisting}[language=oml, caption=Invalid DTDF OML description model excerpt with injected reasoner errors., label=lst:dtdf_desc_oml_error]
instance ThermalSensor : DTDFVocab:SensorComponent [ // TODO: Fix the error on this line using the following error message: `Couldn't resolve reference to Entity 'DTDFVocab:SensorComponent'.'
    base:desc "Thermal sensors measuring system temperature for monitoring",
    DTDFVocab:producedData ThermalSensorDataTransmitted // TODO: Fix the error on this line using the following error message: `Couldn't resolve reference to SemanticProperty 'ThermalSensorDataTransmitted'.'
]
instance ThermalSensorDataTransmitted : DTDFVocab:DataTransmitted [
    DTDFVocab:producedFrom ThermalSensorComponent // TODO: Fix the error on this line using the following error message: `Couldn't resolve reference to NamedInstance 'ThermalSensorComponent'.'
]
\end{lstlisting}
\end{figure}

The generated OML description model is ingested by DTInsight to create a constellation similar to the one showcasing the incubator on \Cref{fig:dt_constellation}. As described in our work~\cite{fiter2025dtinsight}, DTInsight queries the OML description model to build component connections and extract the textual descriptions for the reporting page.

\subsection{Approach Summary}\label{sub:approach_summary}

To answer \textit{RQ1}, we propose DTScribe, a modular technical architecture designed to support local LLM deployment for the secure and automated processing of DT documentation. As illustrated in \Cref{fig:architecture}, the architecture is composed of a three-stage pipeline:

\begin{enumerate}
    \item \textbf{Ingestion and Pre-processing (Component A):} Raw documentation is converted to Markdown and chunked for RAG, creating a clean context for downstream tasks.
    \item \textbf{Characteristic Extraction (Components B, C, D):} We designed this stage to test a \textit{clustered extraction strategy}, grouping the twenty-one DTDF characteristics into six semantic clusters. Although our initial hypothesis posited that clustering would maximize retrieval precision and context management, empirical evaluation (\Cref{sec:evaluation}) demonstrated that it actually degraded extraction accuracy. We document this module to showcase the full pipeline architecture and the rejection of this hypothesis. This phase is further supported by an optional \textit{LLM-as-a-judge} refinement loop, where a secondary model evaluates extraction quality on a Likert scale and triggers re-generation for low-scoring descriptions.
    \item \textbf{Ontology Generation (Components E, F):} To produce valid OML models, we utilize a \textit{split generation} approach. Complex component instances are generated by the LLM via CoT prompting, while simple descriptive attributes are constructed programmatically. This phase includes an iterative \textit{ontology repair loop}, where a symbolic reasoner validates the output and automatically injects error reports back into the LLM context to guide self-correction.
\end{enumerate}

By integrating local LLMs with symbolic verification (reasoners) and iterative refinement loops, this architecture enables the extraction of DT characteristics and the generation of syntactically valid OML description models to create a visualization in DTInsight.

%% file: listings/cluster_extraction.tex
\begin{figure}[htbp]
    \begin{lstlisting}[language=, caption=LLM prompt for clustered extraction., label=lst:extraction_guide]
You are an expert in Digital Twin systems and ontology modeling.
                                                  
You must extract a structured description for the following Digital Twin characteristics for a single Digital Twin system. For each characteristic, write one coherent description in English that is as specific and technical as possible and grounded ONLY in the provided documents. If the information for a characteristic is not available in the documents, return exactly the string "Not in Document" for that field.

CONTEXT DOCUMENTS:
{docs_content}

CHARACTERISTICS TO EXTRACT:
{description}

JUDGE FEEDBACK:
{judge_results}

INSTRUCTIONS:
1. REASONING PHASE: First, analyze the documents step by step:
   - Identify which documents contain relevant information for each characteristic
   - Note any technical details, technologies, or methodologies mentioned
   - Consider the specific context and domain of the use case

2. EXTRACTION PHASE: For each characteristic:
   - Provide specific, detailed descriptions based ONLY on the provided documents
   - Include concrete technical details (tools, technologies, protocols, methods)
   - Be precise about quantities, frequencies, and specifications when mentioned
   - If no evidence is found, state "Not in Document"
   - Implement the information from the judge feedback (if provided) to preserve high-quality characteristics and enhance characteristics scored less than 4.

3. VALIDATION PHASE: Review your extracted information:
   - Ensure all details come from the provided documents
   - Check that technical terms are used correctly
   - Verify completeness of the description
   - Ensure the extraction of characteristics is consistent with the judge feedback (if provided)

IMPORTANT: You MUST respond with ONLY valid JSON. Do not include any explanations, thinking tags, or additional text outside the JSON structure.

EXTRACTED CHARACTERISTICS (JSON FORMAT ONLY):
{format_instructions}

Remember: Be highly specific and technical. Include exact technologies, methods, and specifications mentioned in the documents. Return ONLY the JSON object with no additional text.
\end{lstlisting}
\end{figure}

%% file: text/metrics.tex
\section{Assessing Approach Quality}\label{sec:metrics}

This section answers \textit{RQ2} by a) describing the input papers used to evaluate our approach, including the definition of a \textit{partial ground truth} for characteristic extraction, and then b) defining metrics for LLM-generated textual reports. Together, these elements will be used to answer \textit{RQ3} in the evaluation section (\Cref{sec:evaluation}).

In particular, we use two sets of metrics for evaluating our approach decisions and the produced artifacts: \textit{extraction metrics} (\Cref{subsub:extraction_metrics}) for characteristic extraction, and \textit{ontology metrics} (\Cref{subsub:ontology_metrics}) for ontology generation. In both cases, we track LLM generation time.%

\subsection{DT Exemplar Papers}\label{sub:exemplars}

We benchmark our approach against five DT paper exemplars from the literature: 
\begin{itemize}
    \item \textbf{P1:} \textit{A Digital Shadow for Accurate Robot Motion Control: Integrating Data with Friction Models}~\cite{Heithoff2025ADS}. 
    \item \textbf{P2:} \textit{Engineering Automotive Digital Twins on Standardized Architectures: A Case Study}~\cite{Ramdhan2025EngineeringAD}.
    \item  \textbf{P3:} \textit{Lab-Scale Gantry Crane Digital Twin Exemplar}~\cite{Mertens2025Lab}.
    \item  \textbf{P4:} \textit{Model-driven Digital Twins for AECO}~\cite{Zech2025Model}.
    \item  \textbf{P5:} \textit{The Incubator Case Study for Digital Twin Engineering}~\cite{feng2021incubator}.
\end{itemize}

\textbf{P1} to \textbf{P4} were recently published at the 2nd International Conference on Engineering Digital Twins (\textit{EDTconf 2025})\footnote{\url{https://conf.researchr.org/home/edtconf-2025}}, and thus provide high quality DT exemplars for our pipeline. We use \textbf{P5} as our case study as it presents a recurring example throughout previous papers. Despite \textbf{P1} to \textbf{P4} being written after the DTDF conceptualization, none of them explicitly use the description framework. All these aspects mean the papers are well-suited for evaluating our approach.

\subsection{Establishing Partial Ground Truth}
\label{sec:ground_truth}

\begin{figure*}[htbp]
    \centering
    \includegraphics[width=1\linewidth]{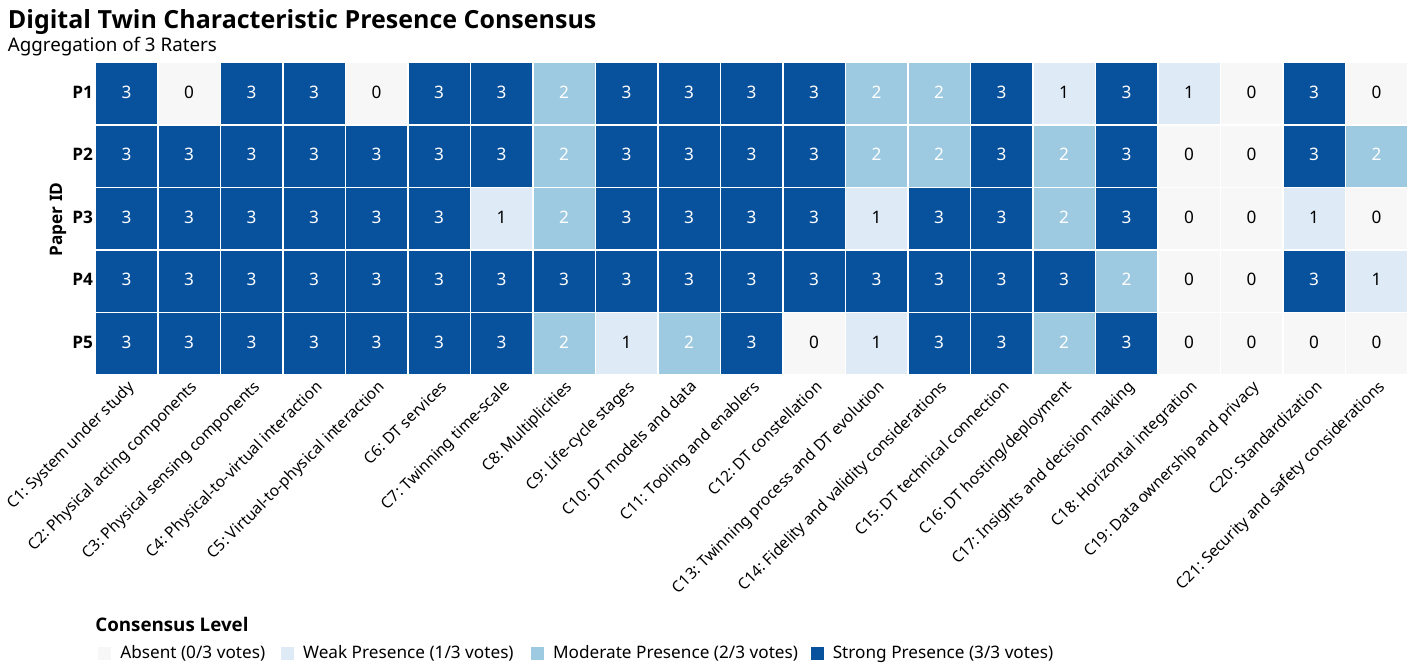}
    \caption{Consensus for DTDF characteristic presence. Papers IDs defined in~\Cref{sub:exemplars}.}
    \label{fig:presence_heatmap}
\end{figure*}

To determine characteristic extraction accuracy, we established a partial ground truth by having all three authors of this article independently rate the presence of twenty-one characteristics across the five DT exemplars used for evaluation. The results are compiled in \Cref{fig:presence_heatmap}.

To validate the reliability of our partial ground truth, we calculated the
inter-rater reliability (IRR) among the three raters. We obtained a Fleiss'
Kappa score of $\kappa = 0.625$, which indicates \textit{substantial agreement}~\cite{fleiss1971measuring,landis1977measurement}.

Finally, we aggregated the ratings using a majority voting scheme: a
characteristic was categorized as \textit{present} if at least two raters
agreed on its presence, and \textit{absent} otherwise. We use these aggregated
labels to determine the extraction accuracy in \Cref{subsub:extraction_accuracy}.

\subsection{Extraction Metrics} \label{subsub:extraction_metrics}

We evaluate the extracted characteristic descriptions using two metrics: \textit{extraction accuracy} and \textit{extraction quality}.

The extraction quality metric ($E_{q}$) is designed as an automated and continuous feedback signal operating within the iterative LLM judge loop to detect potential hallucinations or missing details during generation. In contrast, extraction accuracy ($E_{\text{Acc}}$) represents the primary objective metric evaluated against the partial ground truth (\Cref{sec:ground_truth}). While $E_{q}$ guides the internal refinement process by filtering low-scoring candidate extractions before ontology generation, $E_{\text{Acc}}$ provides the quantitative benchmark reported in our results (\Cref{sec:results}) to answer RQ2.

\textbf{Formal Notation:} Let $D = \{d_1, \ldots, d_n\}$ be the set of $n$ user-provided input documents. For $k \in \{1, \ldots, 21\}$, denote the $k$-th characteristic of the DTDF extracted by the model. The value of $C_k$ is either a textual description or \texttt{NID} (\textit{Not in Document}), indicating the LLM did not extract the characteristic from $D$.

\subsubsection{Extraction Accuracy} \label{subsub:extraction_accuracy}

\textbf{Extraction accuracy} assesses the system's ability to correctly identify the presence or absence of characteristics in the source documents (avoiding hallucinations).

We measure the alignment between the LLM and human annotators regarding the presence of characteristics within the documents based on this  \textit{partial ground truth}. Extraction is treated as a binary classification task for each of the twenty-one characteristics, ignoring the textual quality of the output. Let $\mathbf{p} \in \{0,1\}^{21}$ be the binary ground-truth vector derived from human assessment, and let $\mathbf{t} \in \{0,1\}^{21}$ be the binary LLM-judged extraction vector.  Their components are defined as:

\[
\begin{aligned}
p_k &= \begin{cases}
1 & \text{if characteristic } k \text{ is present in } D \\
  & \text{(human label),} \\
0 & \text{otherwise,}
\end{cases} \\
\\
t_k &= \begin{cases}
1 & \text{if } C_k \neq \texttt{NID} \text{ (any textual description),} \\
0 & \text{if } C_k = \texttt{NID}.
\end{cases}
\end{aligned}
\]

The extraction accuracy $E_{\text{Acc}}$ is calculated as the proportion of concordant pairs between the two vectors:

$$E_{\text{Acc}} = \frac{\text{Agreement count}}{21}=\frac{1}{21} \sum_{k=1}^{21} \mathbf{1}(p_k = t_k)$$
where $\mathbf{1}(\cdot)$ is the indicator function.

\subsubsection{Extraction Quality}

\textbf{Extraction quality} evaluates the semantic precision, relevance, and formulation of extracted descriptions via an LLM-based assessment. Specifically, it measures adherence to the scoring criteria defined in \Cref{lst:scoring_scale}. %

Unlike extraction accuracy, extraction quality is a \textit{reference-free} metric. This approach is necessary because DT reports must serve stakeholders with heterogeneous requirements, ranging from high-level summaries to distinct technical specifications. Consequently, enforcing a single ``ground truth'' reference summary would penalize valid details essential to specific user profiles.

We employ an evaluator LLM to rate each characteristic description $C_k$ given the source document $D$. We define the scoring function $E_{q}(C_k, D) \in \{1, \ldots, 5\}$ as follows:

\begin{itemize}
    \item If $C_k$ contains extracted text, the LLM assigns a score on a 5-point Likert scale (1=lowest quality, 5=highest quality) based on evidence present in $D$.
    \item If $C_k = \texttt{NID}$ (Not in Document), the score is binary: 5 if the absence is verifiable in $D$, and 1 if the omission is incorrect (hallucinated absence).
\end{itemize}

The evaluator is prompted with specific criteria and few-shot examples. For every characteristic, the model generates both the score $E_{q}$ and a textual rationale to ensure interpretability.

\subsection{Ontology Generation Metrics} \label{subsub:ontology_metrics}

We evaluate the pipeline using \textit{success rates} for ontology generation. These metrics quantify the likelihood of the LLM producing a valid OML description at each retry step, as well as the cumulative number of retries required to converge. To determine validity, we use the OWL Reasoner\footnote{\url{https://github.com/opencaesar/owl-tools/tree/master/owl-reason}} to detect syntactic and logical errors.

Formally, we model the generation as a sequential process where a reasoner error triggers a repair loop up to a maximum of $M$ retries. Let the random variable $T \in \{0, 1, \dots\}$ denote the retry index of the first successful validation. $T=0$ represents immediate success, while $T>M$ indicates that the process failed and no valid ontology was generated within the limit $M$. Based on $T$, we define three complementary metrics:
\\

\noindent \textbf{Unconditional Success Rate ($S_i$):} This measures the probability that a random generation run yields a valid result exactly at retry step $i$. It represents the distribution of success across steps: $$S_i = \Pr(T = i) \in [0,1]$$

\noindent \textbf{Cumulative Success Rate ($R_i$):} To assess system reliability up to a given budget, we measure the total probability of obtaining a valid ontology by step $i$. This metric is monotonically increasing: $$R_i = \Pr(T \le i) = \sum_{j=0}^{i} S_j \in [0,1]$$

\noindent \textbf{Conditional Success Rate ($H_i$):} To evaluate the repair mechanism independently of previous runs, we measure the probability of success at step $i$, given that the run has not yet succeeded. While $S_i$ and $R_i$ provide a global view, $H_i$ offers a local view of the repair efficiency at a specific iteration: $$H_i = \Pr(T = i \mid T \ge i) \in [0,1]$$

%% file: text/evaluation.tex
\section{Evaluation}
\label{sec:evaluation}

This section details the evaluation of our approach. In particular, we recall our research questions and introduce our experimental setup.

\Cref{sec:approach} answers RQ1 by defining the technical architecture; \Cref{sec:metrics} answers RQ2 by formalizing the evaluation metrics and establishing the partial ground truth; and the current section establishes the empirical setup to evaluate RQ3. Finally, \Cref{sec:results} presents the empirical benchmark results across 3,000 experiment runs, complemented by an end-to-end demonstration of the complete pipeline deployed with DTInsight.

Our evaluation is guided by the work of Baltes \textit{et al.}~\cite{baltes2025evaluationguidelinesempiricalstudies} by ensuring we follow recommended designing and reporting guidelines of empirical studies using LLMs.

\noindent
We recall our research questions:
\begin{description}
    \item[RQ1:] What is a \textbf{technical architecture with local LLMs} to enable the extraction of DT characteristics and the generation of valid OML description models?
    \item[RQ2:] What \textbf{metrics and methodologies} are required to effectively evaluate completeness and accuracy for LLM-extracted DT characteristics?
    \item[RQ3:] To what extent do iterative reasoner-guided repair and LLM-as-a-judge \textbf{improve the structural validity of the generated OML models} compared to zero-shot extraction?
\end{description}

\noindent In particular, we investigate the following in RQ3:

\begin{description}
    \item[RQ3.1:] Local LLM capability
    \item[RQ3.2:] Impact of clustering and LLM judging
    \item[RQ3.3:] Effectiveness of iterative OML repair
\end{description}

\subsection{Experimental Setup} \label{sub:exp_setup}

As recommended by~\cite{baltes2025evaluationguidelinesempiricalstudies} in Section 4.2 ``Report Model Version, Configuration, and Customizations'', we describe our experimental setup for transparency and reproducibility.

\paragraph{Software Setup}

We evaluated six open-source LLM configurations with varying parameter counts:
\begin{itemize}
    \item {Ministral3 3B Instruct}, {Ministral3 8B Instruct}, and {Ministral3 14B Instruct}\footnote{\url{https://ollama.com/library/ministral-3}}
    \item {Qwen3-Next-80B-A3B}\footnote{\url{https://ollama.com/library/qwen3-next}}
    \item {gpt-oss-120b} and {gpt-oss-20b}\footnote{\url{https://ollama.com/library/gpt-oss}}
\end{itemize}

Vector embeddings were generated using \texttt{EmbeddingGemma}\footnote{\url{https://ollama.com/library/embeddinggemma}}, selected for its state-of-the-art performance among multilingual models under 500M parameters on the Massive Text Embedding Benchmark (MTEB)~\cite{muennighoff2022mteb}. Additionally, we used the open-source \texttt{GLM-4.7}\footnote{\url{https://ollama.com/library/glm-4.7}} cloud model as a judge. LLM deployment characteristics are listed in~\Cref{tab:llm_specs}.

The software environment combined Ollama\footnote{\url{https://ollama.com/}} for local inference, LangChain\footnote{\url{https://www.langchain.com/}} for orchestration, and ChromaDB\footnote{\url{https://www.trychroma.com/}} for vector storage and retrieval. Together these tools form a modular workflow that, apart from the LLM judge, utilizes models of reasonable size for practitioners for local deployment on work computers (under 20B parameters) or servers (80B and 120B parameters).

\begin{table*}[htbp]
    \centering
    \caption{LLM deployment characteristics. Category size is representative for use in our analysis.  Note that the VRAM requirement may be higher in practice due to context overhead and model activations.}
    \label{tab:llm_specs}
    \setlength{\tabcolsep}{4pt}
    \begin{tabular}{ccccccc}
        \toprule
         \makecell{\textbf{Model}} &
         \makecell{\textbf{Parameter}\\\textbf{count}} &
         \makecell{\textbf{Precision}} &
         \makecell{\textbf{Min.}\\\textbf{VRAM}} &
         \makecell{\textbf{Size}\\\textbf{Category}} &
         \makecell{\textbf{Release}\\\textbf{Date}} &
         \makecell{\textbf{Ollama}\\\textbf{ID}}\\
         \midrule
         EmbeddingGemma & 0.3B & BF16 & $\sim$0.6GB & Small & 2025-09-04 & \texttt{85462619ee72}\\
         Ministral3 3B Instruct & 3B & INT4 & $\sim$3GB & Small & 2025-12-02 & \texttt{f04aa1c738f6}\\
         Ministral3 8B Instruct & 8B & INT4 & $\sim$6GB & Small & 2025-12-02 & \texttt{1922accd5827}\\
         Ministral3 14B Instruct & 14B & INT4 & $\sim$9GB & Small & 2025-12-02 & \texttt{4760c35aeb9d}\\
         \hline
         gpt-oss-20b & 20B & INT4 & $\sim$16GB & Medium & 2025-08-05 & \texttt{17052f91a42e}\\
         Qwen3-Next-80B-A3B & 80B & INT4 & $\sim$50GB & Medium & 2025-09-10 & \texttt{b2ebb986e4e9}\\
         \hline
         gpt-oss-120b & 120B & INT4 & $\sim$80GB & Large & 2025-08-05 & \texttt{a951a23b46a1}\\
         GLM-4.7 & 358B & INT4 & $\sim$200GB & Large & 2025-12-22 & \texttt{023608864819}\\
         \bottomrule
    \end{tabular}
\end{table*}

\paragraph{LLM Hyperparameters}

While the ontology repair loop may benefit from the diversity introduced by default sampling parameters, we argue that the evaluation mechanism must remain strictly stable to serve as a reliable benchmark. As such, we use default LLM hyperparameters (temperature, top-k, top-p) for characteristic extraction and ontology generation. To enforce determinism in the LLM judge, we set its temperature to 0 and fix the seed, aligning with practices from domain modeling~\cite{pan2025llm,qualis2025mitigating}. 

\paragraph{Hardware Setup}

All experiments were executed on a Windows 11 laptop with 32 GB of RAM and an NVIDIA GeForce RTX 4060 8GB VRAM GPU, with LLMs running on Ollama Cloud\footnote{\url{https://ollama.com/cloud}} to parallelize experimentation for this article.

\subsection{Experimental Process}\label{sub:exp_process}

To quantify the impact of our architectural components~\cite{giabbanelli2026guidelargelanguagemodels}, we conducted an ablation study by systematically isolating three core features:

\begin{enumerate}
        \item \textbf{Context Management:} We bypassed the clustering module, ingesting the entire document directly into the LLM context to evaluate the performance of organized data retrieval.
        \item \textbf{Iterative Feedback:} We disabled the LLM-as-a-judge refinement loop to evaluate raw extraction accuracy and quality, and the resulting OML generation success rates.
        \item \textbf{Convergence Efficiency:} We analyzed the first-pass success rate by isolating results obtained prior to the OML retry logic. This allows us to measure the base LLM ability to satisfy OML constraints on the first attempt, thereby identifying the improvements provided by the subsequent refinement loops.
\end{enumerate}

This resulted in four experiment configurations: \textbf{Base} with full document ingestion into context, and no judge; \textbf{+Cluster} with clustered extraction with RAG and no judge; \textbf{+Judge} with full document ingestion into context and the LLM-as-a-Judge refinement loop; \textbf{+Cluster+Judge} with both clustered extraction with RAG and the LLM-as-a-judge refinement loop. For each configuration, we isolate success rates obtained before and after the OML refinement loop.

We ran each configuration twenty-five times ($n=25$) to account for the variability of LLM responses. We leverage Ollama Cloud's infrastructure to execute multiple experiments concurrently, but both \textbf{Base} and \textbf{+Cluster} configurations could be ran fully locally by deploying Ministral3 3B and 8B directly on our machine.

Our experimental pipeline was executed three thousand times in total, and ran for around five days. This highlights the necessity to use a cloud infrastructure to parallelize computation and do experiments at scale, even for local models. We also had to develop several strategies to ensure architecture robustness: mitigating network failures from Ollama Cloud servers, handling file locks for concurrent access to the OWL Reasoner, and detecting missing experiment configurations by assigning IDs and tracking experiment runs.

%% file: text/results.tex
\section{Results}
\label{sec:results}

In this section, we present the results of our evaluation. Specifically, we address the three sub-RQs of RQ3 to determine the impact of various features and configurations within our pipeline.

We have created a reporting page\footnote{Available at \url{https://oakeslabmtl.github.io/dtdf-dtscribe/}} to showcase the end result of our pipeline, ingested by DTInsight~\cite{fiter2025dtinsight}. We specifically hand-picked an OML description file from a \textbf{+Judge} configuration for \textbf{P2} with gpt-oss-20b as \Cref{tab:extraction_accuracy} demonstrates this setup contains the most accurate extractions (around 99\% accuracy). 

\subsection{\textit{RQ3.1:} Local LLM Capability}

We first investigate the relationship between model size, extraction quality, and ontology generation quality. Our primary goal is to determine whether \textbf{baseline local LLMs} can produce adequate results without relying on additional pipeline enhancements.

\paragraph{Extraction Accuracy}

\begin{figure*}[htbp]
    \centering
    \includegraphics[width=1\linewidth]{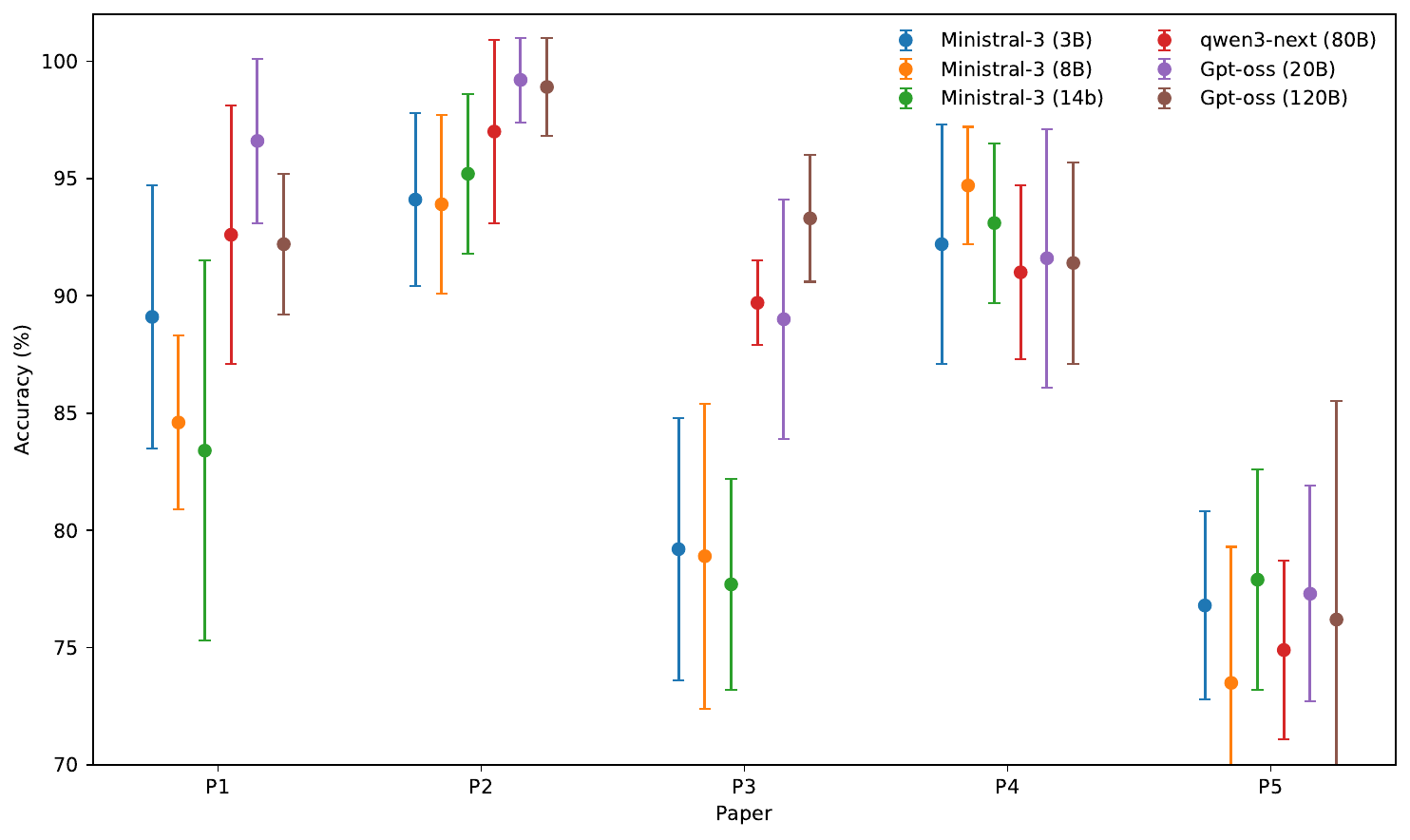}
    \caption{Best \textit{extraction accuracy} results obtained across papers and models.
    }
    \label{fig:extraction_accuracy}
\end{figure*}

Focusing on the extraction accuracy of our \textbf{Base} configuration, \Cref{tab:extraction_accuracy} reveals that all baseline LLMs, including the small Ministral3 3B, achieved an extraction accuracy above 70\%. Second, larger models generally yielded better results, as expected. Third, specific papers, such as \textbf{P5}, present inherent extraction difficulties across all LLMs regardless of size.

We present an alternative view of these results in \Cref{fig:extraction_accuracy}, plotting the best extraction accuracy across papers and models. While the data is noisy and combines non-baseline configurations, it suggests that scaling model size improves accuracy in most cases. However, the performance dips for \textbf{P4} (red diamond) and \textbf{P5} (purple downwards triangle) indicate that larger models do not guarantee improvement. This is likely due to missing information in the underlying papers themselves, as indicated by our partial ground truth (\Cref{fig:presence_heatmap}).

Surprisingly, contrary to the hypothesis that smaller models require RAG to mitigate input complexity, baseline local models with as few as 3B parameters achieved robust extraction accuracy. Notably, these smaller models frequently performed within five points of our largest model.

\paragraph{Ontology Generation}

\begin{figure*}[tbh]
    \centering
    \includegraphics[width=1\linewidth]{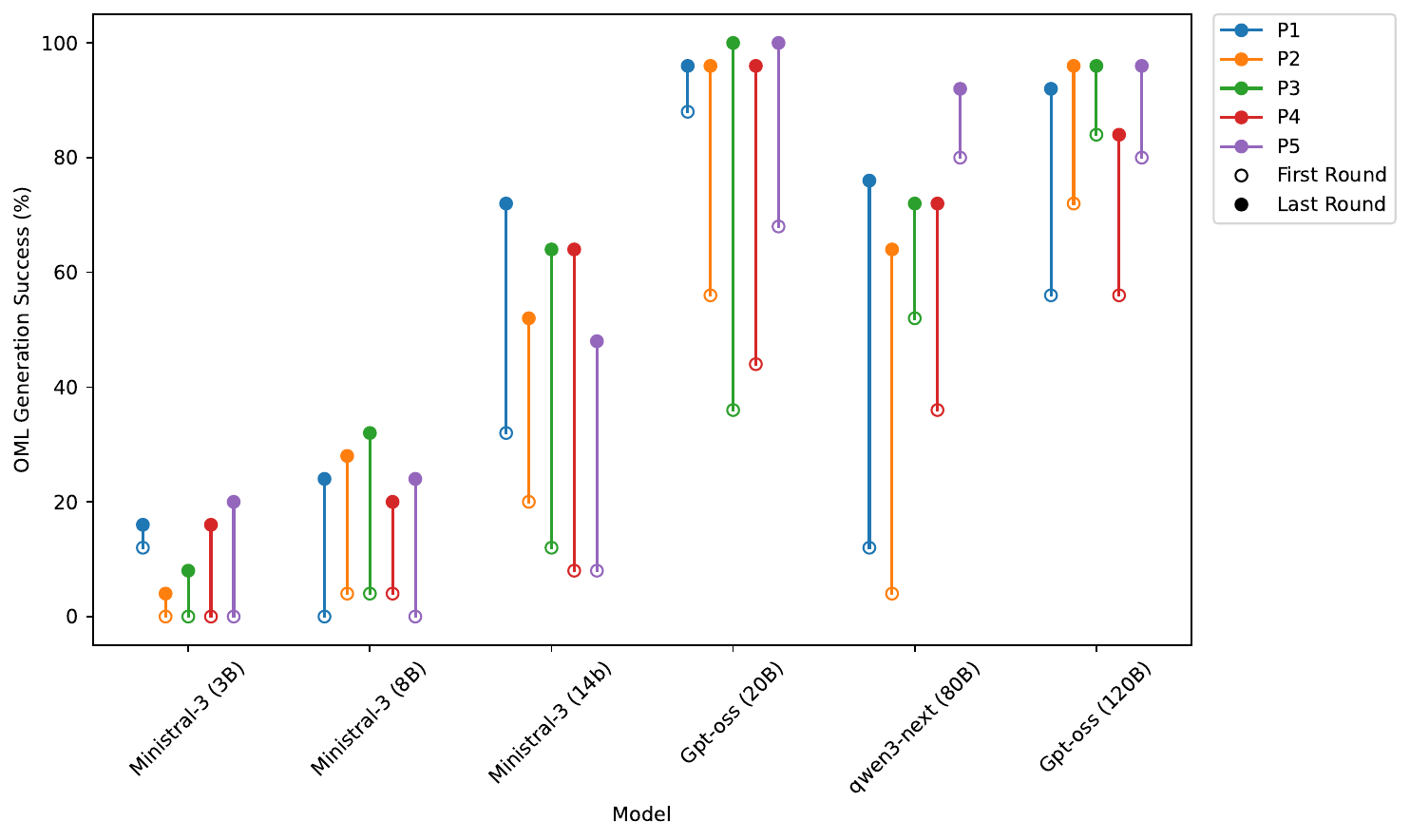}
    \caption{\textit{OML generation success rate} improvement for each model after four retries.}
    \label{fig:oml_plot}
\end{figure*}

To determine baseline generation effectiveness, we evaluate the models' ability to output valid OML models. The comprehensive results are provided in \Cref{tab:success_rates}, with the peak success rate for each model plotted in \Cref{fig:oml_plot}. In this figure, empty circles represent the success rate on the first round of OML generation (prior to reasoner-based refinement), while filled circles indicate the success rate after four rounds of refinement.

For OML generation on the first try, \Cref{tab:success_rates} shows all models except the largest one struggle with around sub-40\% success rate on first try. Only the largest models achieves around 60\% success. \Cref{fig:success_rates_p5} specifically shows for paper \textbf{P5} that the success rate of the \textbf{Base} configuration's first attempt increases with model size, but generally remains low. \Cref{sec:oml_refine_loop_results} discusses the OML generation success after refinement loops.

We also captured the instance count for the OML models generated. To briefly summarize, most configurations and models produced between 25 to 50 instances, with roughly equivalent variance. Thus, we draw no insights from this analysis.

\begin{resultbox}[RQ3.1 Summary]
Small to medium baseline models can be as suitable in terms of extraction accuracy (above 70\%) compared to larger models, depending on the paper.

However, these small to medium models consistently struggle (sub-40\% success) to generate OML description models on the first try. Only our largest model is good enough for OML generation without refinement loops, at around 60\% success.
\end{resultbox}

\subsection{\textit{RQ3.2:} Impact of Clustering and LLM Judging}

Having established baseline model capabilities, we then evaluate the impact of our pipeline enhancements. In \Cref{tab:extraction_accuracy}, we provide the full extraction accuracy results by configuration. These results indicate that, contrary to our expectations, \textbf{extraction configurations involving clustering (with RAG) or judging provide limited accuracy gains relative to their additional computational cost.}

\begin{table*}[htbp]
    \centering
    \small
    \setlength{\tabcolsep}{3pt}
    \caption{\textcolor{black}{Average extraction accuracy relative percentage \textit{change} (\%) and standard deviation across models for the \textbf{+Judge} and \textbf{+Cluster} configurations compared to \textbf{Base}.}}
    \label{tab:judge_cluster_increase}
    \begin{tabular}{lcccccc}
        \toprule
        \textbf{Config.} & \textbf{P1} & \textbf{P2} & \textbf{P3} & \textbf{P4} & \textbf{P5} & \textbf{Overall} \\
        \midrule
        \textbf{+Judge}   & $+1.53 \pm 1.67$ & $+2.83 \pm 2.56$ & $+1.25 \pm 1.42$ & $+4.20 \pm 4.44$ & $+0.72 \pm 1.42$ & $\mathbf{+2.11 \pm 2.71}$ \\
        \textbf{+Cluster} & $-31.19 \pm 18.32$ & $-22.95 \pm 19.17$ & $-13.15 \pm 18.88$ & $-14.59 \pm 12.60$ & $-10.65 \pm 19.41$ & $\mathbf{-18.51 \pm 18.28}$ \\
        \bottomrule
    \end{tabular}
\end{table*}

\begin{table}[htbp]
\centering
\small
\caption{Average processing time for extraction and OML generation (four maximum retries) across configurations. \textit{Note that these results rely on parallelized experiments with \textcolor{black}{local LLMs running on a hosted server}, meaning the absolute processing times should be interpreted as relative estimates of computational efficiency rather than precise benchmarks for local execution.}}
\label{tab:pipeline_costs}
\setlength{\tabcolsep}{1pt}
\begin{tabular}{lccc}
\toprule
\textbf{Configuration} & \makecell{\textbf{Extract}\\\textbf{Time (s)}}& \makecell{\textbf{OML Gen.}\\\textbf{Time (s)}} & \makecell{\textbf{Total}\\\textbf{Time (s)}} \\
\midrule
\textbf{Base} & \textbf{62.5}& 255.0& \textbf{317.5}\\
\textbf{+Cluster} & 208.2 & \textbf{195.8}& 404.0\\
\textbf{+Judge} & 647.9 & 257.2& 905.1\\
\textbf{+Cluster+Judge} & 2636.1& 213.8& 2849.9\\\bottomrule
\end{tabular}%
\end{table}

\input{tables/extraction_accuracy}

Adding a judging step (\textbf{+Judge}) provides the highest extraction accuracy across the majority of models. This means the judge model successfully identifies low-quality characteristics to prompt corrections. However, as shown in \Cref{tab:judge_cluster_increase}, accuracy only improves by an average of 2.11\% and comes at a steep computational price.

As detailed in \Cref{tab:pipeline_costs}, the \textbf{+Judge} configuration increases extraction processing time by an order of magnitude compared to the \textbf{Base} configuration, almost tripling the total experiment duration (from 317.5s to 905.5s). Combining these methods in the \textbf{+Cluster+Judge} configuration creates a bottleneck by increasing processing time by a factor of 9 to over 45 minutes per run,  representing the least efficient trade-off.

Contrary to our initial expectations, clustered extraction (\textbf{+Cluster}) severely degraded performance, returning the lowest accuracy across the board (-18.51\% overall, as shown in \Cref{tab:judge_cluster_increase}). This degradation is most pronounced in larger models (Qwen3-Next-80B-A3B, gpt-oss), which surprisingly under-perform compared to smaller models (Ministral3) in this configuration. We hypothesize that clustering fragments the semantic context: while smaller models may hallucinate or guess over gaps, larger models strictly adhere to the provided (fragmented) context, leading to valid but unhelpful ``Not in Document'' outputs.

\begin{resultbox}[RQ3.2 Summary]
The LLM-as-a-judge loop improves accuracy by a marginal average of 2.11\% but triples processing time. Conversely, clustering fragments document context and degrades accuracy by an average of 18.51\%, hindering larger models. The combination of both methods is infeasible due to significant increases in processing time.
\end{resultbox}

\subsection{\textit{RQ3.3:} Effectiveness of Iterative OML Repair}
\label{sec:oml_refine_loop_results}

As detailed in \Cref{tab:success_rates}, the OML generation results demonstrate that \textbf{iterative repair mechanisms improve OML generation, allowing medium-sized models to compete with larger counterparts.}

\input{tables/oml_generation}

\textbf{Impact of model size.} Retry-based repair loops improve cumulative success rates with four retries ($R_4$) over first-attempt performance ($S_1$), though the utility tends to peak at a certain model size and capability. Smaller models (Ministral3) lack the baseline capability to effectively leverage the repair mechanism. The largest model (gpt-oss-120b) shows diminishing returns (19.0\% improvement) due to its already high initial success rate ($\sim$60\%). Only medium- to large-sized models (gpt-oss-20b, Qwen3-Next-80B-A3B) truly benefit from this mechanism with large improvements (33--39\%). We hypothesize that small differences between these two models emerge from base LLM capabilities and response to prompt structure, which overshadow size differences (20B vs. 80B parameters).

\textbf{Number of retries needed.} \Crefrange{fig:success_rates_p1}{fig:success_rates_p5} illustrate the optimal retry thresholds. The largest model maximizes its benefit after just one or two retries. Medium-sized models (gpt-oss-20B, Qwen3-Next-80B-A3B) require two to three retries to reach their peak success rates. Smaller models show little to no benefit from the retry mechanism at all regardless of the number of attempts.

\textbf{Impact of extraction configuration.} The effectiveness of repair sometimes depends on extraction accuracy: a low accuracy (as is the case of medium to large models as shown in \Cref{tab:success_rates} for the \textbf{+Cluster} configuration) leads to low task complexity, as most of the extracted characteristics are actually \textit{Not in Document}. This explains the high success rates for those models.

\begin{resultbox}[RQ3.3 Summary]
Our repair loop allows 20B parameter models to achieve OML generation success rates comparable to a 120B model, having up to a \textbf{39.6\%} improvement. Practitioners using medium-sized models should plan for three to four generation attempts to reach optimal reliability, while 120B models see diminishing returns beyond two retries. Researchers should also account for input data complexity when benchmarking model generation with LLMs.
\end{resultbox}

\begin{resultbox}[Overall Recommendations]
Based on our evaluation of local LLM extraction and OML generation, we recommend the following configurations depending on hardware and budget constraints:
\begin{itemize}
    \item Small models ($<$14B parameters) are sufficient for data extraction but reliably fail at OML generation. Medium-sized models ($\sim$20B) perform both extraction and generation effectively when paired with an ontology repair loop.
    \item Implement at least two rounds of reasoner-based refinement for OML generation using medium-sized models.
    \item Implement a judge loop only if computational budget and time permit, as the modest accuracy gains ($\sim$2\%) come at a high latency cost. Avoid clustering mechanisms for semantic extraction tasks, as they fracture context and actively degrade performance for larger models.
\end{itemize}
\end{resultbox}

%% file: tables/extraction_accuracy.tex
\begin{table*}[htbp]
    \centering
    \small
    \setlength{\tabcolsep}{3pt}
    \caption{Extraction accuracy ($E_{\text{Acc}}(\%)$, $n=25$), with two judge retries. \colorbox{goodblue}{\textbf{Blue}} denotes best. Papers IDs defined in~\Cref{sub:exemplars}.}
    \label{tab:extraction_accuracy}
    \resizebox{\textwidth}{!}{
    \begin{tabular}{c|l|cccccc}
        \toprule
        \makecell{\textbf{Paper} \\ \textbf{ID}} & \textbf{Configuration}& \makecell{Ministral3 \\ (3B)}
              & \makecell{Ministral3 \\ (8B)}
              & \makecell{Ministral3 \\ (14B)}
              & \makecell{Qwen3-Next \\ (80B)}
              & \makecell{gpt-oss \\ (20B)}
              & \makecell{gpt-oss \\ (120B)} \\
        \hline
        
        \multirow{4}{*}{\textbf{P1}} 
          & \textbf{Base}& 85.7 $\pm$ 7.7 & \textbf{84.6 $\pm$ 3.7} & 82.7 $\pm$ 7.4 & 90.5 $\pm$ 6.0 & 96.2 $\pm$ 5.3 & 90.1 $\pm$ 4.1 \\
          & \textbf{+Cluster}& 73.5 $\pm$ 10.5 & 73.9 $\pm$ 4.2 & 62.3 $\pm$ 4.3 & 45.3 $\pm$ 10.9 & 41.7 $\pm$ 6.6 & 64.0 $\pm$ 16.7 \\
          & \textbf{+Judge}& \textbf{89.1 $\pm$ 5.6} & 84.0 $\pm$ 4.7 & \textbf{83.4 $\pm$ 8.1} & \textbf{92.6 $\pm$ 5.5} & \best{96.6 $\pm$ 3.5} & \textbf{92.2 $\pm$ 3.0} \\
          & \textbf{+Cluster+Judge}& 75.6 $\pm$ 7.7 & 79.4 $\pm$ 7.0 & 72.6 $\pm$ 9.4 & 70.1 $\pm$ 7.2 & 60.4 $\pm$ 8.1 & 68.4 $\pm$ 7.5 \\
        \hline

        \multirow{4}{*}{\textbf{P2}} 
          & \textbf{Base}& 90.9 $\pm$ 1.3 & 87.4 $\pm$ 16.3 & 93.3 $\pm$ 2.4 & 94.3 $\pm$ 5.1 & 98.9 $\pm$ 2.5 & 98.1 $\pm$ 2.7 \\
          & \textbf{+Cluster}& 77.5 $\pm$ 7.3 & 88.8 $\pm$ 7.1 & 81.1 $\pm$ 4.5 & 51.8 $\pm$ 2.9 & 52.6 $\pm$ 5.0 & 78.9 $\pm$ 6.6 \\
          & \textbf{+Judge}& \textbf{94.1 $\pm$ 3.7} & \textbf{93.9 $\pm$ 3.8} & \textbf{95.2 $\pm$ 3.4} & \textbf{97.0 $\pm$ 3.9} & \best{99.2 $\pm$ 1.8} & \textbf{98.9 $\pm$ 2.1} \\
          & \textbf{+Cluster+Judge}& 86.3 $\pm$ 7.6 & 86.5 $\pm$ 8.2 & 84.6 $\pm$ 8.5 & 76.8 $\pm$ 7.4 & 69.3 $\pm$ 9.5 & 78.5 $\pm$ 8.8 \\
        \hline
        
        \multirow{4}{*}{\textbf{P3}} 
          & \textbf{Base}& 72.6 $\pm$ 2.8 & 73.3 $\pm$ 3.6 & 74.9 $\pm$ 4.2 & 88.4 $\pm$ 2.8 & 86.9 $\pm$ 5.7 & 92.4 $\pm$ 2.7 \\
          & \textbf{+Cluster}& 76.2 $\pm$ 9.8 & 75.2 $\pm$ 7.4 & 75.2 $\pm$ 3.6 & 52.8 $\pm$ 3.9 & 61.1 $\pm$ 6.4 & 76.8 $\pm$ 7.1 \\
          & \textbf{+Judge}& 72.2 $\pm$ 1.8 & 73.3 $\pm$ 3.1 & 77.3 $\pm$ 4.8 & \textbf{89.7 $\pm$ 1.8} & \textbf{89.0 $\pm$ 5.1} & \best{93.3 $\pm$ 2.7} \\
          & \textbf{+Cluster+Judge}& \textbf{79.2 $\pm$ 5.6} & \textbf{78.9 $\pm$ 6.5} & \textbf{77.7 $\pm$ 4.5} & 73.3 $\pm$ 7.8 & 77.3 $\pm$ 9.6 & 78.7 $\pm$ 7.4 \\
        \hline
        
        \multirow{4}{*}{\textbf{P4}} 
          & \textbf{Base}& 85.7 $\pm$ 0.0 & 85.7 $\pm$ 0.0 & 88.4 $\pm$ 3.1 & \textbf{91.0 $\pm$ 3.7} & 89.3 $\pm$ 5.9 & 91.0 $\pm$ 4.0 \\
          & \textbf{+Cluster}& 76.2 $\pm$ 6.6 & 85.7 $\pm$ 5.7 & 83.0 $\pm$ 4.8 & 58.1 $\pm$ 19.0 & 76.8 $\pm$ 8.9 & 72.6 $\pm$ 14.0 \\
          & \textbf{+Judge}& \textbf{92.2 $\pm$ 5.1} & \best{94.7 $\pm$ 2.5} & \textbf{93.1 $\pm$ 3.4} & 89.9 $\pm$ 4.0 & \textbf{91.6 $\pm$ 5.5} & \textbf{91.4 $\pm$ 4.3} \\
          & \textbf{+Cluster+Judge}& 81.3 $\pm$ 5.3 & 86.5 $\pm$ 5.9 & 89.9 $\pm$ 5.6 & 73.5 $\pm$ 12.8 & 84.4 $\pm$ 5.9 & 75.4 $\pm$ 10.1 \\
        \hline
        
        \multirow{4}{*}{\textbf{P5}} 
          & \textbf{Base}& 74.3 $\pm$ 3.6 & 71.0 $\pm$ 6.1 & 77.3 $\pm$ 6.3 & \textbf{74.9 $\pm$ 2.9} & \textbf{77.3 $\pm$ 4.6} & 75.8 $\pm$ 2.7 \\
          & \textbf{+Cluster}& \textbf{76.8 $\pm$ 4.0} & 73.0 $\pm$ 3.8 & 77.0 $\pm$ 3.6 & 53.3 $\pm$ 6.7 & 45.3 $\pm$ 7.7 & \textbf{76.2 $\pm$ 9.3} \\
          & \textbf{+Judge}& 74.5 $\pm$ 3.3 & \textbf{73.5 $\pm$ 5.8} & \best{77.9 $\pm$ 4.7} & \textbf{74.9 $\pm$ 3.8} & 77.1 $\pm$ 3.9 & 75.8 $\pm$ 2.3 \\
          & \textbf{+Cluster+Judge}& 74.9 $\pm$ 7.3 & \textbf{73.5 $\pm$ 6.2} & 76.2 $\pm$ 5.0 & 65.5 $\pm$ 12.3 & 66.7 $\pm$ 13.6 & 71.0 $\pm$ 8.1 \\
    
        \hline\hline
        \multicolumn{2}{c|}{\rule{0pt}{4ex}\makecell[c]{\textbf{Best configuration} \\ \textbf{(avg. across papers)}}} & \textbf{+Judge}& \textbf{+Judge}& \textbf{+Judge}&\textbf{+Judge} & \textbf{+Judge}& \textbf{+Judge}\\

        \bottomrule
    \end{tabular}
    }
\end{table*}

%% file: tables/oml_generation.tex
\begin{table*}[htbp]
    \centering
    \small
    \setlength{\tabcolsep}{3pt}
    \caption{Performance comparison on OML generation ($S_1(\%)$ / $R_4(\%)$, $n=25$). $S_1$ denotes the first round success rate, while $R_4$ denotes the cumulative success rate after four retries. \colorbox{goodblue}{\textbf{Blue}} denotes best. Papers IDs defined in~\Cref{sub:exemplars}.}
    \label{tab:success_rates}
    \resizebox{\textwidth}{!}{
    \begin{tabular}{c|l|cccccc}
        \toprule
        \makecell{\textbf{Paper} \\ \textbf{ID}} & \textbf{Configuration} & \makecell{Ministral3 \\ (3B)}
              & \makecell{Ministral3 \\ (8B)}
              & \makecell{Ministral3 \\ (14B)}
              & \makecell{Qwen3-Next \\ (80B)}
              & \makecell{gpt-oss \\ (20B)}
              & \makecell{gpt-oss \\ (120B)} \\
        \hline
        
        \multirow{4}{*}{\textbf{P1}} 
          & \textbf{Base}& 4 $~/~$ 12 & 0 $~/~$ 4 & 4 $~/~$ 24 & 12 $~/~$ \textbf{76} & 28 $~/~$ 84 & 60 $~/~$ 84 \\
          & \textbf{+Cluster}& \textbf{12} $~/~$ \textbf{16} & 0 $~/~$ \textbf{24} & \textbf{32} $~/~$ \textbf{72} & \textbf{64} $~/~$ 72 & \textbf{88} $~/~$ \best{96} & \textbf{92} $~/~$ \textbf{92} \\
          & \textbf{+Judge}& 0 $~/~$ 0 & 0 $~/~$ 4 & 4 $~/~$ 24 & 24 $~/~$ 68 & 48 $~/~$ 88 & 56 $~/~$ \textbf{92} \\
          & \textbf{+Cluster+Judge}& 0 $~/~$ 4 & \textbf{4} $~/~$ \textbf{24} & 20 $~/~$ 56 & 56 $~/~$ \textbf{76} & 72 $~/~$ 92 & 76 $~/~$ \textbf{92} \\
        \hline

        \multirow{4}{*}{\textbf{P2}} 
          & \textbf{Base}& \textbf{0} $~/~$ 0 & \textbf{4} $~/~$ 4 & 4 $~/~$ 32 & 4 $~/~$ 52 & 40 $~/~$ 92 & 64 $~/~$ 76 \\
          & \textbf{+Cluster}& \textbf{0} $~/~$ 0 & \textbf{4} $~/~$ \textbf{28} & \textbf{20} $~/~$ \textbf{52} & \textbf{20} $~/~$ 60 & \textbf{56} $~/~$ \textbf{96} & \textbf{72} $~/~$ \best{96} \\
          & \textbf{+Judge}& \textbf{0} $~/~$ 0 & 0 $~/~$ 4 & 0 $~/~$ 20 & 4 $~/~$ \textbf{64} & 32 $~/~$ 68 & 48 $~/~$ 80 \\
          & \textbf{+Cluster+Judge}& \textbf{0} $~/~$ \textbf{4} & 0 $~/~$ 4 & 16 $~/~$ 32 & 16 $~/~$ 44 & 40 $~/~$ 88 & 56 $~/~$ 76 \\
        \hline
        
        \multirow{4}{*}{\textbf{P3}} 
          & \textbf{Base}& \textbf{0} $~/~$ 0 & 0 $~/~$ 0 & 0 $~/~$ 16 & 24 $~/~$ 64 & 44 $~/~$ 92 & 56 $~/~$ 64 \\
          & \textbf{+Cluster}& \textbf{0} $~/~$ \textbf{8} & \textbf{4} $~/~$ \textbf{32} & \textbf{12} $~/~$ \textbf{64} & \textbf{52} $~/~$ \textbf{72} & \textbf{80} $~/~$ 96 & 64 $~/~$ 80 \\
          & \textbf{+Judge}& \textbf{0} $~/~$ 0 & 0 $~/~$ 4 & 0 $~/~$ 20 & 8 $~/~$ 44 & 36 $~/~$ \best{100} & \textbf{84} $~/~$ \textbf{96} \\
          & \textbf{+Cluster+Judge}& \textbf{0} $~/~$ 0 & 0 $~/~$ 16 & 8 $~/~$ 48 & 28 $~/~$ 56 & 60 $~/~$ 92 & 44 $~/~$ 68 \\
        \hline
        
        \multirow{4}{*}{\textbf{P4}} 
          & \textbf{Base}& \textbf{0} $~/~$ 0 & 0 $~/~$ 12 & 8 $~/~$ 44 & \textbf{36} $~/~$ \textbf{72} & 44 $~/~$ \textbf{96} & 44 $~/~$ 60 \\
          & \textbf{+Cluster}& \textbf{0} $~/~$ 0 & 0 $~/~$ 8 & 8 $~/~$ \textbf{64} & 32 $~/~$ 68 & \textbf{60} $~/~$ 84 & \textbf{56} $~/~$ \textbf{84} \\
          & \textbf{+Judge}& \textbf{0} $~/~$ \textbf{16} & 0 $~/~$ 0 & \textbf{20} $~/~$ 56 & \textbf{36} $~/~$ 60 & \textbf{60} $~/~$ 92 & 48 $~/~$ 72 \\
          & \textbf{+Cluster+Judge}& \textbf{0} $~/~$ 12 & \textbf{4} $~/~$ \textbf{20} & \textbf{20} $~/~$ 44 & 28 $~/~$ 52 & 56 $~/~$ \best{96} & \textbf{56} $~/~$ 68 \\
        \hline

        \multirow{4}{*}{\textbf{P5}} 
          & \textbf{Base}& 0 $~/~$ 8 & \textbf{0} $~/~$ 4 & 4 $~/~$ 28 & 44 $~/~$ 72 & 28 $~/~$ 80 & \textbf{80} $~/~$ \textbf{96} \\
          & \textbf{+Cluster}& 0 $~/~$ 0 & \textbf{0} $~/~$ 8 & \textbf{8} $~/~$ \textbf{48} & \textbf{80} $~/~$ \textbf{92} & \textbf{68} $~/~$ \best{100} & 56 $~/~$ 80 \\
          & \textbf{+Judge}& \textbf{4} $~/~$ 8 & \textbf{0} $~/~$ 0 & 0 $~/~$ 8 & 32 $~/~$ 72 & 28 $~/~$ 96 & 64 $~/~$ 84 \\
          & \textbf{+Cluster+Judge}& 0 $~/~$ \textbf{20} & \textbf{0} $~/~$ \textbf{24} & 0 $~/~$ 12 & 40 $~/~$ 72 & 56 $~/~$ 88 & 68 $~/~$ 84 \\
        \hline

        \hline\hline
      \multicolumn{2}{r|}{\rule{0pt}{3ex}\textbf{Avg. Improv. ($R_4 - S_1$)}} 
        & {4.4\%} & {10.2\%} & {28.8\%} & {33.4\%} & \textbf{39.6\%} & {19.0\%} \\

    \bottomrule
    \end{tabular}
    }
\end{table*}

%% file: text/discussion.tex
\section{Discussion}
\label{sec:discussion}

This section discusses our approach, including the modeling pain points we address, lessons learned from preliminary experimentation, and our threats to validity.

\subsection{Addressing Pain Points}
\label{sec:pain-points}

Prior studies~\cite{sousa2025mapping,sousa2025unraveling} identified 71 pain points associated with domain modeling, and synthesized the 16 most significant and frequently encountered ones. By employing local LLMs to automatically generate accurate DT reports aligning with the DTDF ontology, DTScribe attempts to address several of those pain points:
\begin{description}
\item[(PP04) Effortful Negotiation:] This pain point refers to the time-consuming process of reaching agreement among domain experts. The automatically generated DT characteristics table produced by our LLM-based system provides initial concept definitions, serving as a shared foundation upon which experts can iterate, thereby reducing negotiation effort.%

\item[(PP07) Unclear Conceptualization:] Writing clear and complete model definitions is often considered difficult and tedious. In DTInsight~\cite{fiter2025dtinsight}, we simplified conceptualization by formalizing the DTDF in OML, which acts as a \textit{domain-specific language} for ontology definition. In this paper, we extend that solution by automatically generating an initial OML description that the user can then edit to streamline modeling and reduce cognitive load.

\item[(PP11) Effortful Diagramming:] Creating comprehensible visual representations of ontologies can be labor-intensive. By generating an ontology compatible with DTInsight, our approach produces clear and structured visualizations automatically. This eases the burden of manual diagramming.

\item[(PP16) Unclear Documentation:] Documentation in DT modeling is often sparse or ambiguous. The automatic generation of ontologies, especially when integrated with DTInsight, enables clearer, more interactive, and continuously updated documentation.
\end{description}

Previous research has suggested that ``the additional time required for certain modeling tasks, such as negotiating concept definitions and generating verbalizations, [is] necessary rather than wasteful''~\cite{sousa2025mapping}, and thus automating reporting may prove counterproductive as definitions may not have been properly accepted by all parties. However, we still believe having content to discuss rather than starting from scratch helps modelers and experts reach agreement faster~\cite{lippolis2025ontology}.

\subsection{Expected Benefits}

While conducting a formal empirical evaluation of practical benefits remains challenging due to the emerging, ad-hoc nature of current DT engineering and a lack of standardized case studies, we anticipate that DTScribe will offer significant advantages in real-world industrial scenarios.

First, DTScribe reduces the human effort and technical expertise required for DT reporting. The traditional, manual approach to DT reporting is highly demanding: practitioners must install an OML editor, learn the modeling language, understand the DTDF vocabulary, and manually author the DT description. Previous work indicates that manually populating the DTDF characteristics table takes approximately 50 minutes, even for users already familiar with the framework \cite{gilSystematicReportingFramework2024}. 
We also have anecdotal evidence from a DT engineering course taught by the third author\footnote{\url{https://bentleyoakes.com/digital-twin-engineering-course/}} which suggests a steep learning curve for the ontology and tooling techniques. Students required at least two hours of dedicated tutorials to learn the material. Thus, we conclude that manual reporting takes on the order of several hours, even assisted by our DTInsight tool.

In contrast, the current DTScribe approach enables users to simply provide their existing documentation, processing it in minutes to automatically generate and update the constellation model. We thus can claim that this approach significantly lowers the time and effort barrier to entry to obtain an initial draft model.

Second, this reduction in effort translates to concrete benefits for industrial usage contexts. A challenge in real-world DT engineering is enabling diverse stakeholders to understand the conceptual architecture, as well as the inputs and outputs of services for a DT currently in development. Our vision is that practitioners can use DTScribe to rapidly generate a baseline draft from their existing materials. Even if internal industrial documentation is not as comprehensive as published academic articles, automatically transforming these documents into explicit, interactive, and continuously accessible artifacts provides immediate value~\cite{fiter2025dtinsight}. The resulting DT constellation diagrams establish a common language and shared visualization to better align stakeholders as discussed in previous work~\cite{goffi2025beerdt}.

Finally, industrial DTs are not static and evolve continuously as practitioners add or modify services~\cite{alskaif2025evolution}. The reporting artifacts generated by our pipeline serve as living documents capable of evolving alongside the system. While we leave precise empirical measurements of usability gains (such as exact time savings and reduced effort) and the implementation of incremental documentation updates to future work, DTScribe represents a first step toward accessible, automated DT reporting in practice.
\\

\subsection{Lessons Learned}\label{subsec:lessons_learned}

This section details some lessons learned during our preliminary experiments and potential future approach improvements.

\subsubsection{Multimodal Document Pre-processing}

We experimented with local multimodal pre-processing with a vision LLM (Qwen2.5-VL-7B) to simultaneously extract text and generate textual descriptions for figures. However, we discarded this method due to significant limitations: (a) reliability issues, as the LLM frequently hallucinated content or omitted sentences; (b) high latency, requiring approximately one minute per page; (c) instability, characterized by occasional infinite loops; and (d) system complexity, which introduced an unnecessary point of failure. While general-purpose vision LLMs proved unsuitable, specialized multimodal parsing tools such as Docling\footnote{\url{https://www.docling.ai/}} may offer a more robust solution.

\subsubsection{Hyperparameter Tuning} \label{sub:hyperparameter_tuning}

Despite the literature suggesting hyperparameter tuning improves domain modeling with LLMs~\cite{bulhakov2025investigating}, our preliminary experiments did not generate substantial benefits. As such, we do not utilize finetuning in our approach, but instead highlight that there may be potential in improving LLM output quality further.

\subsubsection{Local/Hosted Approach Tradeoff}

A local deployment setup on consumer GPU with 8 GB of VRAM enabled characteristic extraction and OML generation to be performed directly within our pipeline, providing data privacy and full control over parameters and input data. However, we observed occasional slowdowns under repeated retry loops and large context windows proportional to those shown in \Cref{tab:pipeline_costs}. Preliminary experiments showed certain smaller models, such as LLama3 8B, showed inconsistencies when forced to comply with structured outputs.

In contrast, cloud-based services such as Claude\footnote{\url{https://claude.ai/}} or ChatGPT\footnote{\url{https://chatgpt.com/}} provide high-quality outputs even from unstructured inputs, due to the robustness of their proprietary architectures and larger model size. This reduces the need for aggressive pre-processing and strict prompting strategies. Additionally, external API services such as Ollama Cloud provide lower latency and access to bigger models, reducing the computational burden. However, relying on cloud services introduces limitations related to data privacy, reproducibility risks due to provider-side updates or pricing changes, and availability constraints linked to network connectivity and provider reliability.

In practice, the deployment depends on operational constraints. When sufficient hardware is available and continuous integration is required, a locally deployed setup using API-accessible models can balance performance and control. In contrast, when privacy requirements are minimal and hardware capacity is limited, a fully cloud-based solution may offer higher baseline performance. As open-source models continue to improve, the performance gap between local and hosted approaches is expected to narrow, potentially favoring fully local deployments in future industrial settings.

\subsubsection{LLM-as-a-Judge}

As shown in \Cref{tab:extraction_accuracy}, the pipeline including the LLM-as-a-judge achieves higher extraction accuracy on average, indicating that the judge is generally capable of detecting the hallucinations in the generated characteristics. However, improvement does not consistently translate into better performance in the subsequent OML generation task. One possible explanation is that LLMs tend to perform more accurately on relatively simple evaluation tasks, such as identifying whether a generated statement is grounded in the reference text, than on more complex judgment tasks that require assigning a quality score based on multiple evaluation criteria. As the latter demands  balancing several dimensions simultaneously, which may reduce the alignment of assigned scores.

Furthermore, there is room for experimentation on the judge scoring prompt. Although the prompt was designed to capture the desired qualities of extracted characteristics, the LLM remains constrained by its pre-trained knowledge and may also be biased toward polished writing styles, limiting its ability to accurately interpret domain-specific terminology and capture complex technical relationships in the text. Further prompt engineering may bridge that gap to better align the scoring process with the characteristics that lead to more functional OML generation.

These observations indicate that, despite deterministic configurations and the capabilities of modern LLMs, several factors may still compromise the accuracy of LLM-based judges, producing stable yet misaligned evaluations with respect to human expert judgment. Because of this, its assessments should not be considered equivalent to those of a human domain expert, but rather used as an automated proxy to support the iterative refinement process.

LLM-as-a-judge also introduces substantially high processing times and token cost overhead. As shown in \Cref{tab:pipeline_costs}, adding the judge loop increases processing time from 317.5 seconds (\textbf{Base}) up to 905.1 seconds (\textbf{+Judge}). Given this high overhead for a marginal average accuracy gain of around $2\%$, we recommend against employing LLM-as-a-judge in latency-sensitive scenarios such as automated continuous reporting loops that prioritize execution speed and efficiency. However as LLMs increase in capability and get cheaper, we expect that this tradeoff will diminish, leading us to keep LLM-as-a-judge in our architecture.\\

\subsection{Threats to Validity}

Here we present the threats to validity of our article, based on Wohlin \textit{et al.}~\cite{Wohlin2024Experimentation}.

\subsubsection{Internal Validity}

Threats to internal validity refer to possible errors and biases in our experimentation.

Despite its effectiveness, LLM-based evaluation presents limitations such as lack of reliability and accuracy. Reliability issues arise from the inherent stochastic nature of LLMs. To mitigate this, we enforce deterministic settings by fixing the random seed and setting the temperature to zero to improve consistency across runs within our experimental settings.

The output quality of LLMs also heavily relies on prompts~\cite{crupi2025effectiveness} and architecture. To mitigate the effects of these prompts, we follow the guidelines of Baltes \textit{et al.}~\cite{baltes2025evaluationguidelinesempiricalstudies} and report LLM versioning and the main prompt used in \Cref{lst:extraction_guide}. Similarly, retrieval performance also relies on the queries used, as very detailed and complex queries return low semantically similar fragments. Therefore, we chose to use associated keywords, keeping these as broad as possible while still preserving the meaning of each cluster.

Internal validity is also threatened when we use cloud-hosted LLMs, as the throughput can vary over time due to shared resources and provider-side changes. However, we considered that a cloud-based platform was more stable and performant than running our large and multi-run experimentation on a laptop machine.

\subsubsection{Construct Validity}

Construct validity addresses the choice of evaluation metrics. 

From the literature, suitable metrics for our precise problem definition were not found. Thus, we defined \textit{RQ2} to determine these metrics, as discussed in \Cref{sec:metrics}. These include LLM-judging and comparison against a partial human-defined ground truth, to provide qualitative and quantitative results of the output. Note that a partial ground truth was defined in this work as we lack a robust mapping of DTs and their descriptions in the DTDF. Thus we were unable to use standard metrics such as recall and accuracy.

For the definition of the partial ground truth (\Cref{sec:ground_truth}), we acknowledge the threats to validity introduced by subjective human evaluation. The three raters annotated the characteristics independently, but they are also co-authors of this paper, which introduces a risk of shared bias. To limit this risk, the raters worked independently before their labels were aggregated. The resulting Fleiss' Kappa score of $\kappa = 0.625$ corresponds to substantial agreement~\cite{fleiss1971measuring,landis1977measurement}, indicating that the criteria were applied consistently across raters. While this does not eliminate the possibility of systematic bias inherited from the authors' shared perspective, it supports the reliability of the annotations as a partial ground truth. Replication with independent, external annotators would further strengthen the validity of these results and is left to future work.

Another limitation is our focus on OML validity from the reasoner as a measure of success. In some rare cases, we noticed the LLM could ``cheat'' by generating over-simplified or partial OML to pass this validation. While we already check for empty outputs and treat that as a failure, adding a strict deterministic metric alongside reasoner validity, such as targeting a specific model size or counting OML instances, is challenging because there is no baseline for the required complexity of an early-stage draft model. Instead, we expect this ``lazy LLM'' behavior to be mitigated by the iterative refinement loops that prompts the model to restore missing elements while attempting to maintain well-formedness. In future work, this limitation could be further addressed by introducing stricter constraints directly into the OML vocabulary, though this must be carefully balanced against the flexibility required for generating initial draft models.

As mentioned earlier, a lower characteristic extraction accuracy also tends to produce a higher OML generation success rate, as fewer model elements are produced. This metric therefore somewhat depends on the extraction accuracy and cannot be measured independently. Thus, we acknowledge that this OML generation metric has issues, but it does produce a quantifiable result for pipeline integration and analysis.

\subsubsection{Conclusion Validity}

Conclusion validity concerns whether the conclusions we draw are supported by the results.

Despite the stochastic nature of LLMs, we are confident in the results produced by our rigorous experimentation. Based on these results, we have drawn solid insights and recommendations on the application of LLMs to characteristic extraction and model generation.

\subsubsection{External Validity}

The threat of external validity relates to the generalizability of our conclusions.

The time- and compute-intensive nature of our experimentation necessarily limited the number of papers, LLMs, and configurations that we could investigate.

For our experimentation, we chose six LLMs with a variety of parameter sizes and model families. We believe that this mitigates the threat of using other LLMs in the approach. We also note that LLM performance is increasing, so future models are likely to perform better. For the configurations, we intended to investigate the effects of clustering and LLM-as-a-judge, which are the major decisions of our approach.

For paper selection, we targeted high-quality scientific publications from the DT engineering domain. This sampling strategy was chosen to ground our analysis on well-structured and comparable inputs with consistent target audiences and document lengths.

However, we recognize that academic publications represent a relatively homogeneous benchmark. Real-world industrial DT documentation is often fragmented, unstructured, and distributed across legacy specifications and code repositories, while peer-reviewed DT descriptions remain scarce. Using academic papers allowed us to rigorously evaluate our extraction and generation mechanisms, but generalizability to less structured industrial datasets remains a key limitation and an important direction for future work.

Jin \textit{et al.}~\cite{jin2024evaluation} note that internal company documents tend to be more complex than the cases used in this study. LLMs may struggle to effectively extract information from extensive and highly complex documents, particularly when data is distributed across many sources. Thus, our results may provide an upper bound on the extraction quality compared to use in practice. However, our core contributions of the approach and insights still provide a baseline approach and evaluation for further development.

%% file: text/related.tex
\section{Related Work}\label{sec:related_work}

This section reports on related work concerning LLMs as applied to DTs and model/ontology generation.

\subsection{LLMs for DTs}

There is some initial work on applying LLMs to DTs, primarily to reduce the technical barrier in understanding and integrating technical formats. As far as we are aware, we are the first to target the DT reporting challenge.

Giabbanelli \textit{et al.} suggested using LLMs for integrating modeling and simulation tools~\cite{giabbanelli2025over,giabbanelli2026guidelargelanguagemodels}. They regard ontologies as potential candidates to formally specify concepts to be shared among models, thus mitigating LLM hallucinations. They also consider employing LLMs ``for generating human-readable explanations of source code, simulation models, and simulation results at various levels of technical depth''~\cite{giabbanelli2025over}. However, our goal is \textit{conceptual} DT characteristic report generation. In particular we target a pre-established and OML-formalized DTDF, to facilitate high-level communication and planning among stakeholders.

Gebreab \textit{et al.}~\cite{Gebreab2024DTGenAI} investigated the use of generative AI for data integration in DTs using ontologies, as well as for generating 3D models. Koskinen \textit{et al.}~\cite{koskinen2024challenges} investigated the use of LLMs for generating DT simulation code. Our work instead focuses on DT reporting with LLMs rather than actual construction and integration of DT components.

Other approaches have also considered using LLMs to provide natural language interaction with DTs. In particular, Zhang \textit{et al.}~\cite{zhang2024large} provide a framework for utilizing LLMs to a) explain the DT's decisions, b) transform the user's natural language commands into precise DT actions, and c) with the LLM as a planner itself. Kirner \textit{et al.}~\cite{kirner2025developing} use a locally-hosted LLM, a RAG approach, and a chat-based UI to allow the user to easily query information about a construction site. Again, while these approaches share techniques, we address the problem of extracting  and presenting concrete DT characteristics from text reports.

\subsection{LLMs for Domain Modeling}

The literature has also investigated the application of LLMs to model and ontology generation. 

\subsubsection{Guiding LLMs}
Despite promising success in generation tasks, the well-known issues of semantic misalignment and hallucinations means that LLMs are not perfect and require guidance in their generation~\cite{camara2023assessment}.

For example, Pan \textit{et al.}~\cite{pan2025llm} utilize an Ecore meta-model to guide the LLM to produce valid JSON files, which are then transformed into XMI format. Bozyigit \textit{et al.}~\cite{bozyigit2024generating} investigated automated conceptual model generation from requirements, comparing an LLM with a grammatical rule-based technique. Qualis examined repairing SysML v2 generation errors using LLMs, and argued that future work could focus on ``ontology-backed verification [to] improve efficiency, reproducibility, and the reliability of benchmarking outcomes''~\cite{qualis2025mitigating}. We in fact explicitly use OML with its integrated semantic reasoner, using it to benchmark the improvements of multiple rounds of LLM-based but reasoner-backed repairs.

Chen \textit{et al.} focused on \textit{fully automated} domain modeling without any human intervention using LLMs~\cite{chenAutomatedDomainModeling2023a}. They evaluate the generated results manually with a consensus approach, comparing it to the reference solution of ten problem descriptions from an undergraduate course.
Chen \textit{et al.} later developed a framework for improving the consistency and quality of LLM-generated graphs using the Mermaid diagramming language. It synthesizes candidate graphs to generate the optimal one. It is particularly efficient the smaller the LLM, with five candidates needed for optimal quality improvement using Llama3.1-8B~\cite{chen2025accurate}.
In contrast, our focus is on fully automated \textit{DT characteristics extraction} that we extended to domain modeling using ontologies to allow the modeler to visualize the result as reported by DTInsight.

Mehenni and Zouaq~\cite{mehenniOntology2024} propose an approach close to ours, where they  extract medical concepts and attributes from clinical notes. They use ontology-augmented LLM prompts with a custom constrained decoding strategy, such that the LLM produces summaries using concepts described in the ontology. In contrast, our approach enforces that the OML description model produced by the LLM must conform to the OML vocabulary, and directs the LLM to fix any detected non-conformity.

\subsubsection{Requirements and Architecture Models}

Beyond system models, LLMs have also been used for requirement specification and architecture description. For example, Jin \textit{et al.}~\cite{jin2024evaluation} performed requirements modeling for CPSs using LLMs to generate a \textit{problem diagram}. Hatahet \textit{et al.}~\cite{hatahet2025generating} generate software architecture diagrams and state machines by providing an LLM with a class diagram and associated code. 

We instead focus on DT reporting, by targeting DT characteristics modeling with ontologies to represent a \textit{conceptual architecture diagram} (DT constellation) and plain-text characteristics.

\subsubsection{Ontologies and Knowledge Graphs}

In the literature, research on LLM-generated ontologies has focused on OWL generation
\cite{saeedizade2024navigating,lippolis2025ontology,zhang2024ontochat} rather than OML~\cite{elaasarOpenCAESARBalancingAgility2023} generation. The exception is two unpublished works from NASA JPL. OML Code\footnote{\url{https://github.com/opencaesar/oml-code}} is a VSCode extension being developed for supporting OML generation with LLMs, with a graphical viewer and editor. Another project\footnote{\url{https://www.opencaesar.io/projects/2024-04-12-Document-Analyzer-Using-LLM.html}} is similar to our work, where documents are analyzed based on an OML vocabulary. However, these works are unpublished and do not tackle the precise case of DT reporting that we do.

Yang \textit{et al.} leverage LLMs for ontology and knowledge graph construction in the medical domain \cite{yang2024llm}. They use a hand-built ontology and LLM-generated competency questions (CQs) to expand the knowledge graph, using RAG on PubMed literature data. An LLM judge evaluates responses against the ground truth from a one to five scale using four criteria: \textit{accuracy}, \textit{completeness}, \textit{relevance}, and \textit{consistency}. Kommineni \textit{et al.}~\cite{kommineni2024towards} take a very similar approach, and applies this approach to deep learning methodologies. In our case, rather than using an ontology to scope and structure the LLM output, the DTDF vocabulary guides DTDF characteristics extraction, later used for OML description generation.

Lippolis et al.~\cite{lippolis2025ontology} generated OWL ontology drafts using LLMs based on ontology requirements such as user stories and competency questions (CQs). They proposed two methods, independent and incremental ontology generation, evaluated with OOPS!~\cite{poveda2014oops} and qualitatively assessed by domain experts.  Their work builds on that of  Saeedizade and Blomqvist~\cite{saeedizade2024navigating} who find that smaller LLMs are not able to consistently generate OWL without requiring fine-tuning and extensive prompt engineering, as in~\cite{babaei2023llms4ol,mai2024llms}. In contrast, our approach employs LLMs to generate complete OML ontologies directly from documents and iteratively refine them using reasoner feedback.

Chi{\textcommabelow s} \textit{et al.}~\cite{chisGRAPHxxDSMLEngineering2025} developed a ``low-code knowledge graph building'' tool for easily creating ontologies from a visual drag-and-drop interface, and thus democratize the use of knowledge graphs. Similar to our proposal, they avoid RDF/OWL jargon and encourage users to think about component relationships. However, they do not target DTs nor use OML, and instead propose a general solution for representing flexible graph structures. However, this makes KG creation complex, and does not align with the DTDF.

Chat interfaces for building ontologies with LLMs have also been proposed. Zhang \textit{et al.}~\cite{zhang2024ontochat} employed chat-based LLMs to generate CQs from a domain-expert user story, which were used by ontology engineers to model the ontology. In contrast, we use pre-existing documents to fully automate DT reporting in the DTDF ontology. Similarly, 
Coutinho \textit{et al.}~\cite{Coutinho2025} provide an LLM assistant to generate ontologies based on UFO. In contrast, our work does not have the user in the loop, other than to indicate the source documents and potentially correct the outputted DT description model.

%% file: text/conclusion.tex
\section{Conclusion}\label{sec:conclusion}

This article has described our DTScribe approach to automate the reporting of Digital Twins (DTs). Our approach ingests written descriptions of DTs, and then extracts (using local LLMs) textual and structure characteristics based on our Digital Twin Description Framework (DTDF)~\cite{gilSystematicReportingFramework2024}. LLMs are also used to generate \textit{description models} in the Ontological Modeling Language (OML) to formally represent the DT. These models are queried by our previously-developed DTInsight tool~\cite{fiter2025dtinsight} to produce an interactive visualization and reporting page for the DT\footnote{Such as \url{https://oakeslabmtl.github.io/dtdf-dtscribe/}}.

From our evaluation and results, we show that it is feasible to automate this process with local LLMs. We define quality metrics for our approach including extraction accuracy and generation success, and correlate the decisions in our approach decisions with increases or decreases in these qualities. We then provide specific recommendations on this topic, such as utilizing medium sized models, employing two or three model refinement loops, and avoiding an LLM-as-a-judge or clustering approach.

\subsection*{Future Work}

On the technical side, there are further directions to improve the capabilities and quality of DTScribe. 

For example, one of our next steps is to expand the input data for DTScribe to include multi-modal local or hosted documents from diverse sources. This includes videos, audio recordings, meeting notes, images, etc. which may contain valuable information about the DT.

Another possible improvement is to investigate fine-tuning of the LLMs, as recommended by \cite{babaei2023llms4ol}. In particular, using different fine-tuned LLMs for characteristic extraction and OML ontology generation may improve quality and reliability.%

The framework itself could use further detailing, with
SHERPA~\cite{chen2025sherpa} or Graph of Thoughts (GoT)~\cite{besta2024graph} employed to improve LLM prompting, and automatically search existing knowledge to build the ontology.

One aspect of our current approach is that we do not customize the produced report to individual stakeholders. While this is technically feasible through prompt engineering, we decided to not pursue this customization. Instead, the reports are generated as a general representation since the DT constellation and the reports are intended to support a shared understanding across stakeholders.

Future work could define quality metrics and model generation based on predefined personas~\cite{shankar2024validates}. The goal is to address the ``Effortful Verbalization''~\cite{sousa2025unraveling,sousa2025mapping} pain point (PP05) by automatically generating DT reporting pages that meet the user at their level of expertise, tailored to different project stakeholders such as a manager or an engineer. We could enable weighting of the final output to generate more detailed or broader reports. This would resemble the collection of ontology requirements from user stories~\cite{zhang2024ontochat}, but here we would collect DT characteristics from documents to generate an \textit{ontology-based DT report}.

The user could also be involved to co-construct the DTDF ontology with an LLM agent. Here, the agent could a) create competency questions to evaluate the knowledge it has formulated with regards to the documents it has, or b) directly ask the user for details when missing information about a characteristic, and then update the DT model over time to ensure alignment with the actual DT. This resembles past work to help ontology engineers model ontologies~\cite{zhang2024ontochat}, and other recent works~\cite{cabot2025vibe,silva2025towards,burgueno2024human,braun2020crowd}.

We are also in discussions with larger research networks to develop DT descriptions for their DTs. We plan to conduct experiments to build DT reporting pages for their DTs, and validate our approach and tooling at a wider scale. Doing so would also likely involve aligning the DTDF with other emerging standards~\cite{vicat2026cross}.

%% file: text/acknowledgment.tex
\section{Acknowledgments}

We would like to thank Boqi (Percy) Chen for his feedback and insights for domain modeling with LLMs. This research was supported by a MITACS Globalink Research Internship and the NSERC Discovery project \textit{Accelerating DT Construction} (no. RGPIN-2024-05622).

%% file: text/appendix.tex
\clearpage
\onecolumn
\appendix
\section{Appendix} \label{sec:appendix}

\small
\renewcommand{\arraystretch}{1.25}
\setlength{\tabcolsep}{3pt}

\begin{longtable}{@{} p{0.24\linewidth} p{0.70\linewidth} @{}}
    
    \caption{DTDF~\cite{gilSystematicReportingFramework2024} characteristic extraction prompts across all clusters.}
    \label{tab:characteristics_combined}                                                                                                                                                                                                                                                                  \\
    \toprule
    \multicolumn{2}{c}{\textbf{$\mathcal{CC}_1$: Purpose Cluster}}                                                                                                                                                                                                                                        \\ 
    \midrule
    \endfirsthead
    
    \caption[]{DTDF characteristic extraction prompts (continued).}                                                                                                                                                                                                                                                   \\
    \toprule
    \multicolumn{2}{c}{\textit{Continued from previous page}}                                                                                                                                                                                                                                             \\
    \midrule
    \endhead
    
    \midrule
    \multicolumn{2}{r}{\textit{Continued on next page...}}                                                                                                                                                                                                                                                \\
    \endfoot
    
    \bottomrule
    \endlastfoot

    $C_1$: System under study              & Describe the Physical Twin system being studied. Include the type of system, its main components, operational domain, and key characteristics that make it suitable for digital twinning.                                                                    \\ \addlinespace
    $C_6$: DT services                     & Detail the specific services that the Digital Twin provides to users and the physical system. Include service types (monitoring, optimization, prediction, control, visualization), target users, and service capabilities.                                  \\ \addlinespace
    $C_{11}$: Tooling and enablers         & Describe the specific tools, technologies, frameworks, and enablers used to implement the Digital Twin. Include software platforms, development tools, simulation engines, databases, and any specialized technologies with their roles and functionalities. \\ 
    
    \midrule[\heavyrulewidth] %
    \multicolumn{2}{c}{\textbf{$\mathcal{CC}_2$: Orchestration Cluster}}                                                                                                                                                                                                                                  \\
    \midrule
    $C_7$: Twinning time-scale             & Describe the temporal aspects of the Digital Twin including synchronization frequencies between physical and digital components, time scales for different services, latency requirements, and temporal granularity needs.                                   \\ \addlinespace
    $C_8$: Multiplicities                  & Describe the multiplicities and hierarchical structure including multiple digital twin instances, centralized vs decentralized architectures, coordination mechanisms between twins, and scope of responsibilities for each instance.                        \\ \addlinespace
    $C_{12}$: DT constellation             & Describe the overall orchestration and coordination of the Digital Twin system including architecture patterns, component integration, resource management, and system-wide coordination mechanisms.                                                         \\ \addlinespace
    $C_{18}$: Horizontal integration       & Describe integration with external systems including data exchange protocols, interoperability standards, dependencies on external systems, and integration patterns with other Digital Twins or enterprise systems.                                         \\
    
    \midrule[\heavyrulewidth]
    \multicolumn{2}{c}{\textbf{$\mathcal{CC}_3$: Components Cluster}}                                                                                                                                                                                                                                     \\
    \midrule
    $C_{10}$: DT models and data           & Describe the computational models and data components including model types (geometric, behavioral, analytical), data sources, data management approaches, model relationships, and their specific roles in the Digital Twin constellation.                  \\ \addlinespace
    $C_2$: Physical acting components      & Describe the physical actuators and control mechanisms including types of actuators, control capabilities, remote control interfaces, actuation ranges and limitations, safety constraints, and addressing mechanisms.                                       \\ \addlinespace
    $C_3$: Physical sensing components     & Describe the sensing infrastructure including sensor types, measurement capabilities, spatial distribution, sampling frequencies, data transmission rates, accuracy and precision specifications, and data collection mechanisms.                            \\ \addlinespace
    $C_{14}$: Fidelity and validity        & Describe model fidelity requirements, validation and verification methods, uncertainty sources and quantification, error detection and correction mechanisms, and quality assurance processes for models and data.                                           \\
    
    \midrule[\heavyrulewidth]
    \multicolumn{2}{c}{\textbf{$\mathcal{CC}_4$: Connectivity Cluster}}                                                                                                                                                                                                                                   \\
    \midrule
    $C_4$: Physical-to-virtual interaction & Describe data flows from physical to digital including data types transmitted, communication protocols, transmission frequencies, event triggers, data quality assurance, preprocessing steps, and data integration mechanisms.                              \\ \addlinespace
    $C_5$: Virtual-to-physical interaction & Describe control flows from digital to physical including command types, control message formats, validation and authentication mechanisms, feedback systems, conflict resolution, and command execution confirmation.                                       \\ \addlinespace
    $C_{15}$: DT technical connection      & Describe the technical network infrastructure including communication protocols, network topology and architecture, bandwidth and latency requirements, reliability and redundancy mechanisms, and security protocols for network communication.             \\ \addlinespace
    $C_{16}$: DT hosting/deployment        & Describe the deployment infrastructure including hosting platforms (cloud/edge/on-premise), computational and storage requirements, scalability mechanisms, deployment models supported, and availability/reliability assurance.                             \\
    
    \midrule[\heavyrulewidth]
    \multicolumn{2}{c}{\textbf{$\mathcal{CC}_5$: Life-cycle Cluster}}                                                                                                                                                                                                                                     \\
    \midrule
    $C_9$: Life-cycle stages               & Describes the lifecycle phases in which the DT takes place. It also informs which representation phase the DT covers of its physical counterpart, i.e., as designed (ideal), as manufactured, or as operated.                                                \\ \addlinespace
    $C_{13}$: Twinning process             & Describe the engineering and development processes including methodologies used, quality assurance practices, testing and validation approaches, version control, and evolution management strategies.                                                       \\ \addlinespace
    $C_{17}$: Insights and decision making & Describe the decision support capabilities including types of insights generated, analytics and processing capabilities, decision making processes, user interaction methods, and confidence measures for recommendations.                                   \\ \addlinespace
    $C_{20}$: Standardization              & Describe standards compliance including relevant industry standards, specifications followed, certification requirements, compliance frameworks, and interoperability standards adopted.                                                                     \\
    
    \midrule[\heavyrulewidth]
    \multicolumn{2}{c}{\textbf{$\mathcal{CC}_6$: Governance Cluster}}                                                                                                                                                                                                                                     \\
    \midrule
    $C_{19}$: Data ownership and privacy   & Describe data governance including data ownership policies, privacy protection measures, compliance with data protection regulations, consent management, data anonymization, and data sharing agreements.                                                   \\ \addlinespace
    $C_{21}$: Security and safety          & Describe security and safety measures including cybersecurity protocols, access control mechanisms, authentication systems, threat mitigation strategies, fail-safe mechanisms, risk assessment approaches, and safety-critical system considerations.       \\
    
\end{longtable}

\clearpage
\twocolumn

\begin{table*}[tb]
    \caption{RAG querying keywords for each characteristic cluster}
    \label{tab:cluster_queries}
    \centering
    \small
    \renewcommand{\arraystretch}{1.25}
    \resizebox{\textwidth}{!}{
    \begin{tabular}{@{}p{0.13\linewidth}p{0.83\linewidth}@{}}
        \toprule
        \textbf{Cluster}                        & \textbf{RAG Querying Keywords}                                                              \\ 
        \midrule
        
        $\mathcal{CC}_1$\newline(Purpose)       & 
        \textbf{System under study:} system purpose and objectives, twin description, operational context, performance objectives, stakeholder goals. \par\smallskip
        \textbf{DT services:} monitoring and simulation, prediction and optimization, decision-support tools, control and actuation, visualization and interaction. \par\smallskip
        \textbf{DT enablers:} technologies, tools, data acquisition pipeline, modeling and simulation tools, AI/ML components.                \\ 
        \midrule
        
        $\mathcal{CC}_2$\newline(Orchestration) & 
        \textbf{DT temporal aspects:} synchronization frequencies, time scales, and temporal requirements. \par\smallskip
        \textbf{System multiplicities:} multiple twins, distributed architectures, coordination mechanisms. \par\smallskip
        \textbf{DT constellation orchestration:} system-wide coordination, component integration, service orchestration. \par\smallskip
        \textbf{External integration:} data exchange with other systems, interoperability, horizontal integration patterns.                   \\ 
        \midrule
        
        $\mathcal{CC}_3$\newline(Components)    & 
        \textbf{DT models and data:} computational models, data sources, model types, data management, model validation. \par\smallskip
        \textbf{Physical acting components:} actuators, control mechanisms, remote control capabilities, actuation systems. \par\smallskip
        \textbf{Physical sensing components:} sensors, measurement systems, data collection mechanisms, sensing technologies. \par\smallskip
        \textbf{Fidelity and validation:} accuracy requirements, validation methods, uncertainty quantification, verification processes.      \\ 
        \midrule
        
        $\mathcal{CC}_4$\newline(Connectivity)  & 
        \textbf{Physical-to-digital flows:} sensor data transmission, communication protocols, data preprocessing, event handling. \par\smallskip
        \textbf{Digital-to-physical flows:} command transmission, control signals, feedback mechanisms, command validation. \par\smallskip
        \textbf{Network infrastructure:} communication protocols, network topology, bandwidth requirements, connection reliability. \par\smallskip
        \textbf{Deployment infrastructure:} hosting platforms, computational resources, cloud/edge deployment, scalability considerations.    \\ 
        \midrule
        
        $\mathcal{CC}_5$\newline(Life-cycle)    & 
        \textbf{DT lifecycle phases:} development stages, deployment phases, evolution processes, maintenance activities. \par\smallskip
        \textbf{Engineering processes:} development methodologies, quality assurance, testing approaches, version control. \par\smallskip
        \textbf{Decision support:} insight generation, analytics capabilities, decision making processes, user interactions. \par\smallskip
        \textbf{Standards and compliance:} industry standards, certification requirements, compliance frameworks, interoperability standards. \\ 
        \midrule
        
        $\mathcal{CC}_6$\newline(Governance)    & 
        \textbf{Data governance:} data ownership policies, privacy regulations, data protection measures, consent management. \par\smallskip
        \textbf{Security considerations:} cybersecurity measures, access control, authentication, threat mitigation, security protocols. \par\smallskip
        \textbf{Safety requirements:} safety-critical considerations, fail-safe mechanisms, risk assessment, hazard analysis. \par\smallskip
        \textbf{Compliance frameworks:} regulatory compliance, data protection laws, industry regulations, audit requirements.                \\ 
        \bottomrule
        
    \end{tabular}
    }
\end{table*}

\input{figs/results/oml_generation_success}

%% file: figs/results/oml_generation_success.tex
\begin{figure*}[!htbp]
    \centering
    \captionsetup[subfigure]{justification=centering}
    
    \begin{subfigure}[b]{0.30\textwidth}
        \centering
        \includegraphics[width=\textwidth]{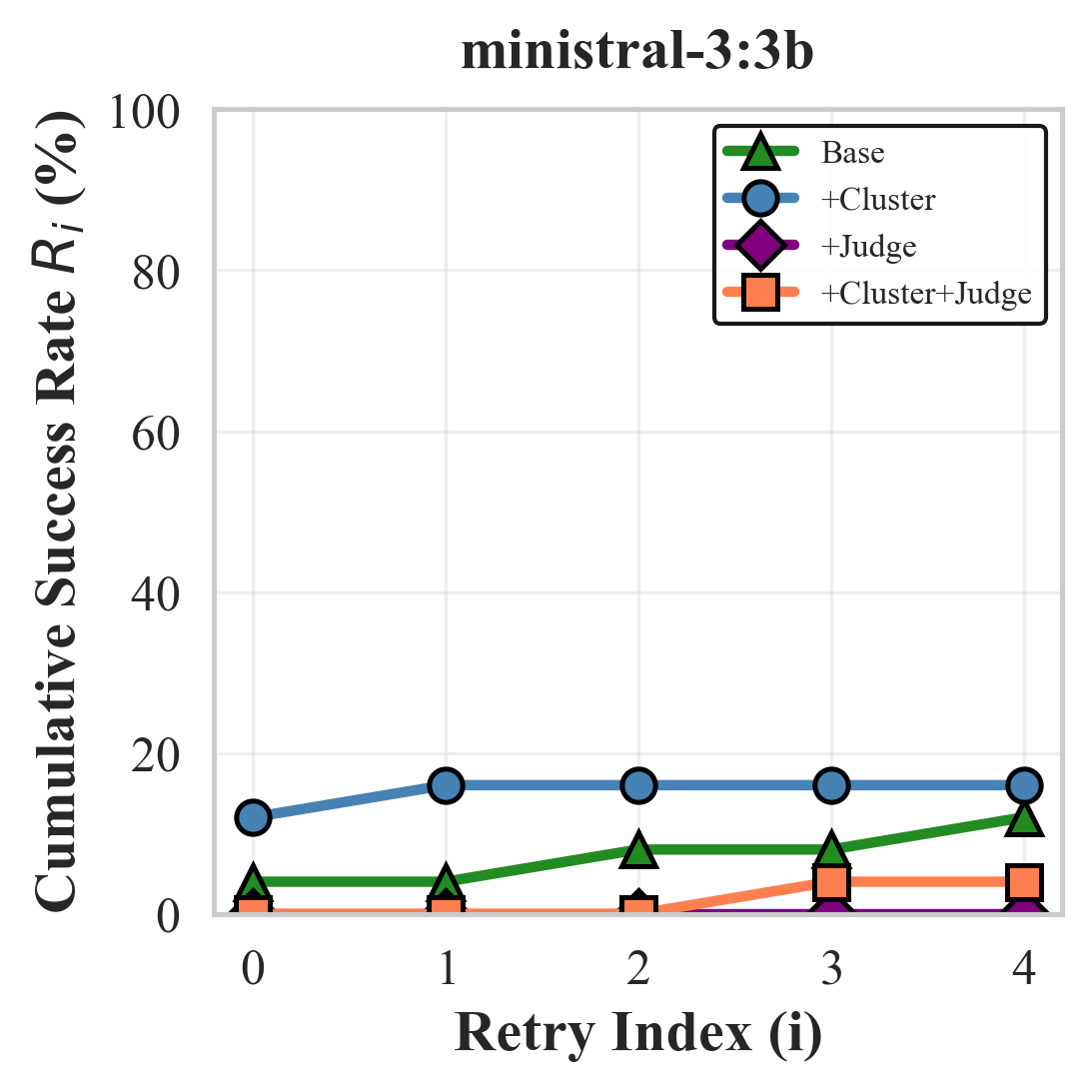}
        \label{fig:p1_1}
    \end{subfigure}
    \hfill
    \begin{subfigure}[b]{0.30\textwidth}
        \centering
        \includegraphics[width=\textwidth]{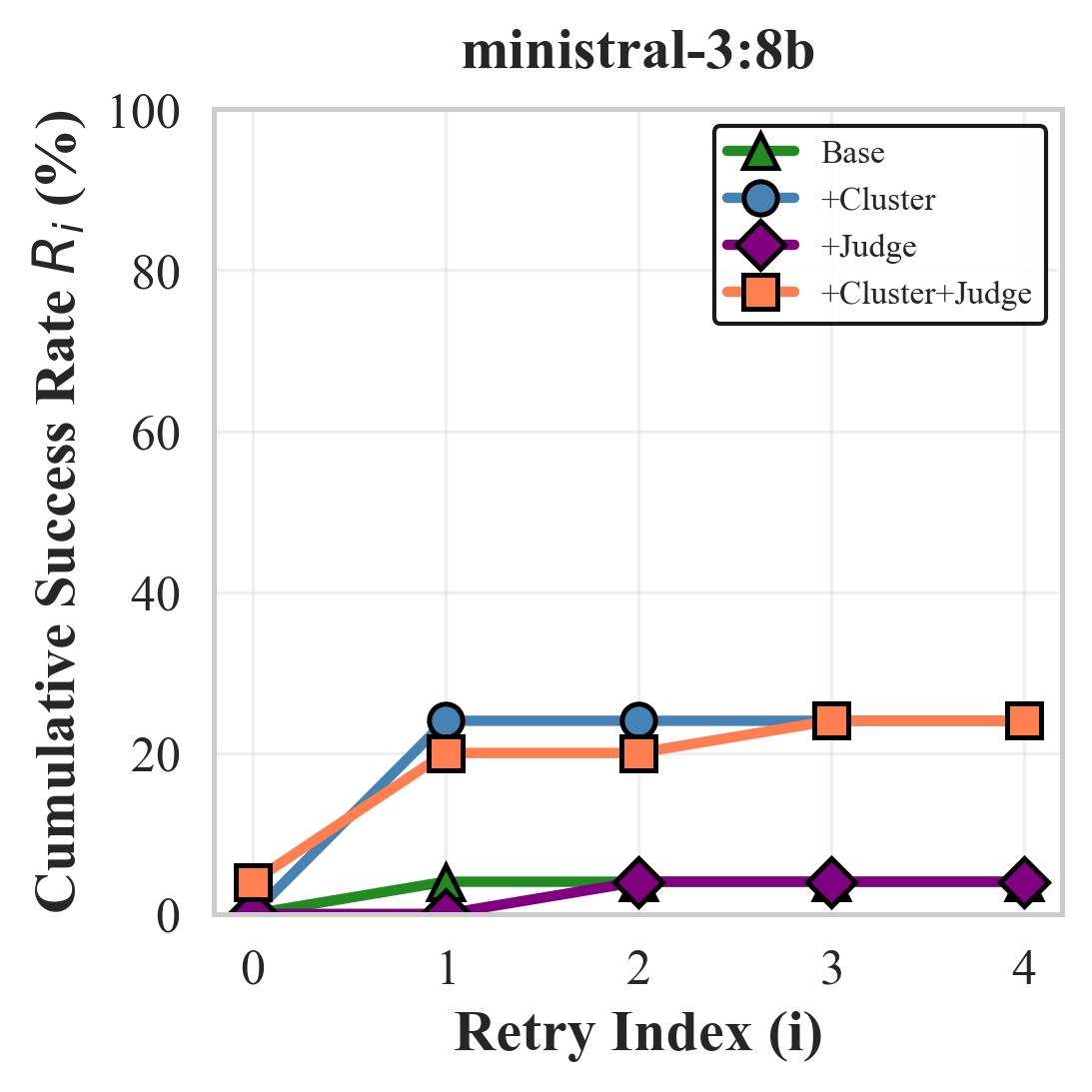}
        \label{fig:p1_2}
    \end{subfigure}
    \hfill
    \begin{subfigure}[b]{0.30\textwidth}
        \centering
        \includegraphics[width=\textwidth]{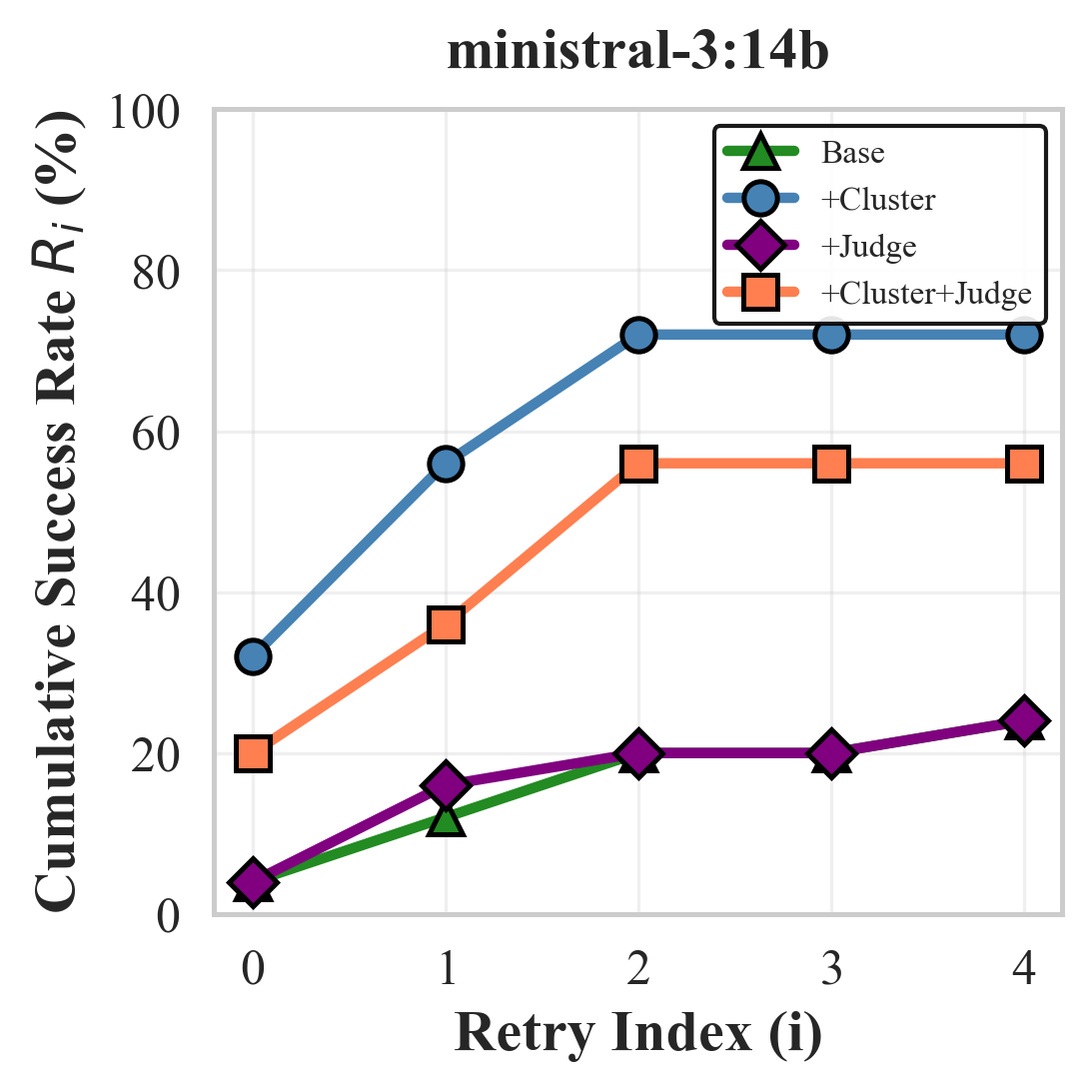}
        \label{fig:p1_3}
    \end{subfigure}

    \vspace{0.2cm}

    \begin{subfigure}[b]{0.30\textwidth}
        \centering
        \includegraphics[width=\textwidth]{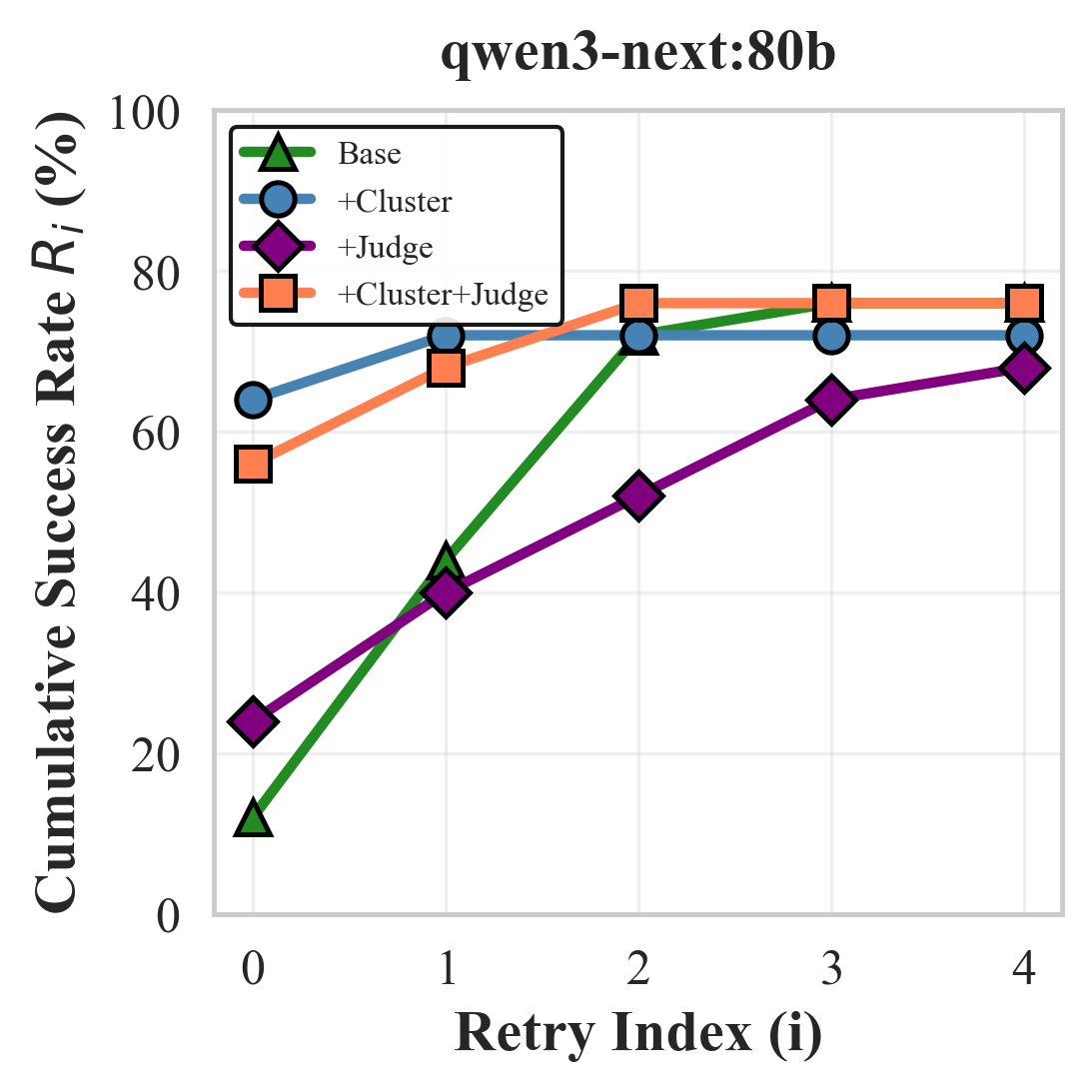}
        \label{fig:p1_4}
    \end{subfigure}
    \hfill
    \begin{subfigure}[b]{0.30\textwidth}
        \centering
        \includegraphics[width=\textwidth]{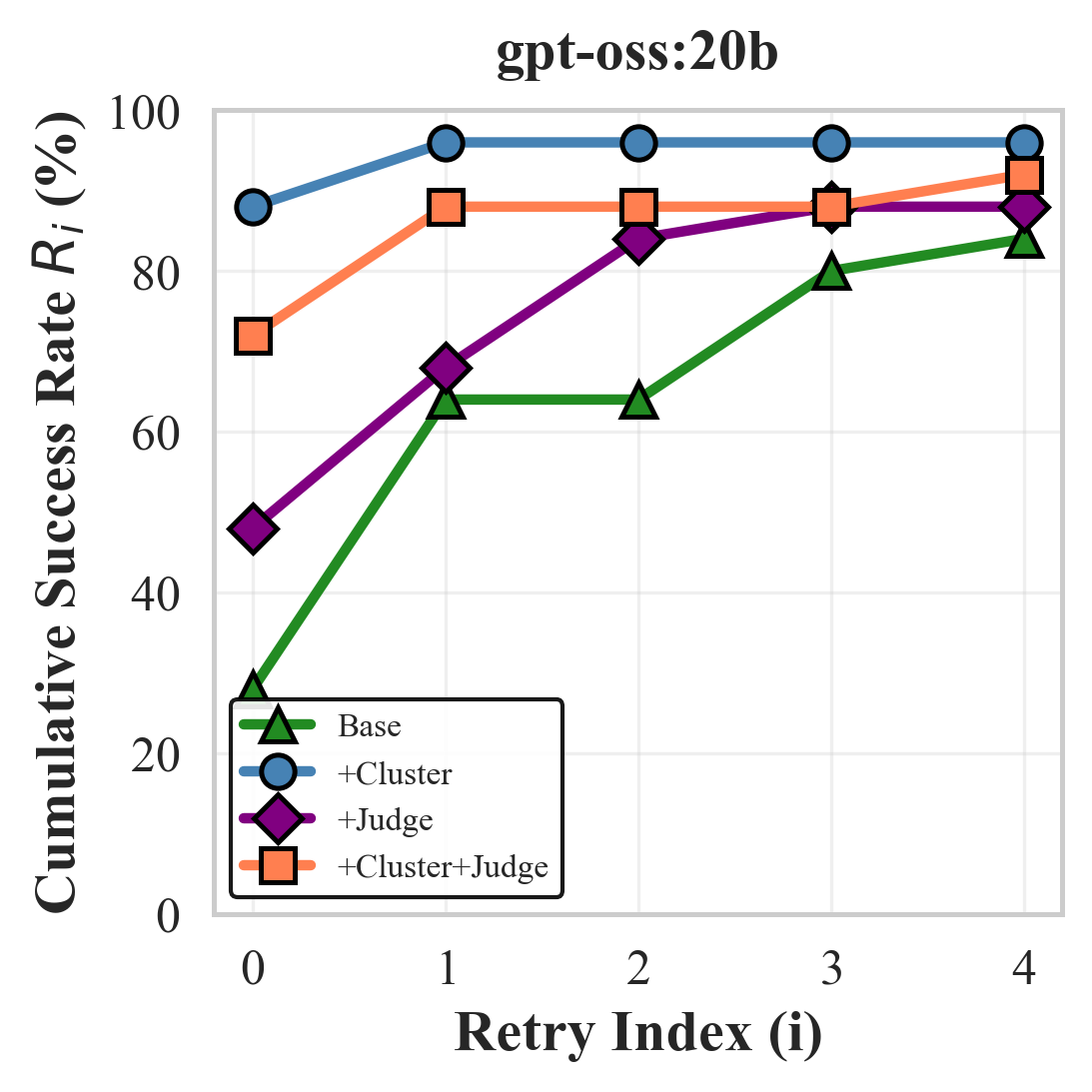}
        \label{fig:p1_5}
    \end{subfigure}
    \hfill
    \begin{subfigure}[b]{0.30\textwidth}
        \centering
        \includegraphics[width=\textwidth]{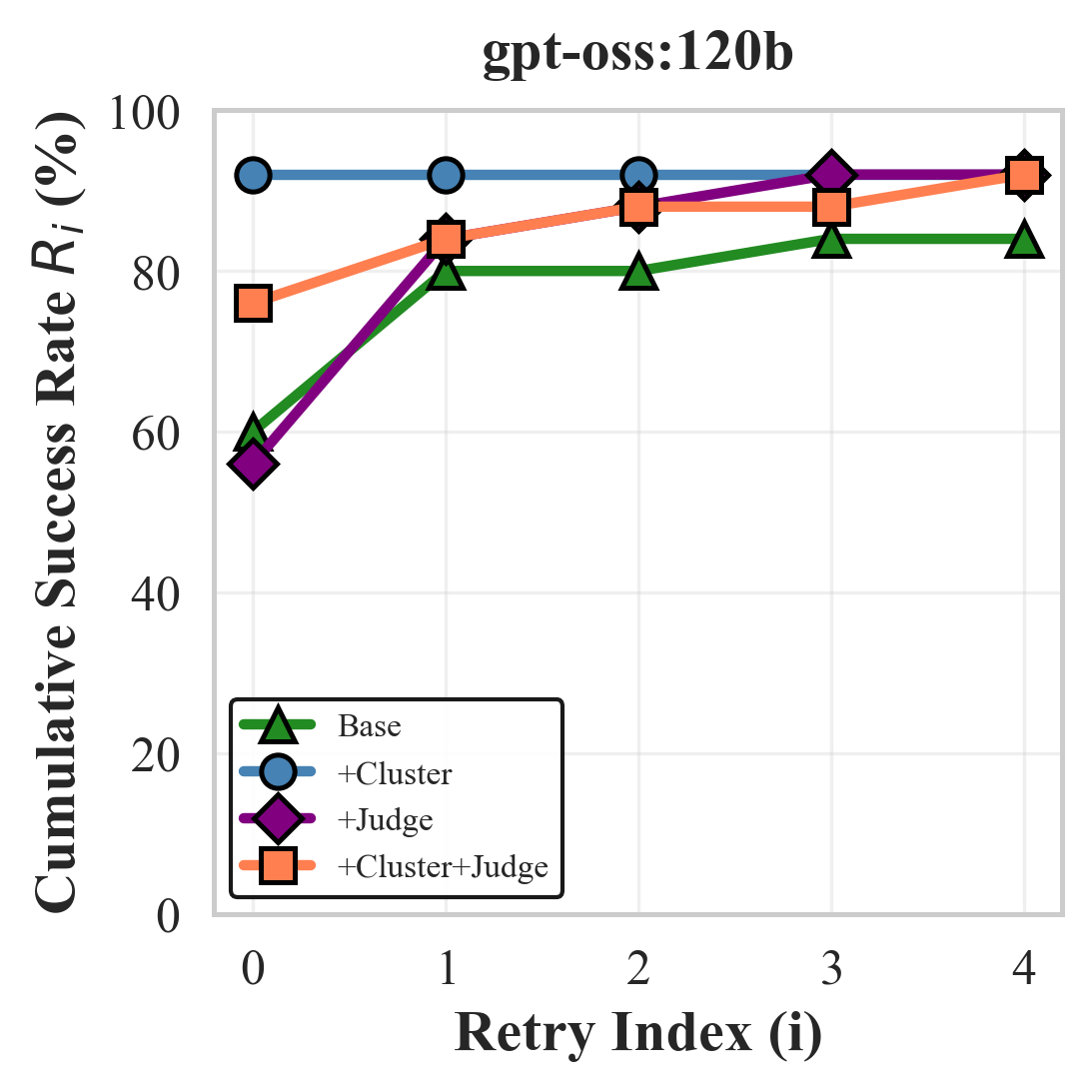}
        \label{fig:p1_6}
    \end{subfigure}

    \caption{Cumulative success rate of OML generation across four repair attempts ($n=25$, \textbf{P1}).}
    \label{fig:success_rates_p1}

    \vspace{0.6cm} 

    \begin{subfigure}[b]{0.30\textwidth}
        \centering
        \includegraphics[width=\textwidth]{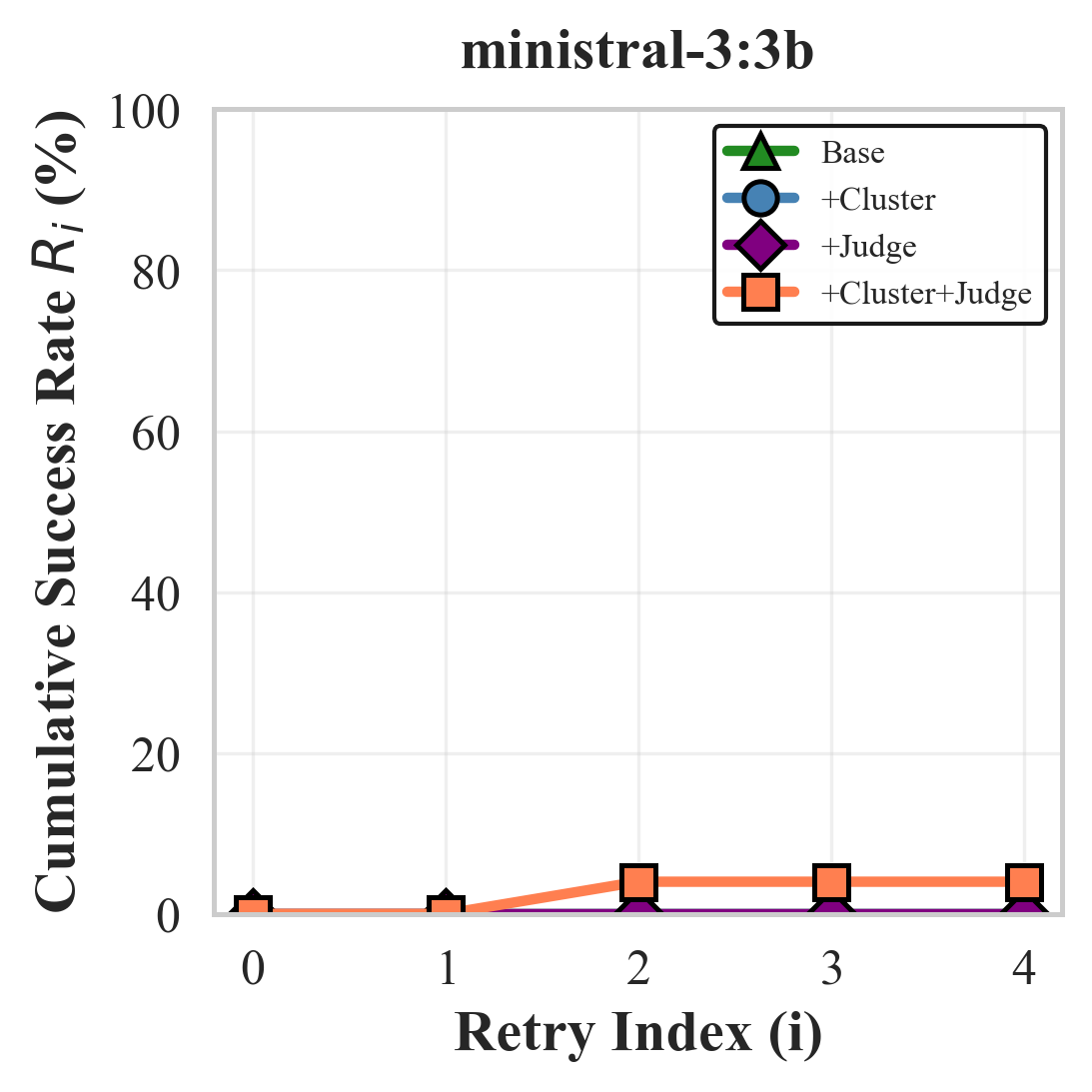}
        \label{fig:p2_1}
    \end{subfigure}
    \hfill
    \begin{subfigure}[b]{0.30\textwidth}
        \centering
        \includegraphics[width=\textwidth]{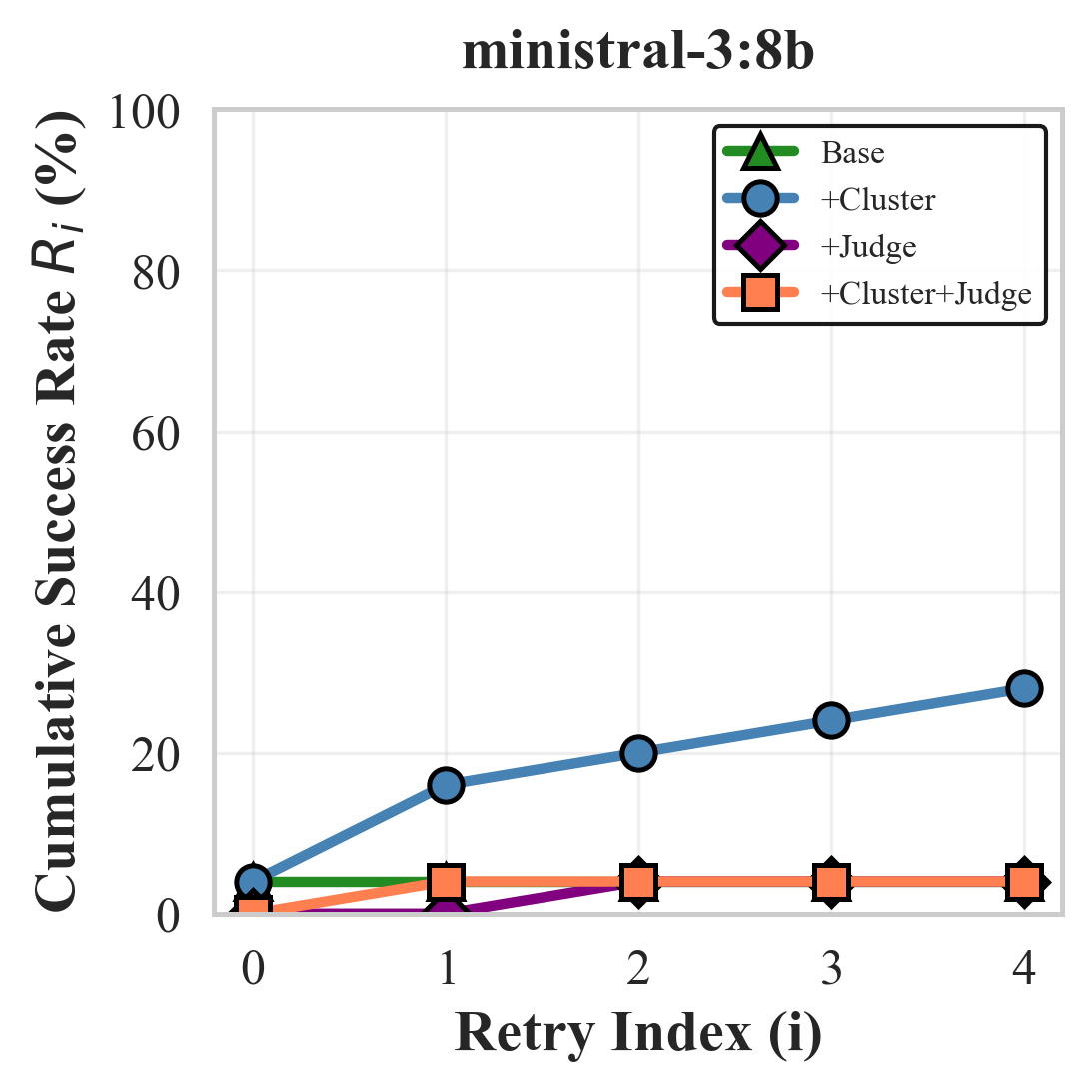}
        \label{fig:p2_2}
    \end{subfigure}
    \hfill
    \begin{subfigure}[b]{0.30\textwidth}
        \centering
        \includegraphics[width=\textwidth]{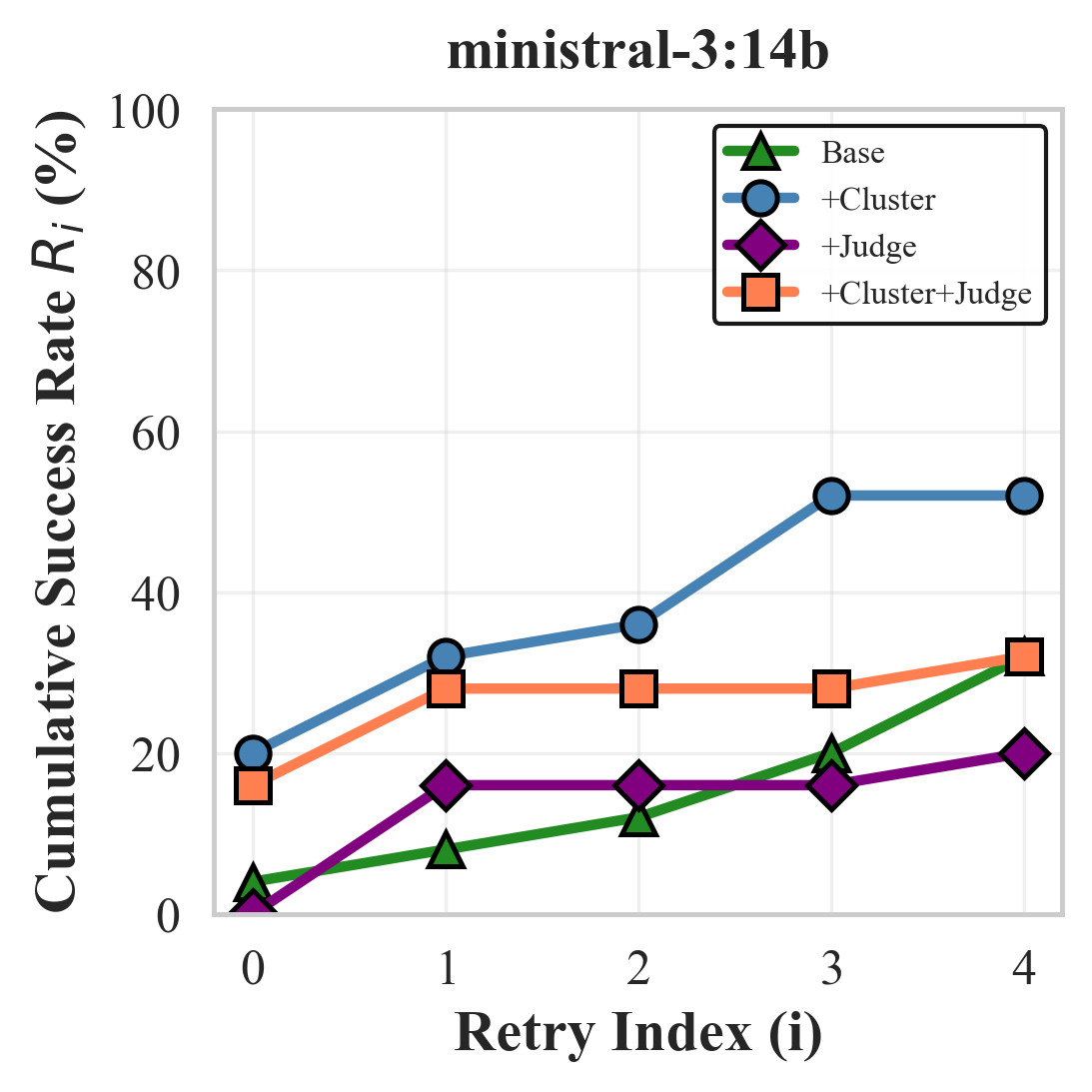}
        \label{fig:p2_3}
    \end{subfigure}

    \vspace{0.2cm} 

    \begin{subfigure}[b]{0.30\textwidth}
        \centering
        \includegraphics[width=\textwidth]{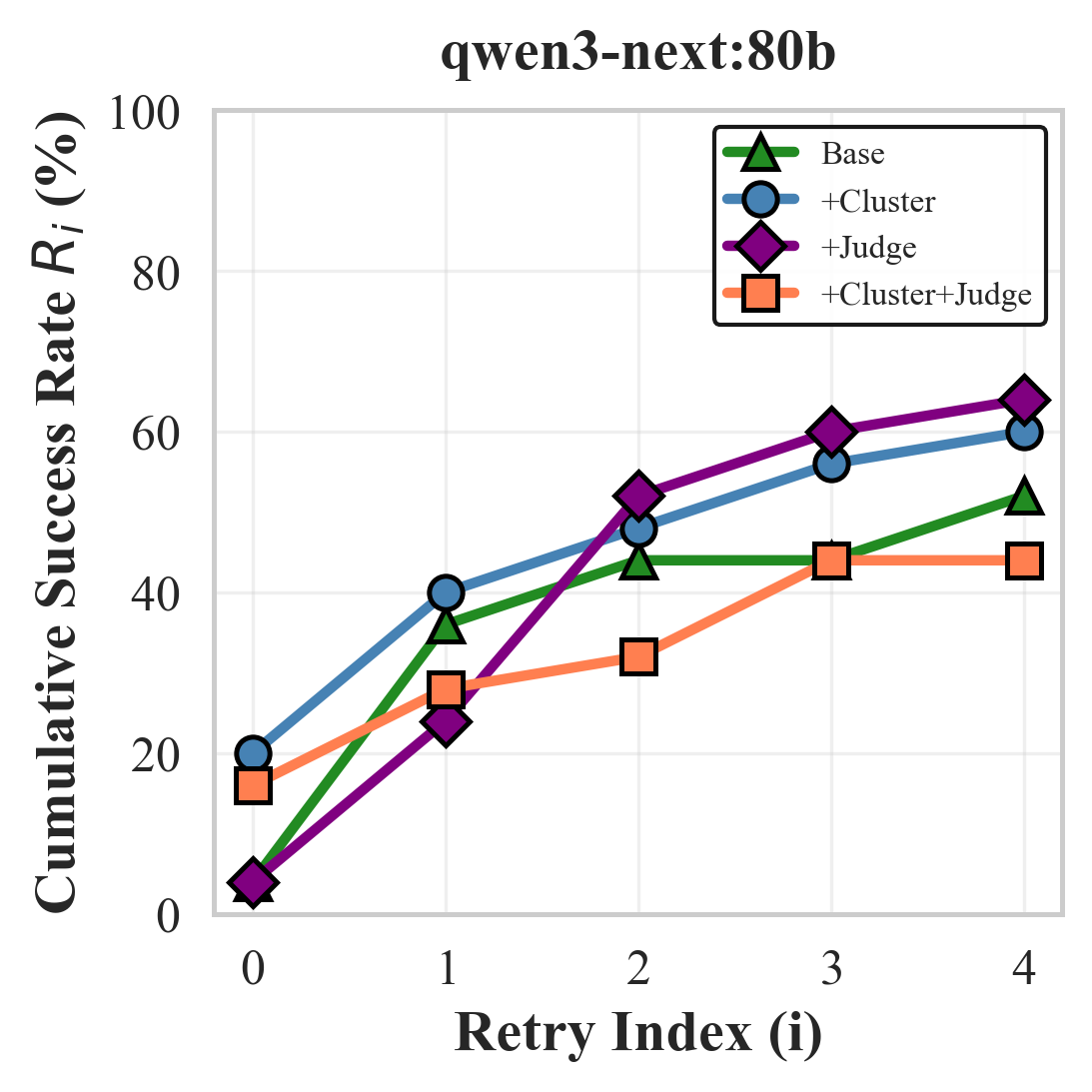}
        \label{fig:p2_4}
    \end{subfigure}
    \hfill
    \begin{subfigure}[b]{0.30\textwidth}
        \centering
        \includegraphics[width=\textwidth]{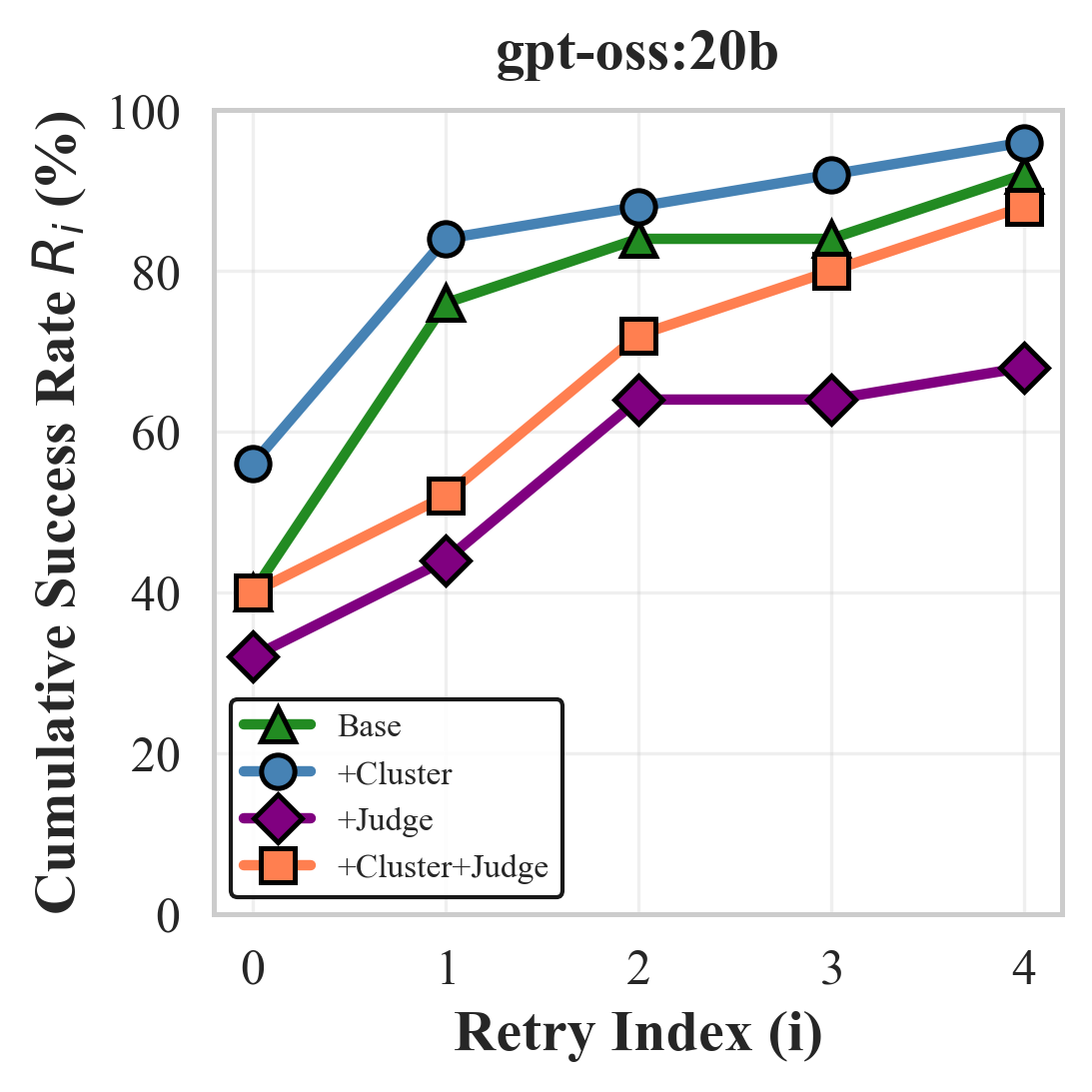}
        \label{fig:p2_5}
    \end{subfigure}
    \hfill
    \begin{subfigure}[b]{0.30\textwidth}
        \centering
        \includegraphics[width=\textwidth]{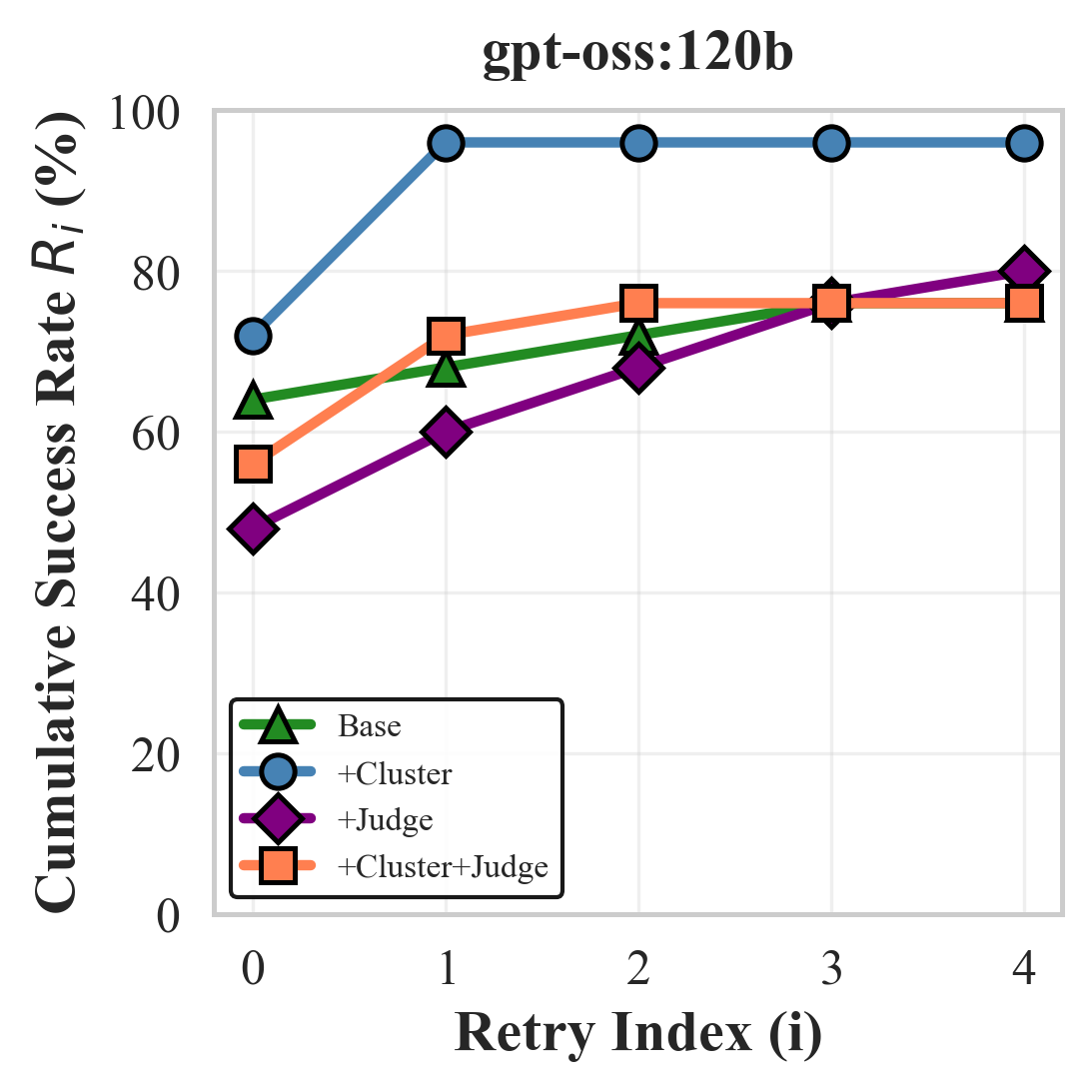}
        \label{fig:p2_6}
    \end{subfigure}

    \caption{Cumulative success rate of OML generation across four repair attempts ($n=25$, \textbf{P2}).}
    \label{fig:success_rates_p2}
\end{figure*}

\begin{figure*}[!htbp]
    \centering
    \captionsetup[subfigure]{justification=centering}
    
    \begin{subfigure}[b]{0.30\textwidth}
        \centering
        \includegraphics[width=\textwidth]{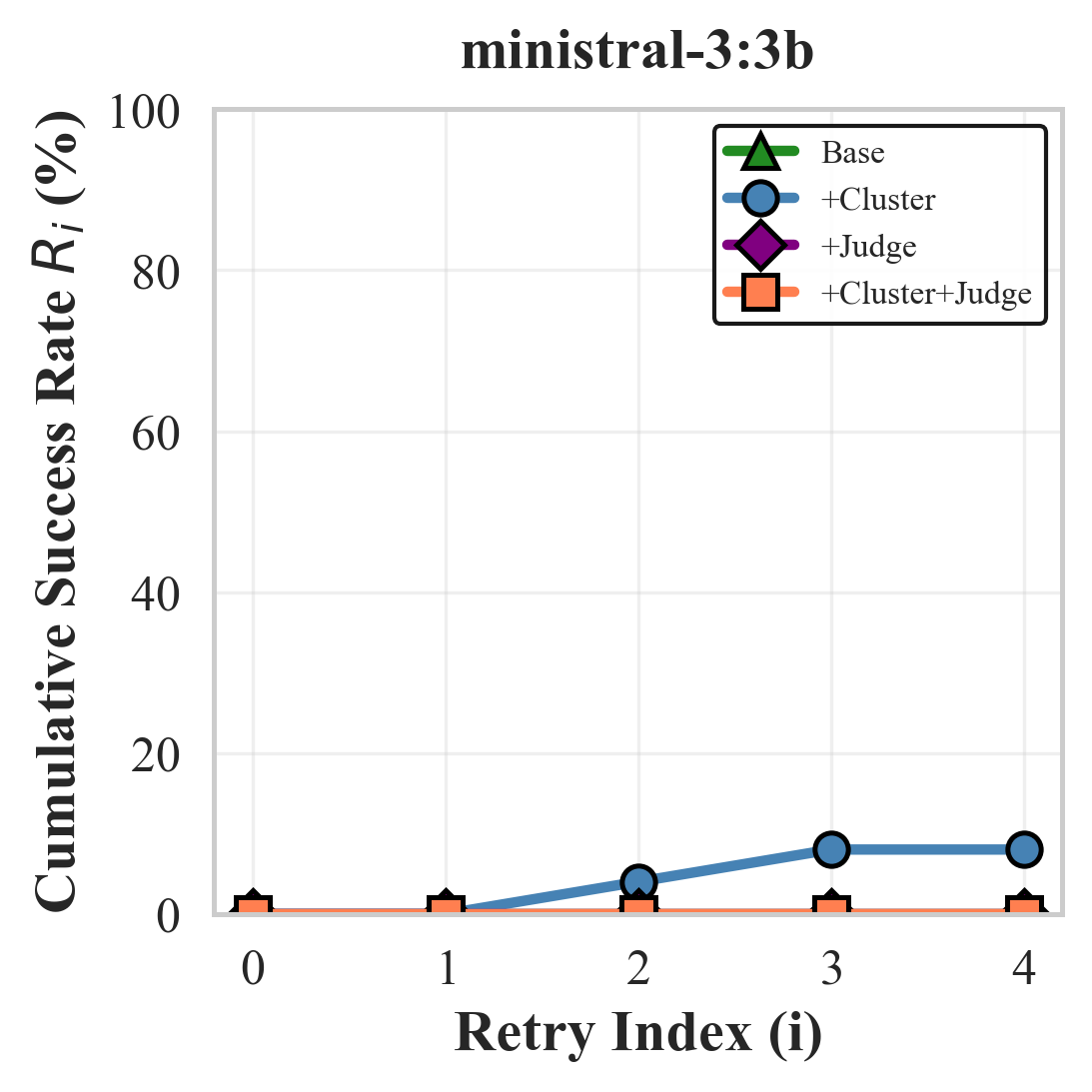}
        \label{fig:p3_1}
    \end{subfigure}
    \hfill
    \begin{subfigure}[b]{0.30\textwidth}
        \centering
        \includegraphics[width=\textwidth]{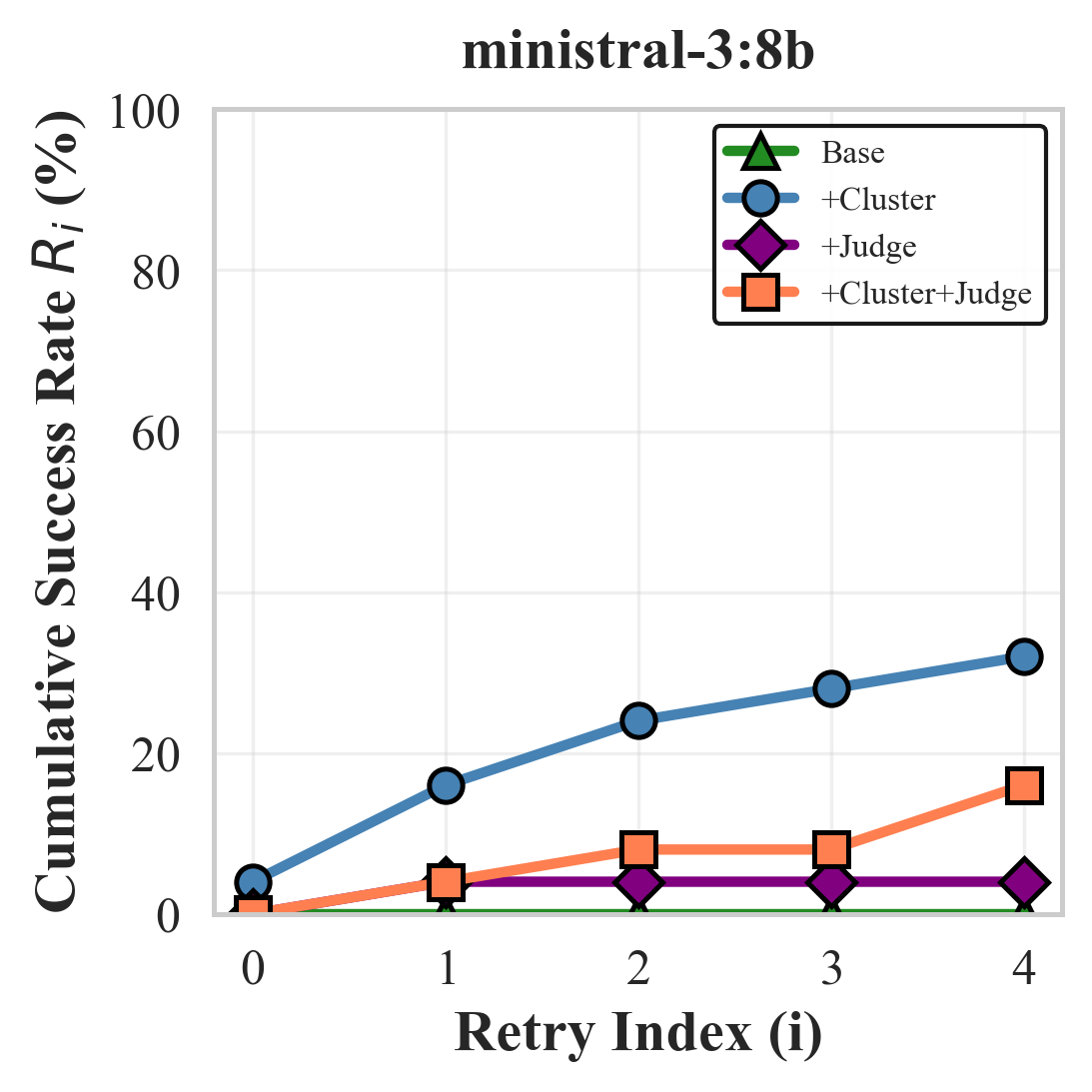}
        \label{fig:p3_2}
    \end{subfigure}
    \hfill
    \begin{subfigure}[b]{0.30\textwidth}
        \centering
        \includegraphics[width=\textwidth]{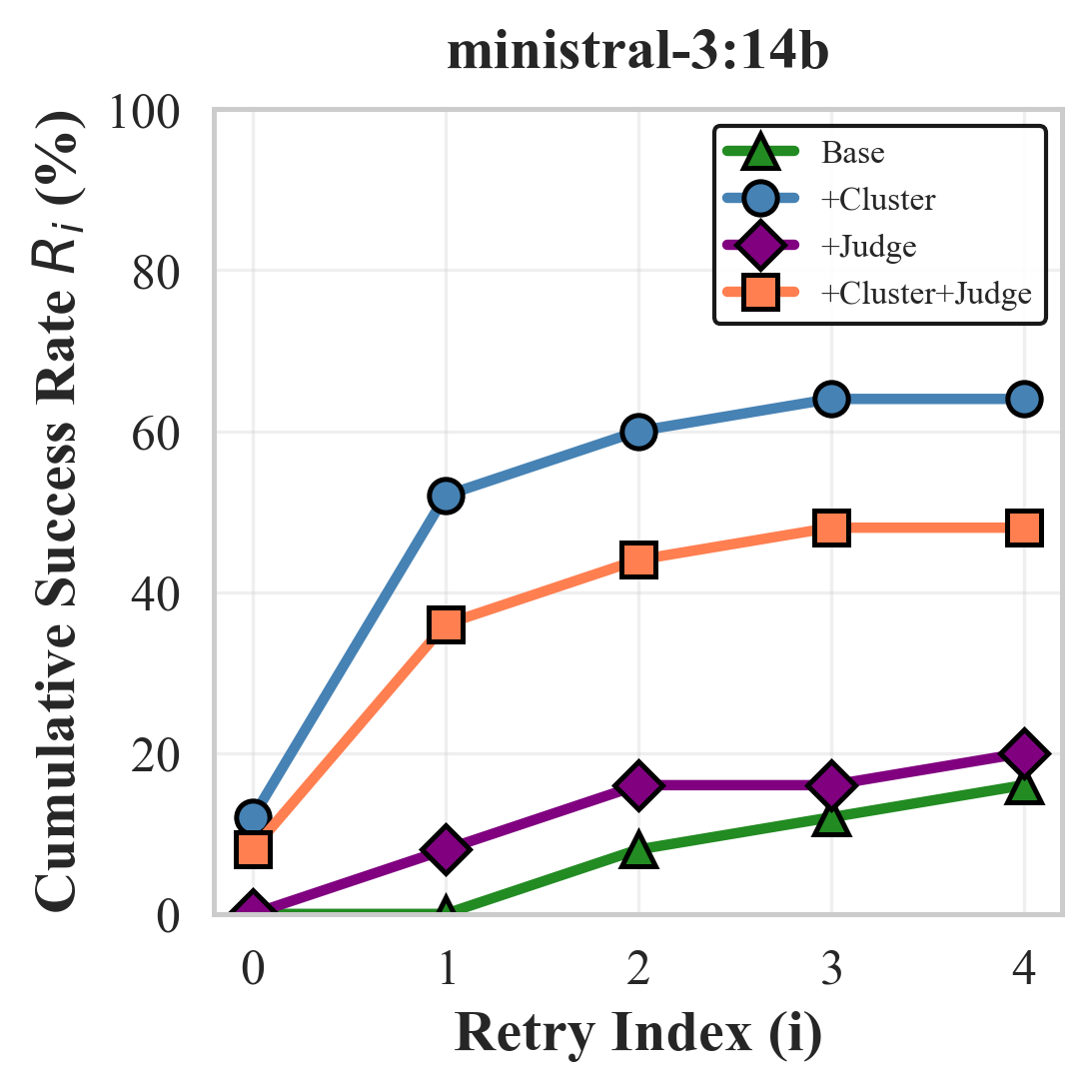}
        \label{fig:p3_3}
    \end{subfigure}

    \vspace{0.2cm}

    \begin{subfigure}[b]{0.30\textwidth}
        \centering
        \includegraphics[width=\textwidth]{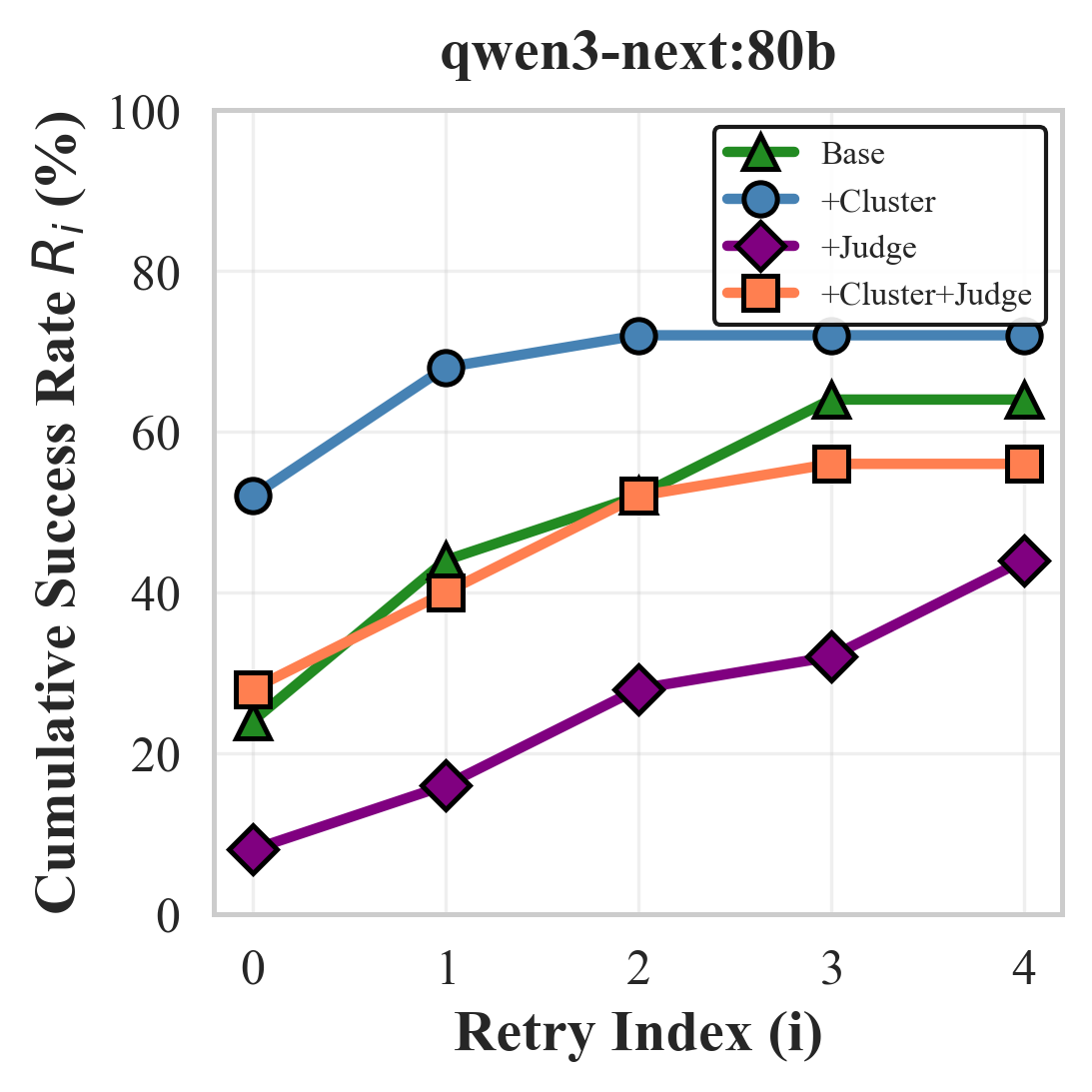}
        \label{fig:p3_4}
    \end{subfigure}
    \hfill
    \begin{subfigure}[b]{0.30\textwidth}
        \centering
        \includegraphics[width=\textwidth]{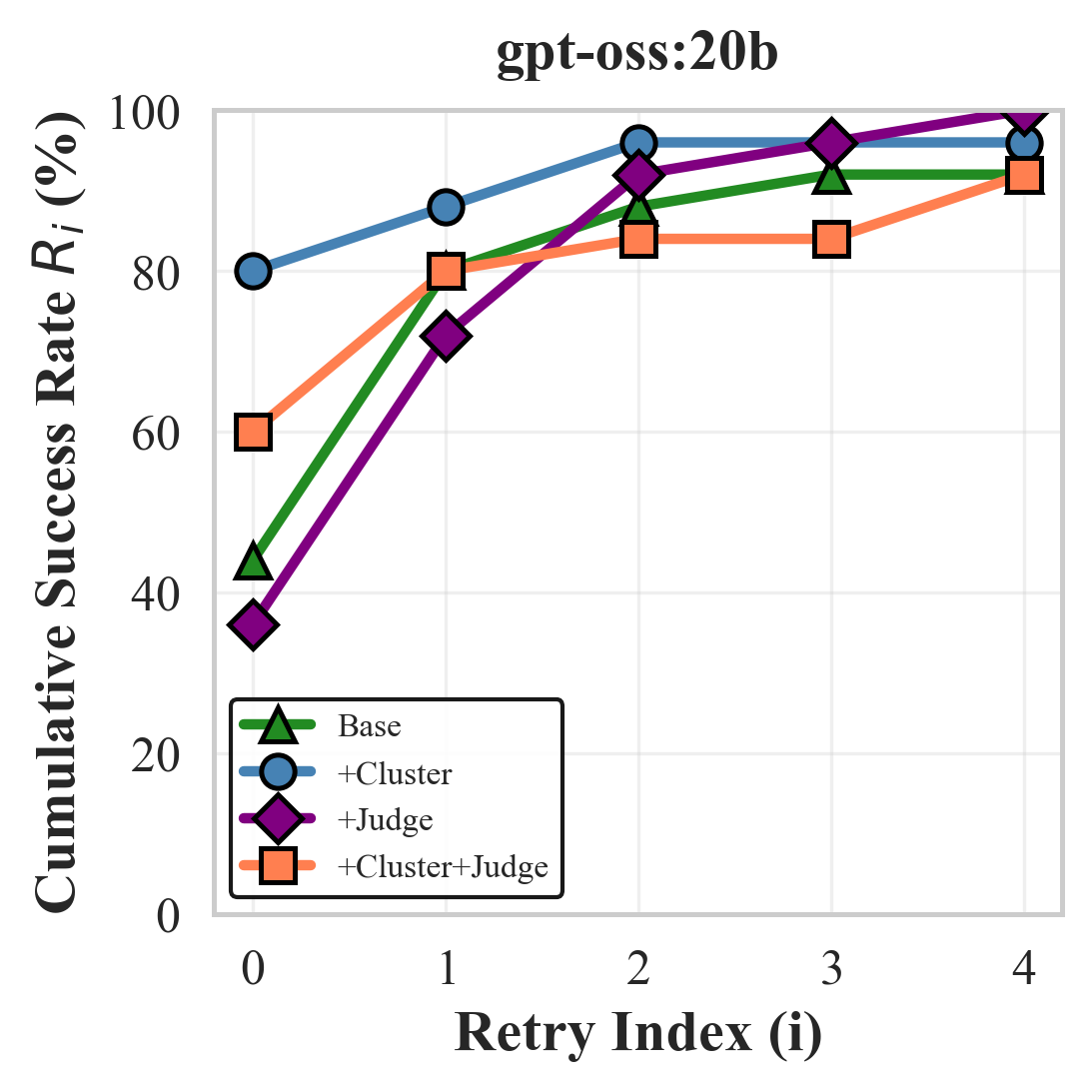}
        \label{fig:p3_5}
    \end{subfigure}
    \hfill
    \begin{subfigure}[b]{0.30\textwidth}
        \centering
        \includegraphics[width=\textwidth]{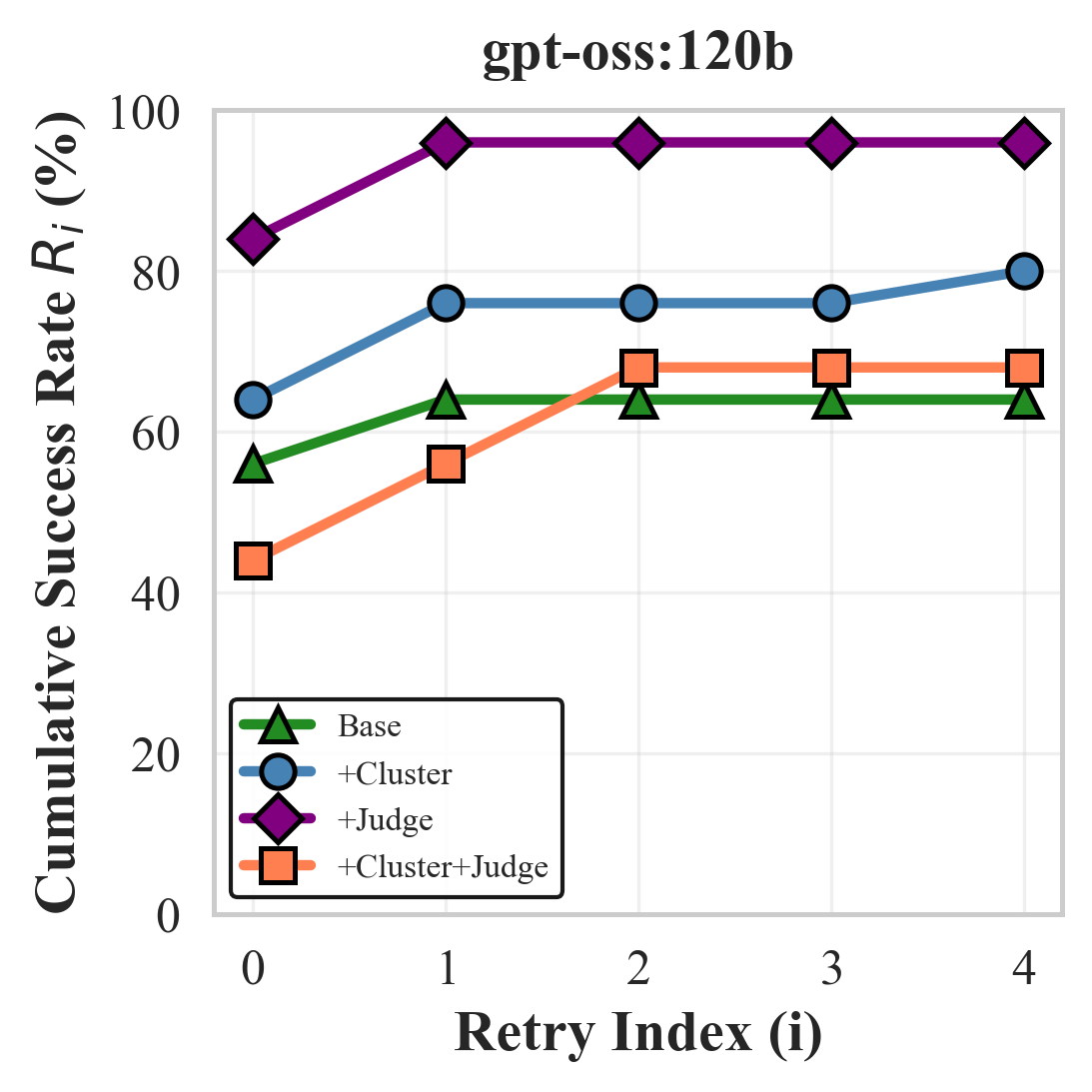}
        \label{fig:p3_6}
    \end{subfigure}

    \caption{Cumulative success rate of OML generation across four repair attempts ($n=25$, \textbf{P3}).}
    \label{fig:success_rates_p3}

    \vspace{0.6cm}

    \begin{subfigure}[b]{0.30\textwidth}
        \centering
        \includegraphics[width=\textwidth]{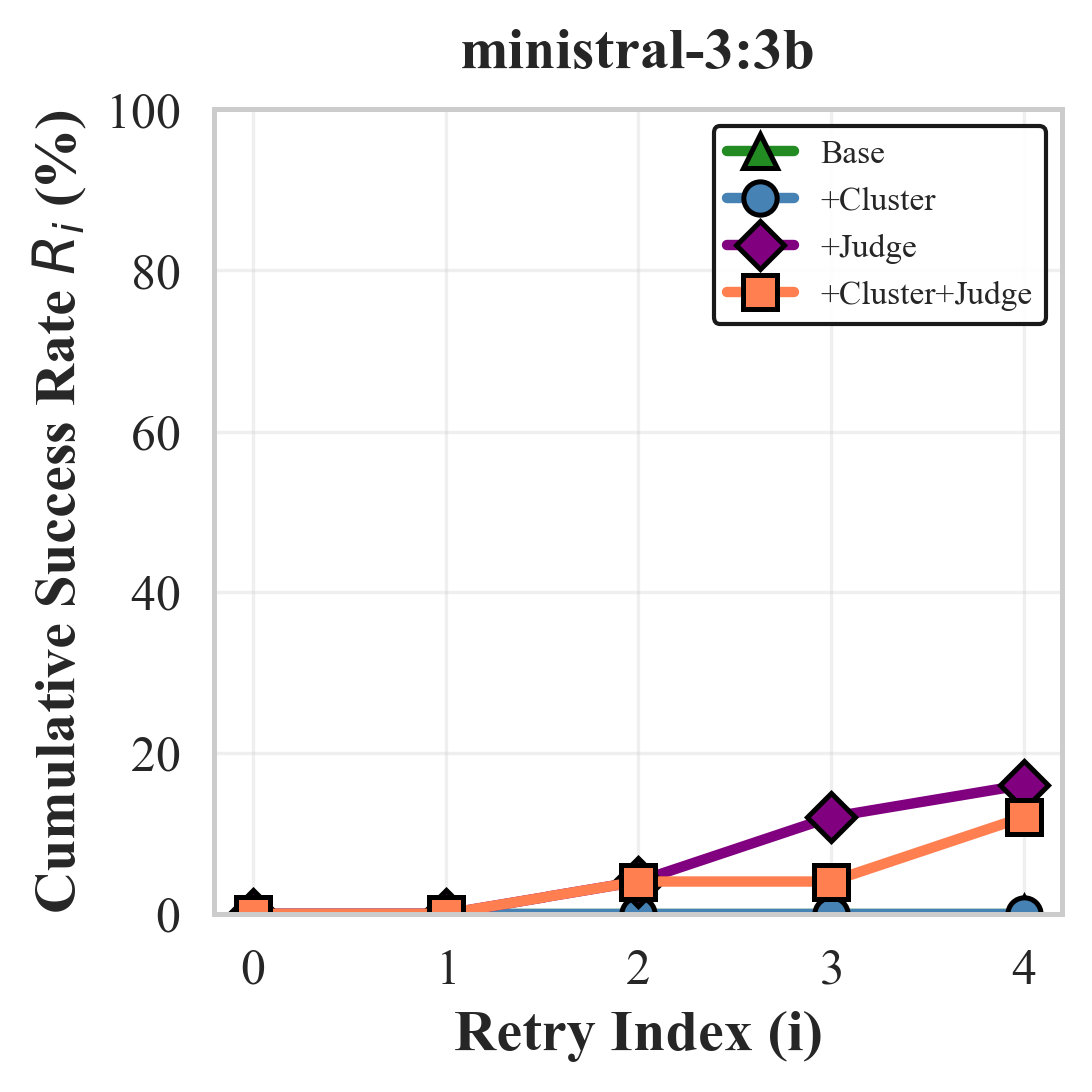}
        \label{fig:p4_1}
    \end{subfigure}
    \hfill
    \begin{subfigure}[b]{0.30\textwidth}
        \centering
        \includegraphics[width=\textwidth]{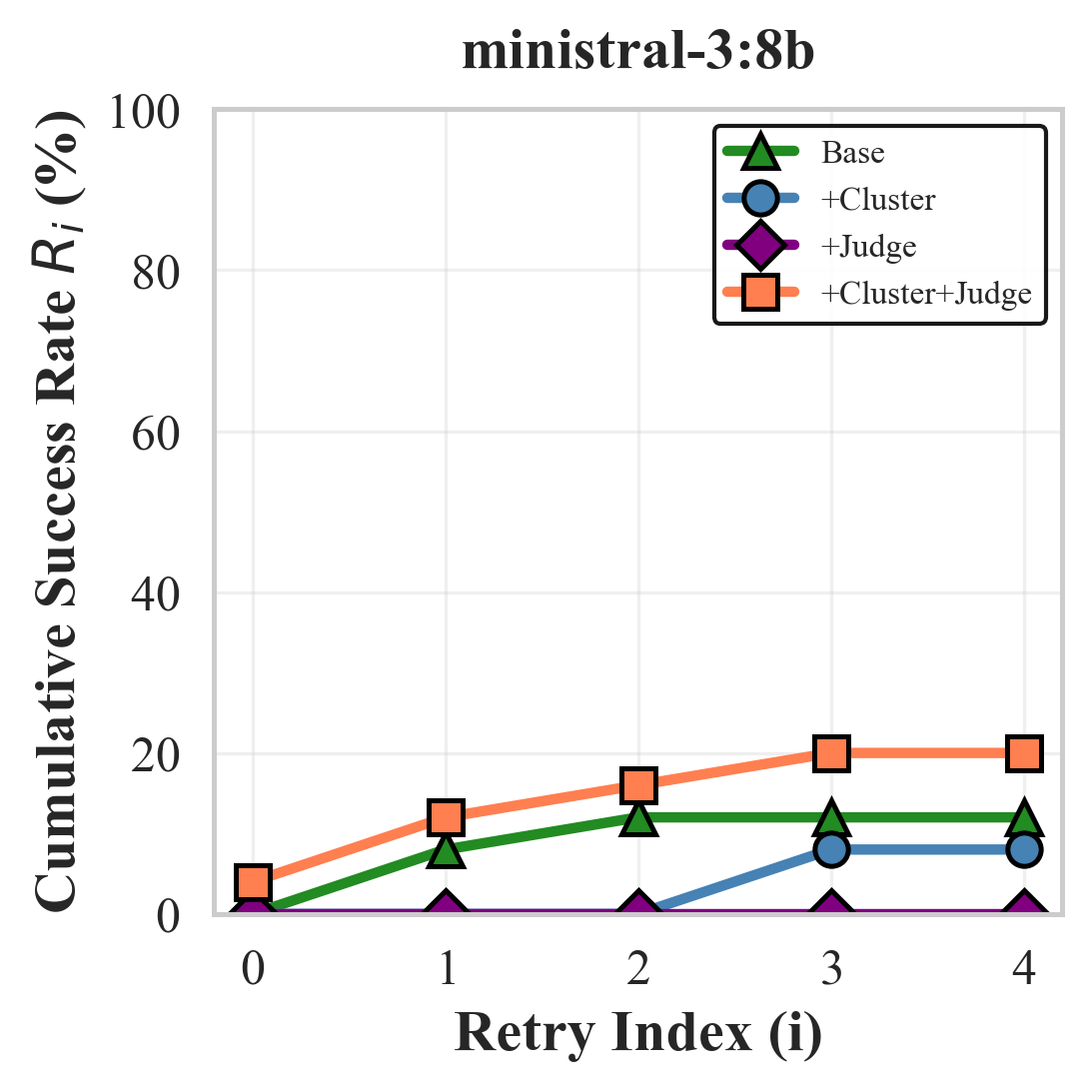}
        \label{fig:p4_2}
    \end{subfigure}
    \hfill
    \begin{subfigure}[b]{0.30\textwidth}
        \centering
        \includegraphics[width=\textwidth]{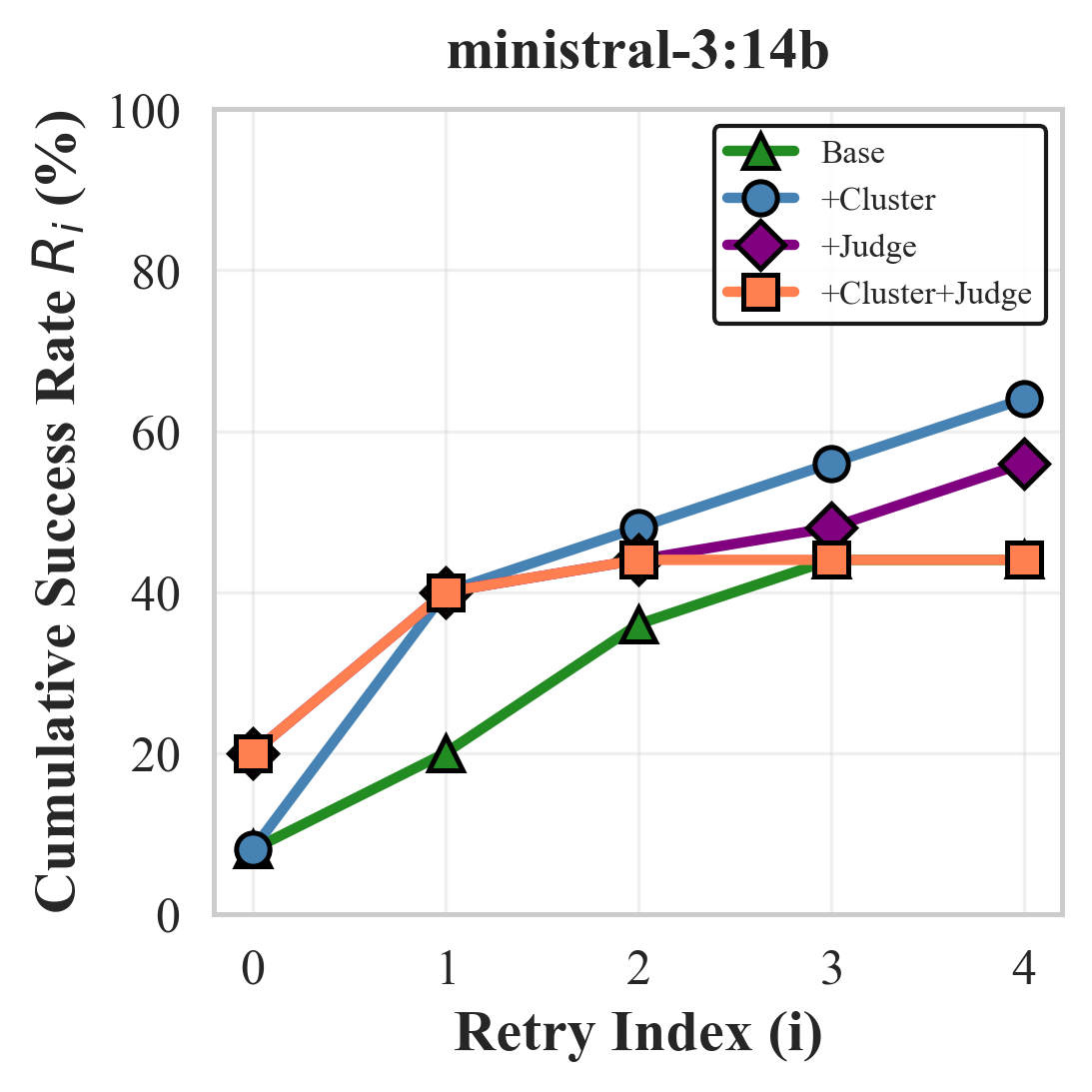}
        \label{fig:p4_3}
    \end{subfigure}

    \vspace{0.2cm}

    \begin{subfigure}[b]{0.30\textwidth}
        \centering
        \includegraphics[width=\textwidth]{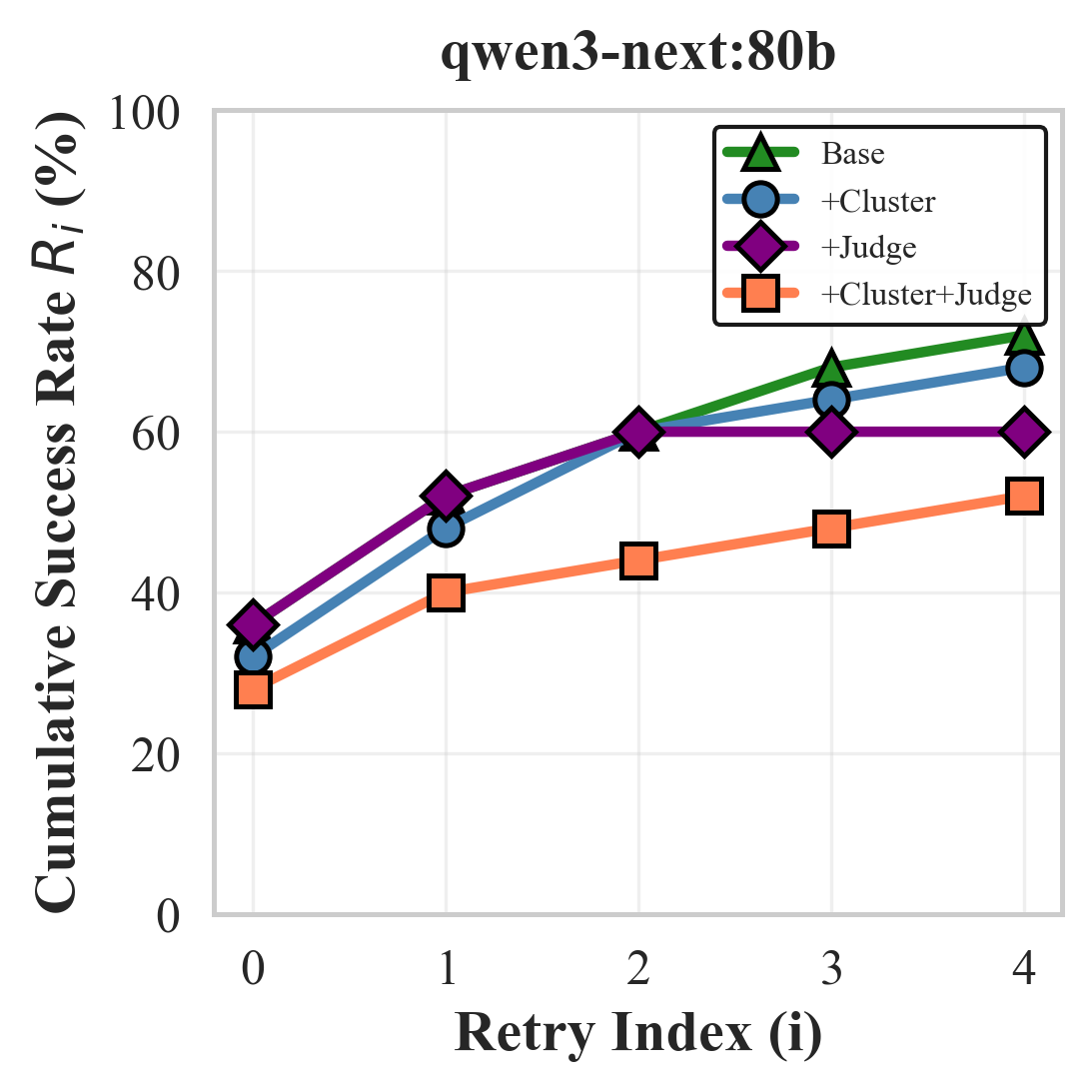}
        \label{fig:p4_4}
    \end{subfigure}
    \hfill
    \begin{subfigure}[b]{0.30\textwidth}
        \centering
        \includegraphics[width=\textwidth]{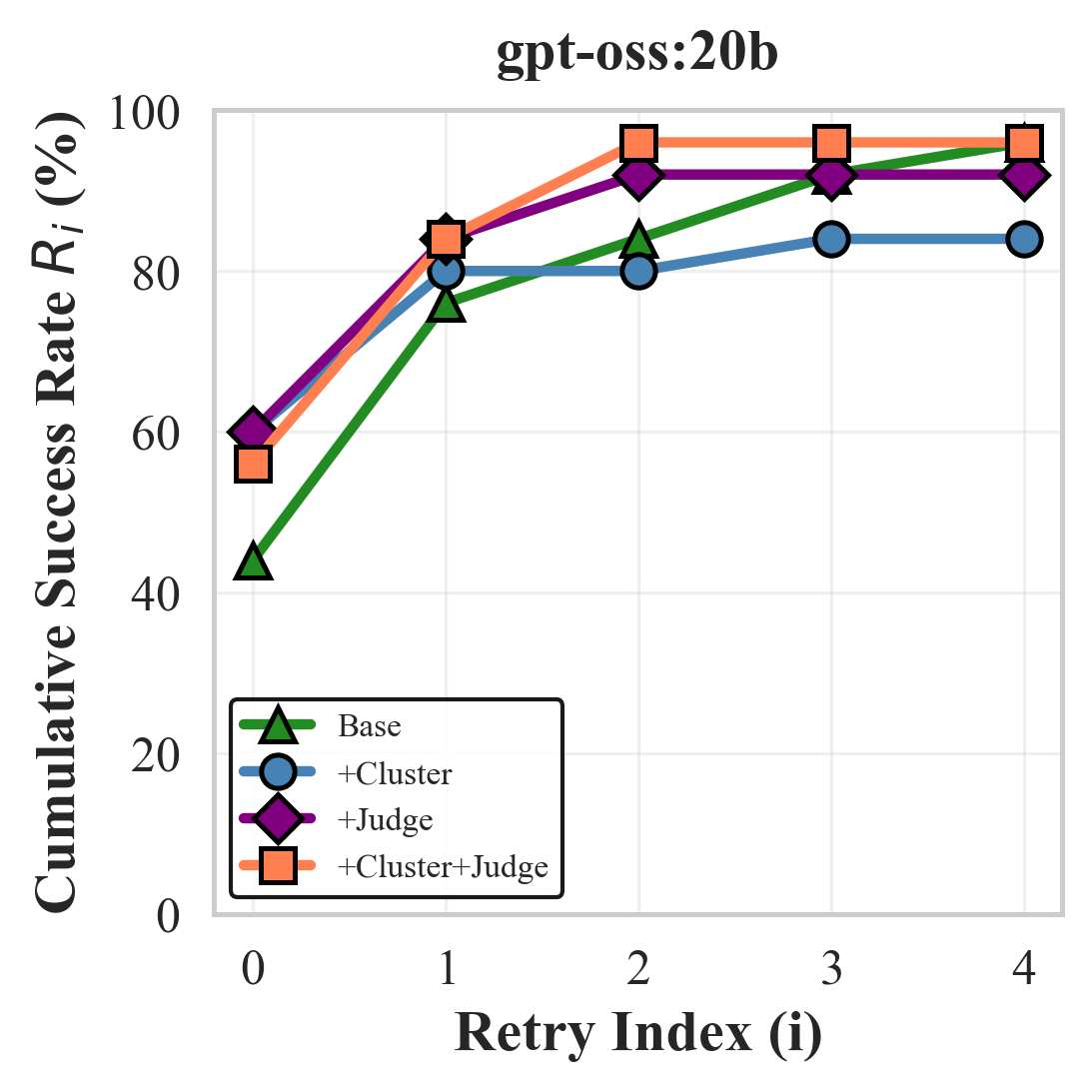}
        \label{fig:p4_5}
    \end{subfigure}
    \hfill
    \begin{subfigure}[b]{0.30\textwidth}
        \centering
        \includegraphics[width=\textwidth]{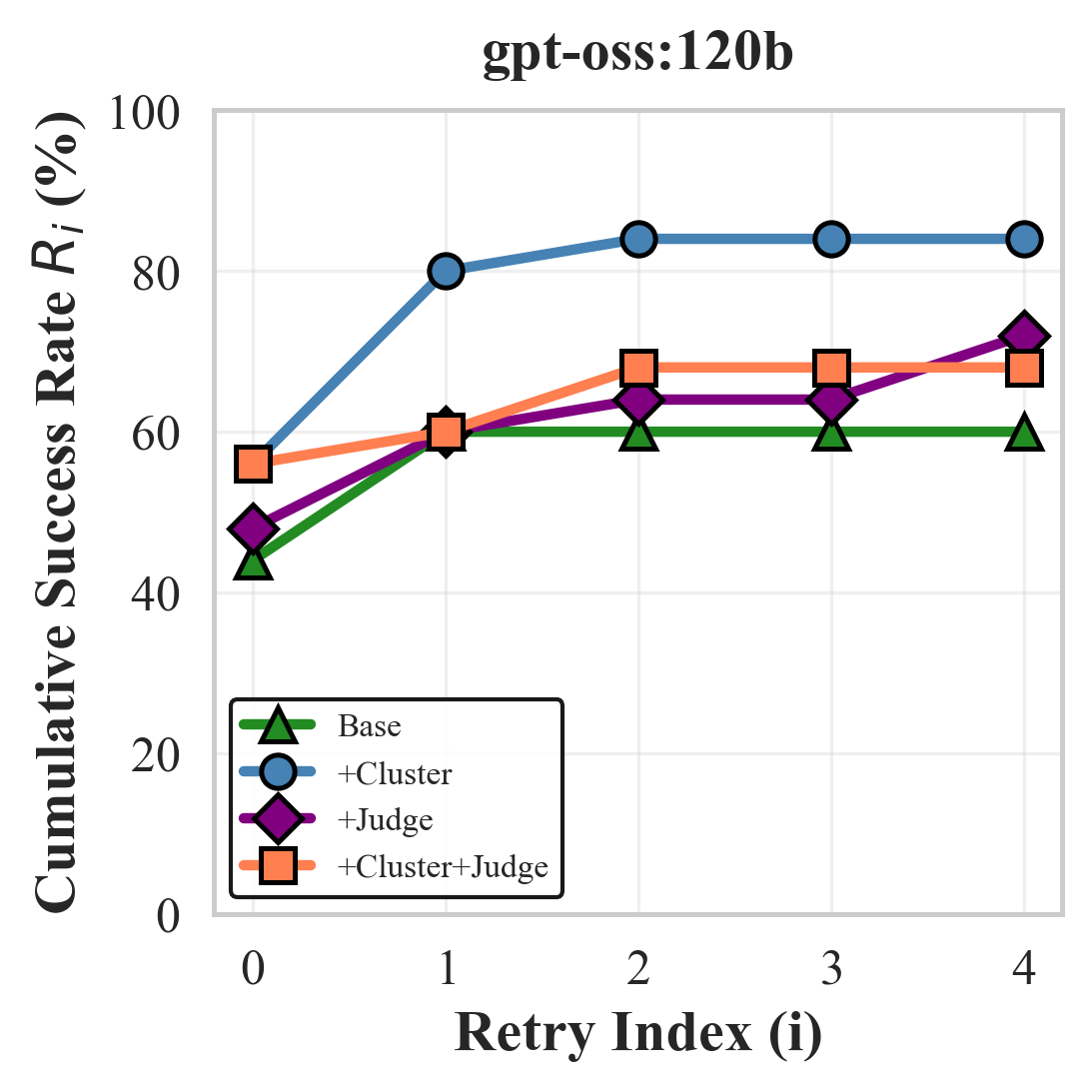}
        \label{fig:p4_6}
    \end{subfigure}

    \caption{Cumulative success rate of OML generation across four repair attempts ($n=25$, \textbf{P4}).}
    \label{fig:success_rates_p4}
\end{figure*}

\begin{figure*}[!htbp]
    \centering
    \captionsetup[subfigure]{justification=centering}
    
    \begin{subfigure}[b]{0.30\textwidth}
        \centering
        \includegraphics[width=\textwidth]{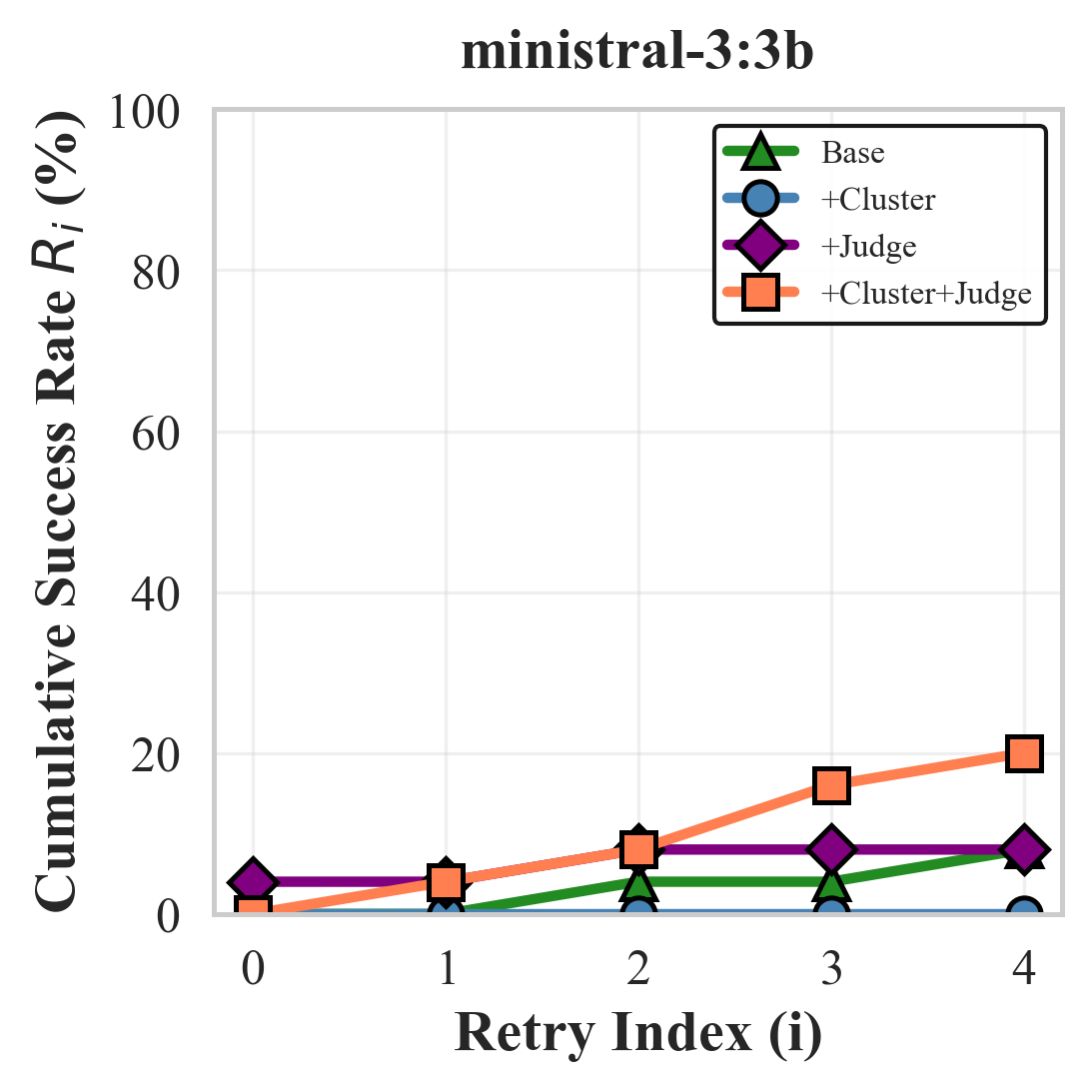}
        \label{fig:p5_1}
    \end{subfigure}
    \hfill
    \begin{subfigure}[b]{0.30\textwidth}
        \centering
        \includegraphics[width=\textwidth]{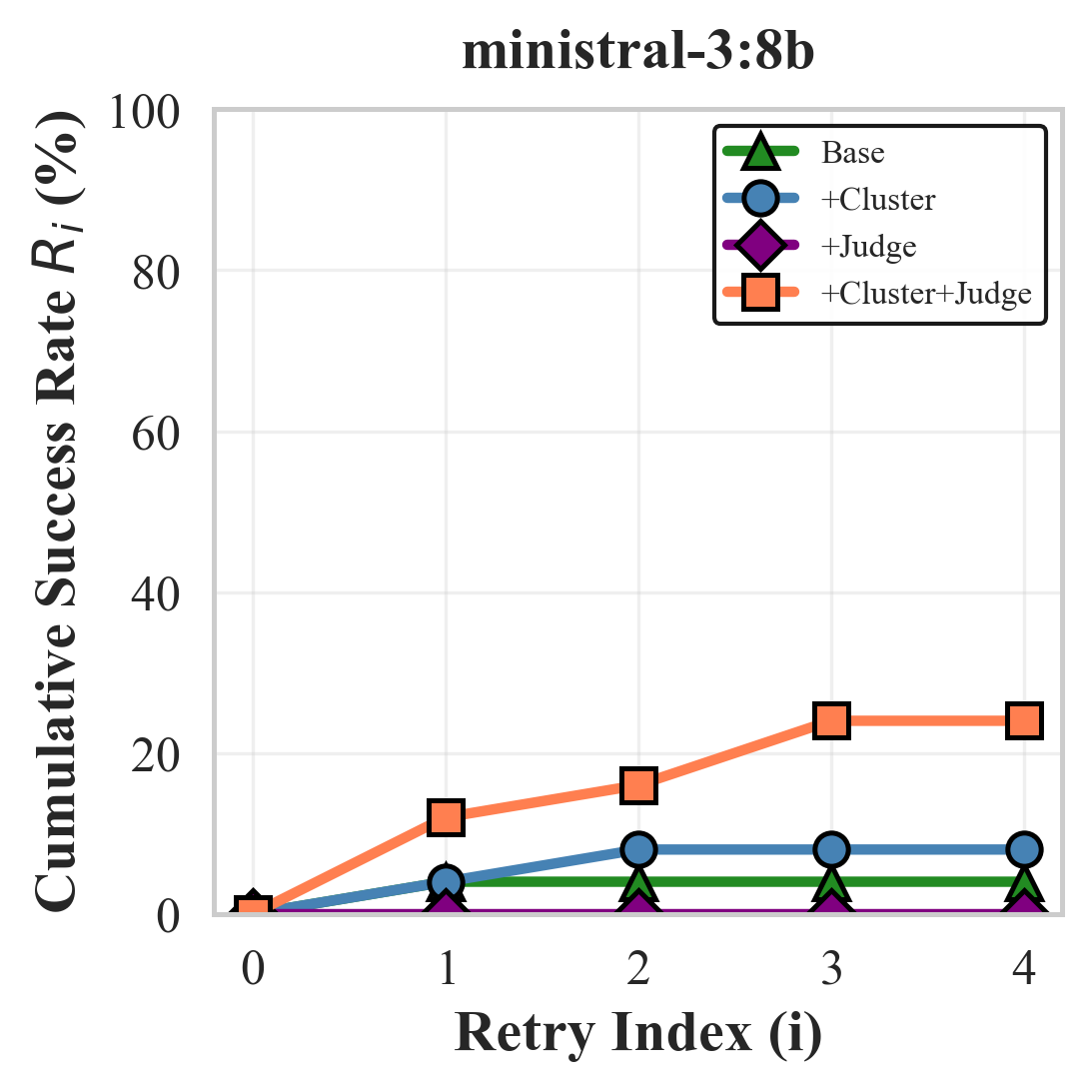}
        \label{fig:p5_2}
    \end{subfigure}
    \hfill
    \begin{subfigure}[b]{0.30\textwidth}
        \centering
        \includegraphics[width=\textwidth]{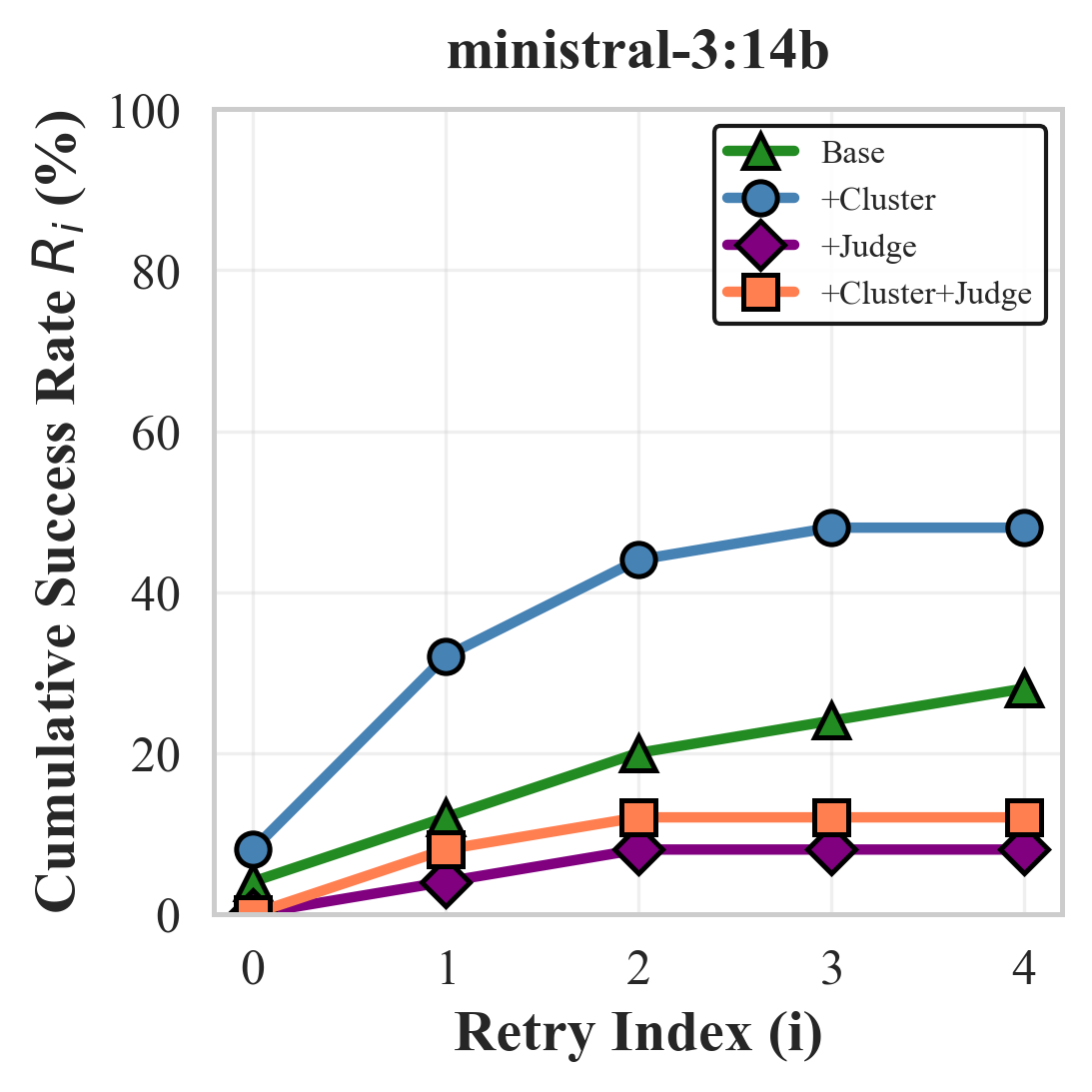}
        \label{fig:p5_3}
    \end{subfigure}

    \vspace{0.2cm}

    \begin{subfigure}[b]{0.30\textwidth}
        \centering
        \includegraphics[width=\textwidth]{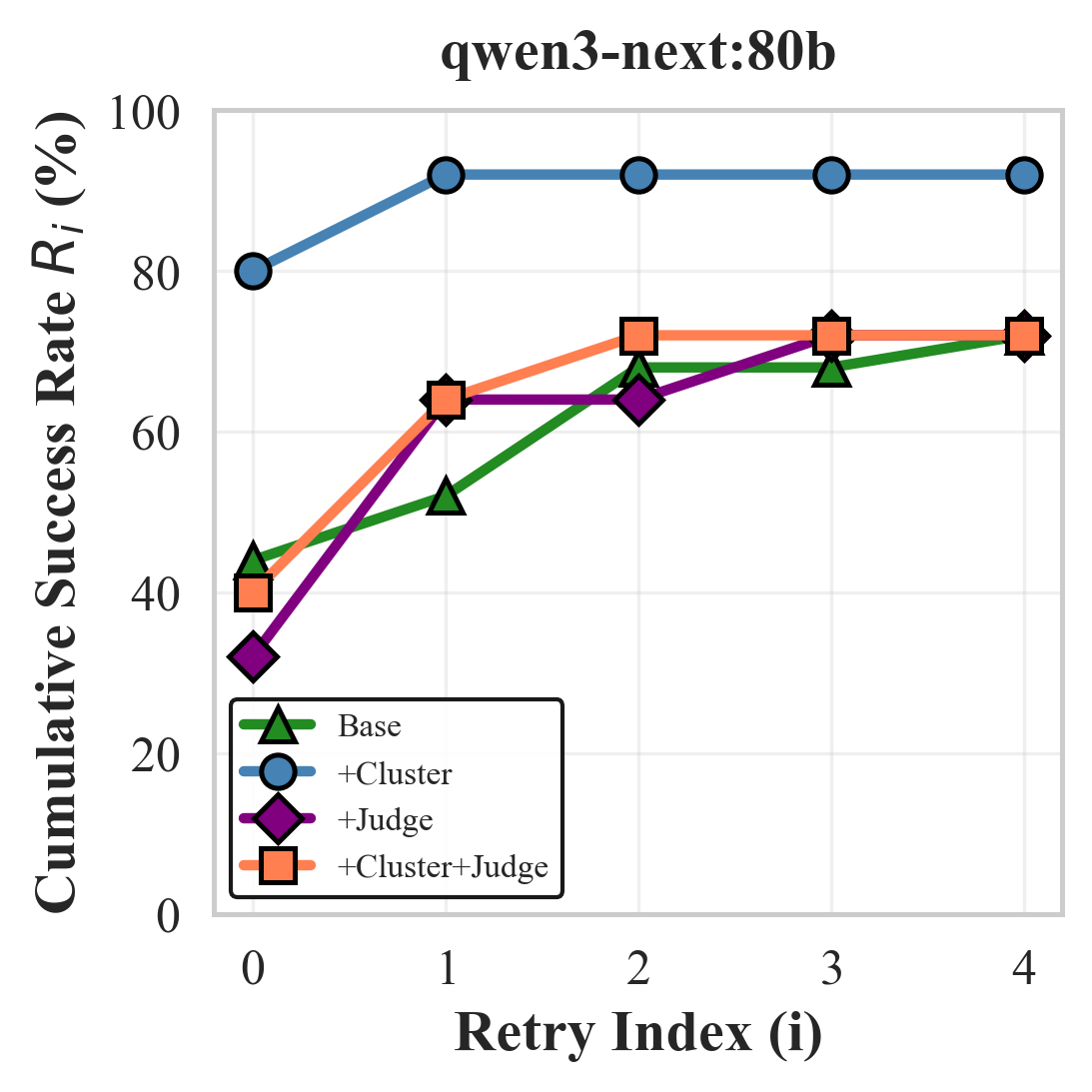}
        \label{fig:p5_4}
    \end{subfigure}
    \hfill
    \begin{subfigure}[b]{0.30\textwidth}
        \centering
        \includegraphics[width=\textwidth]{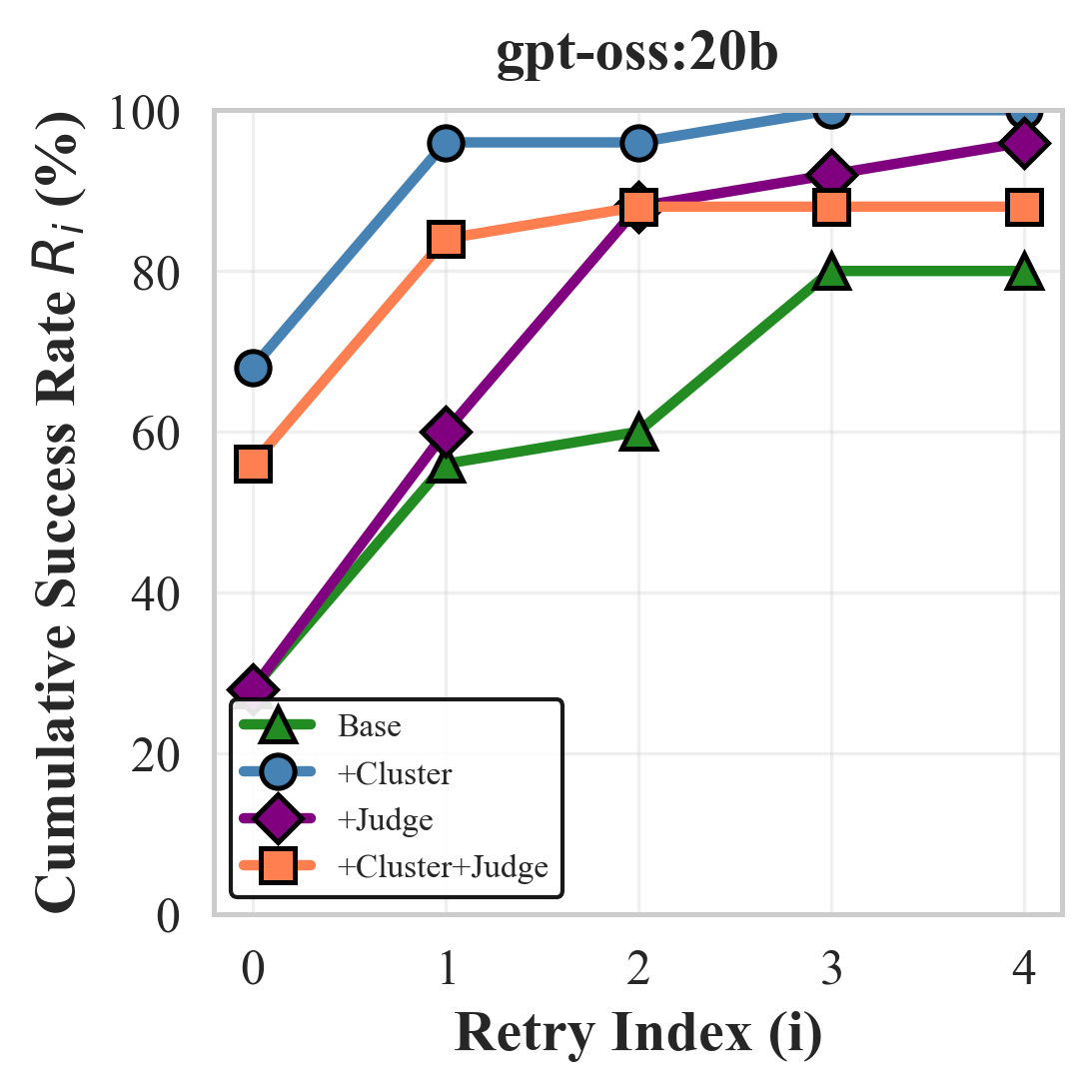}
        \label{fig:p5_5}
    \end{subfigure}
    \hfill
    \begin{subfigure}[b]{0.30\textwidth}
        \centering
        \includegraphics[width=\textwidth]{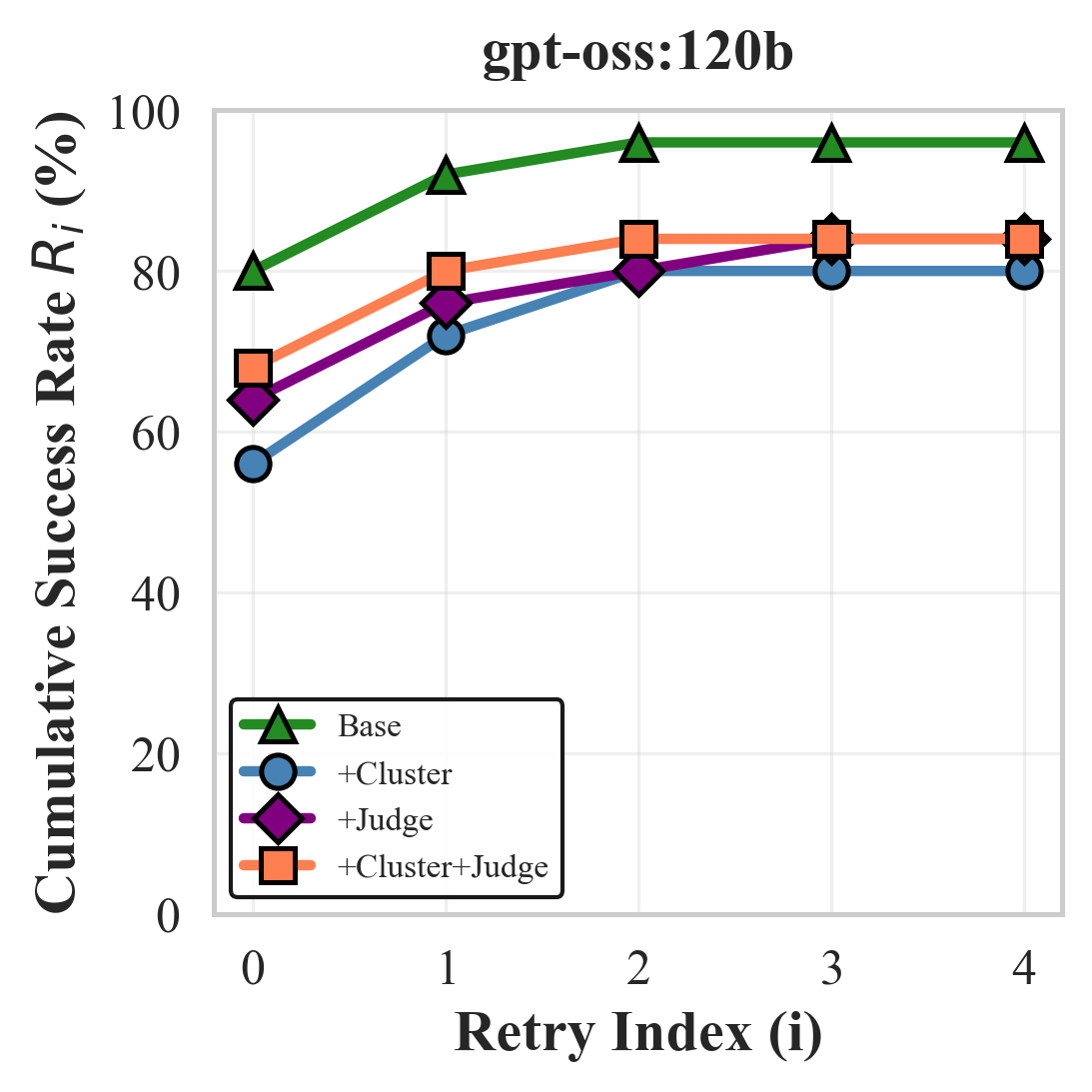}
        \label{fig:p5_6}
    \end{subfigure}

    \caption{Cumulative success rate of OML generation across four repair attempts ($n=25$, \textbf{P5}).}
    \label{fig:success_rates_p5}
\end{figure*}